\documentclass[final,5p,times,twocolumn,sort&compress]{elsarticle}

\usepackage{epsfig}
\usepackage{amssymb}
\usepackage{amsthm}
\usepackage{color}
\usepackage{graphicx}
\usepackage[table]{xcolor}
\usepackage{lscape}
\usepackage{caption}
\usepackage{widetext}
\usepackage{flushend}
\usepackage{cuted}
\usepackage{soul}

\journal{Journal of Magnetism and Magnetic Materials}

\begin{document}

\begin{frontmatter}



\title{Reentrant phase transitions of a coupled spin-electron model on doubly decorated planar lattices with two or three consecutive critical points}


\author{Hana \v Cen\v carikov\'a\corref{cor1}\fnref{label1}}
\cortext[cor1]{Corresponding author:}
\ead{hcencar@saske.sk}
\author{Jozef Stre\v{c}ka\fnref{label2}}
\author{Marcelo L. Lyra\fnref{label3}}

\address[label1]{Institute  of  Experimental  Physics,  Slovak   Academy   of Sciences, Watsonova 47, 040 01 Ko\v {s}ice, Slovakia}
\address[label2]{Department of Theoretical Physics and Astrophysics, Faculty of Science, P.~J. \v{S}af\' arik University, Park Angelinum 9, 040 01 Ko\v{s}ice, Slovakia}
\address[label3]{Instituto de Fisica, Universidade Federal de Alagoas, 57072-970 Macei\'o, AL, Brazil}
\begin{abstract}
The generalized decoration-iteration transformation is adapted for the exact study of a coupled spin-electron model on 2D lattices in which localized Ising spins reside on nodal lattice sites and mobile electrons are delocalized  over pairs of decorating sites. The model takes into account a hopping term for mobile electrons, the Ising coupling between mobile electrons and localized spins as well as the Ising coupling between localized spins ($J'$). The ground state, spontaneous magnetization and specific heat are examined for both ferromagnetic ($J'>0$) as well as antiferromagnetic ($J'<0$) interaction between the localized spins. Several kinds of reentrant transitions between the paramagnetic ($P$), antiferromagnetic ($AF$) and ferromagnetic ($F$) phases have been found either with a single critical point, or with two consecutive critical points ($P-AF$/$F-P$) and three successive critical points $AF/F-P-F/AF-P$. Striking thermal variations of the spontaneous magnetization depict a strong reduction due to the interplay between annealed disorder and quantum fluctuations in addition to the aforementioned reentrance. It is shown that the specific heat displays diverse thermal dependencies   including finite cusps at the critical temperatures.
\end{abstract}


\end{frontmatter}


\section{Introduction}
\label{s1}
During the past decades the more interesting and unconventional phenomena like the giant magnetoresistivity~\cite{McWhan,Helmolt,Chahara,Schiffer}, metal-insulator transitions~\cite{Wachter}, itinerant ferromagnetism~\cite{Kanamori,Koster,Li}, metamagnetic transitions~\cite{Honecker, Siemens, Kikuchi},  mixed-valence phenomena~\cite{Wachter},  enhanced magnetocaloric effect~\cite{Gschneidner1,Gschneidner2,Gschneidner3,Bruck}, superconductivity~\cite{Takada,Kamihara,Kamihara2}, electronic ferroelectricity~\cite{Kimura, Ikeda}, or multiferroicity~\cite{Cheong,Brink}  have attracted much attention of experimental as well as theoretical physicists  with the aim to describe and explain the origin of these intriguing phenomena.  In spite of enormous efforts, some of these phenomena still lack full understanding and have not been reliably explained so far. Most of the aforementioned  collective phenomena arise from the mutual interaction between mobile and localized electrons, which is of a highly cooperative nature with many active degrees of freedom. In general, it is a very complex task to solve this kind of interacting many-body problem, which requires application of various alternative approaches. One of them, currently in foreground, is the application of an appropriate type of a numerical method, which can be either the exact (e.g., the exact diagonalization~\cite{Lanczos, Dagotto}) or approximate (e.g., the Density Matrix Renormalization Group ~\cite{Malvezzi,White1,White2} or Monte Carlo methods~\cite{Newman,MC2}). Another approach, which enables to investigate the physical properties of mutually coupled electron and spin subsystems, is based on  first-principle calculations  known as a Density Functional Theory~\cite{Engel}. Although both aforementioned approaches are very useful, there are serious limitations when applying them to coupled spin-electron systems owing to 
restrictions related to CPU time, machine memory or other computational difficulties associated with finite-size effects. 

Recently, another fascinating approach has been suggested by Pereira {\it et al}.~\cite{Pereira1,Pereira2} when  applying a relatively simple analytical method based on the generalized decoration-iteration transformation to an interacting spin-electron system on a diamond chain. Following Fisher's ideas~\cite{Fisher}, an arbitrary statistical-mechanical system (even of quantum nature), which merely interacts with either two or three outer Ising spins, may be replaced with the effective interactions between the outer Ising spins through the generalized decoration-iteration or star-triangle mapping transformations~\cite{Fisher,Syozi}. The procedure elaborated by
Pereira {\it et al}.~\cite{Pereira1,Pereira2} has been later adopted to other interacting spin-electron models in  one~\cite{Cisarova,Cisarova2,Lisnyi1,Lisnyi2,Nalbandyan,Galisova3,Galisova4,Galisova5} or two dimensions~\cite{Strecka0,Galisova,Galisova2,Doria}. The interest in this field of study has two different  reasons. The first, very prosaic, reason is that such relatively simple procedure allows us to obtain ground states  as well as thermodynamic characteristics for an  interesting class of two-component spin-electron systems.  These systems has gained a new direction with the realization of optical lattices~\cite{Greiner}, namely, they represent a theoretical counterpart of possible experimental realizations of  coupled spin-electron systems on decorated lattices that  allow us to study many types of collective phenomena in their pure nature~\cite{Mukherjee,Vicencio,Noda}. The second, more crucial, reason  relates to the possibility to provide important hints on fundamental questions concerning the origin of phenomena such as the giant magnetocaloric effect, metamagnetic transitions, reentrant phase transitions, etc.%

    The organization of this paper is as follows. In Sec.~\ref{model}, we will first introduce  a coupled spin-electron model on 2D lattices in which localized Ising spins reside on nodal lattice sites and mobile electrons are delocalized  over pairs of decorating sites, together with the crucial steps of an exact mapping procedure, which has been used to obtain exact  closed-form expressions for the critical temperature, order parameter and other relevant thermodynamic characteristics. The most interesting results for the ground state and  the finite-temperature phase diagrams are collected in Sec.~\ref{Results} along with thermal dependencies of magnetization and specific heat. In this section, our attention will be also focused on the possibility of observing  reentrant phase transitions. Finally, some concluding remarks are drawn in Sec.~\ref{Conclusion}.

\section{Model and Method}
\label{model}
\begin{figure}[h!]
\begin{center}
\includegraphics[width=7cm,height=3.5cm]{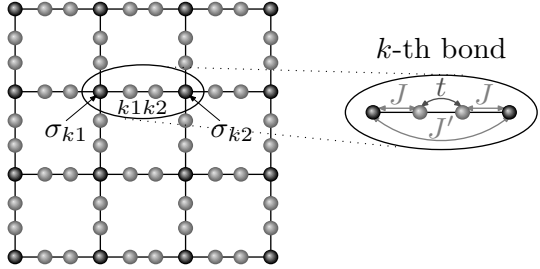}
\caption{\small A schematic representation of the studied spin-electron model on a doubly decorated square lattice. Black balls represent the nodal sites occupied by the localized Ising spins and gray balls represent the decorating sites occupied by at most four mobile electrons per dimer. The ellipse illustrates the $k$-th bond.}
\label{fig1}
\end{center}
\end{figure}

Let us investigate an interacting spin-electron system on doubly decorated 2D lattices, as displayed in Fig.~\ref{fig1}. The investigated model contains one localized Ising spin at each nodal lattice site and a set of mobile electrons delocalized over the pairs of decorating sites (dimers) placed at each bond (see Fig.~\ref{fig1}). From the experimental point of view, such coupled spin-electron models could capture physical properties of polymeric coordination compounds like [Ru$_2$(OOCtBu)$_4$]$_3$[M(CN)$_6$] (M=Fe, Cr)~\cite{Polymeric}.
\\
In the present model, each decorating site involves a $s$-type orbital, which can be occupied by at most two electrons with opposite spins  in accordance with the Pauli's exclusion principle. The electron motion described by the hopping amplitude $t$ is allowed just between nearest-neighbour decorating sites. Each localized Ising spin interacts with the  nearest-neighbour mobile electrons through the Ising coupling $J$. In the present work we go beyond the previous studies~\cite{Doria,cenci1} by taking into account the additional Ising coupling $J'$ between nearest-neighbour localized Ising spins.  We suppose that this interaction is not negligible and it can lead to a new physics in the coupled spin-electron systems due to a competition with the effective coupling originating from the hopping process of the mobile electrons. The total Hamiltonian of the interacting spin-electron system on such doubly decorated 2D lattice can be written as follows
\begin{eqnarray}
\hat{\cal H}\!\!\!&=&\!\!\!-t\sum_{k=1}^{Nq/2}\sum_{\gamma=\uparrow,\downarrow}(\hat{c}^\dagger_{k1,\gamma}\hat{c}_{k2,\gamma}+h.c.) 
\nonumber\\
\!\!\!&-&\!\!\!J\sum_{\langle ik\rangle}\hat{\sigma}^z_{i}(\hat{n}_{k\alpha,\uparrow}-\hat{n}_{k\alpha,\downarrow})
-J'\sum_{\langle kl\rangle}\hat{\sigma}^z_{k}\hat{\sigma}^z_{l}
\; ,
\label{eq1}
\end{eqnarray}
where $\hat{c}^\dagger_{k\alpha,\gamma}$ and $\hat{c}_{k\alpha,\gamma}$ ($\alpha$=1,2) are the creation and annihilation fermionic operators for the mobile electrons delocalized over the $k$-th decorating dimer, $\hat{n}_{k\alpha,\gamma}=\hat{c}^\dagger_{k\alpha,\gamma}\hat{c}_{k\alpha,\gamma}$ and $\hat{n}_{k\alpha}=\hat{n}_{k\alpha,\uparrow}+\hat{n}_{k\alpha,\downarrow}$ are the respective number operators. The operator $\hat\sigma^z_{i}$ denotes the $z$-component of the spin-1/2 operator with eigenvalues $\sigma^z_{i}=\pm1$. The first term in Eq.~(\ref{eq1}) describes  the quantum-mechanical hopping of mobile electrons delocalized over a couple of decorating  sites $k1$ and $k2$ from the $k$-th dimer. The second term represents the Ising-type exchange interaction between the mobile electrons and their nearest Ising neighbours. The last term  in  Eq.~(\ref{eq1}) represents a further-neighbour Ising-type exchange interaction between the nearest-neighbour localized Ising spins.
  The symbol $N$  denotes the total number of localized Ising spins and $q$ is their coordination number. For a practical reason, we may rewrite the total Hamiltonian~(\ref{eq1}) in the form of a sum over bond Hamiltonians $\hat{\cal H}=\sum_{k=1}^{Nq/2}\hat{\cal H}_k$, where each bond Hamiltonian $\hat{\cal H}_k$ accounts for the kinetic energy of mobile electrons from the $k$-th decorating dimer, the exchange interaction between the mobile electrons and  their nearest-neighbour Ising spins, and finally, the exchange interaction between two  localized Ising spins attached to the $k$-th decorating dimer
\begin{eqnarray}
\hat{\cal H}_k=\!\!\!&-&\!\!\!t(\hat{c}^\dagger_{k1,\uparrow}\hat{c}_{k2,\uparrow}+\hat{c}^\dagger_{k1,\downarrow}\hat{c}_{k2,\downarrow}+
\hat{c}^\dagger_{k2,\uparrow}\hat{c}_{k1,\uparrow}+\hat{c}^\dagger_{k2,\downarrow}\hat{c}_{k1,\downarrow})
\nonumber
\\
\!\!\!&-&\!\!\!J\hat\sigma^z_{k1}(\hat{n}_{k1,\uparrow}-\hat{n}_{k1,\downarrow})-
J\hat\sigma^z_{k2}(\hat{n}_{k2,\uparrow}-\hat{n}_{k2,\downarrow})
\nonumber\\
\!\!\!&-&\!\!\!J'\hat\sigma^z_{k1}\hat\sigma^z_{k2}\;.
\label{eq2}
\end{eqnarray}
Let us calculate the  grand-canonical partition function $\Xi$ of the correlated spin-electron system, 
which will allow us to rigorously analyze the ground-state as well as the thermodynamic properties. In general, the calculation of the grand-canonical partition function is a difficult problem, but for the model defined by Eq.~(\ref{eq1}) there exists an elegant way due to the commutativity between  different bond Hamiltonians $[\hat{\cal H}_{i},\hat{\cal H}_j]$=0. This fact allows us to  partially factorize the grand-canonical partition function into the product of bond partition functions $\Xi_k$
\begin{eqnarray}
\Xi=\sum_{\{\sigma_i\}}\prod_{k=1}^{Nq/2}\mbox{Tr}_k\exp(-\beta\hat{\cal H}_k)\exp(\beta\mu\hat{n}_k)=\sum_{\{\sigma_i\}}\prod_{k=1}^{Nq/2}\Xi_k\;,
\label{eq3}
\end{eqnarray}
where $\beta=1/(k_BT)$, $k_B$ is the Boltzmann's constant, $T$ is the absolute temperature, $\hat{n}_k=\hat{n}_{k1}+\hat{n}_{k2}$ is the total number operator for mobile electrons delocalized over the $k$-th decorating dimer and $\mu$ is their chemical potential. The summation $\sum_{\{\sigma_i\}}$ in Eq.~(\ref{eq3}) runs over all possible configurations of the nodal Ising spins and the symbol Tr$_k$ stands for the trace over the degrees of freedom of the mobile electrons from the $k$-th decorating dimer. It is evident from this notation that it is necessary to diagonalize the bond Hamiltonian $\hat{\cal H}_k$ in order  to find the grand-canonical partition function. Apparently, the bond Hamiltonian $\hat{\cal H}_k$ commutes with the total number of mobile electrons per bond ($\hat{n}_k$) as well as the $z$-component of the total spin of the mobile electrons  $\hat{S}_k^z=\sum_{\alpha={\{1,2\}}}(\hat{n}_{k\alpha,\uparrow}-\hat{n}_{k\alpha,\downarrow})$. Hence, it follows that the matrix form of the bond Hamiltonian $\hat{\cal H}_k$ can be divided  into several disjoint blocks ${\cal H}_k(n_k,S^z_k)$ corresponding to orthogonal Hilbert subspaces characterized by different numbers of mobile electrons ($n_k$) and different values of the total spin $S^z_k$. Since the further-neighbour exchange coupling $J'$ between the localized Ising spins (the last term in Eq.~(\ref{eq2})) is independent of the mobile electrons,  it is sufficient to diagonalize the reduced form of the bond Hamiltonian ${\cal \widetilde{H}}_k$ including only the first two terms depending on the mobile electrons, whereas the obtained eigenvalues  ($\widetilde{E}_k$) must be subsequently extended by an additional term  $-J'\sigma_{k1}\sigma_{k2}$ in order to obtain the eigenvalues $E_k$ of the full bond Hamiltonian ${\cal H}_k$. \\
\underline{\it (a) The subspace with $n_k=0$.}\\
This subspace includes only one basis state, and namely, $|0,0\rangle_k$ (the notation $|0,0\rangle_k$ denotes the vacuum state of the pair of decorating sites on the $k$-th bond)  with $S^z_k=0$. Thus, the corresponding block Hamiltonian has the form ${\cal \widetilde{H}}_k(0,0)=0$, which gives the eigenvalue
\begin{eqnarray}
E_{k1}=-J'\sigma_{k1}\sigma_{k2}\;.
\label{eq4}
\end{eqnarray}
\\
\underline{\it (b) The subspace with $n_k=1$.}\\
In this case, four basis states $\hat{c}^\dagger_{k1,\uparrow}|0,0\rangle_k$, $\hat{c}^\dagger_{k2,\uparrow}|0,0\rangle_k$ $\hat{c}^\dagger_{k1,\downarrow}|0,0\rangle_k$, $\hat{c}^\dagger_{k2,\downarrow}|0,0\rangle_k$ lead to two different $2\times2$  block Hamiltonians with the total spin $S^z_k=\pm1$
\begin{eqnarray}
{\cal \widetilde{H}}_k(1,\pm1)=
\left( 
\begin{array}{cc}
\mp J\sigma_{k1} & -t\\
-t & \mp J\sigma_{k2} 
\end{array}
\right)\;,
\label{eq5}
\end{eqnarray}
which give, after the direct diagonalization, the following eigenvalues for $S^z_k=1$
\begin{eqnarray}
E_{k2,k3}\!\!\!&=&\!\!\!-\frac{J(\sigma_{k1}+\sigma_{k2})}{2}\pm\frac{\sqrt{J^2(\sigma_{k1}-\sigma_{k2})^2+4t^2}}{2}
\nonumber\\
\!\!\!&-&\!\!\!J'\sigma_{k1}\sigma_{k2} 
\;,
\label{eq6}
\end{eqnarray}
and for $S^z_k=-1$
\begin{eqnarray}
E_{k4,k5}\!\!\!&=&\!\!\!\frac{J(\sigma_{k1}+\sigma_{k2})}{2}\pm\frac{\sqrt{J^2(\sigma_{k1}-\sigma_{k2})^2+4t^2}}{2}
\nonumber\\
\!\!\!&-&\!\!\!J'\sigma_{k1}\sigma_{k2} 
\;.
\label{eq7}
\end{eqnarray}
\underline{\it (c) The subspace with $n_k=2$.}\\
In this Hilbert subspace, three different block Hamiltonians can be discerned according to the value of the total spin $S^z_k$ available to the basis states  $\hat{c}^\dagger_{k1,\gamma}\hat{c}^\dagger_{k2,\gamma}|0,0\rangle_k$. For two mobile electrons with equally oriented spins $S^z_k=\pm2$  the bond Hamiltonians ${\cal \widetilde{H}}_k$ take the following form 
\begin{eqnarray}
{\cal \widetilde{H}}_k(2,\pm2)=
\mp J(\sigma_{k1}+\sigma_{k2}) \;,
\label{eq8}
\end{eqnarray}
which directly give other two eigenvalues
\begin{eqnarray}
E_{k6,k7}=\mp J(\sigma_{k1}+\sigma_{k2})-J'\sigma_{k1}\sigma_{k2}\;.
\label{eq9}
\end{eqnarray}
The Hilbert subspace spanned over the four basis states $\hat{c}^\dagger_{k1,\uparrow}\hat{c}^\dagger_{k2,\downarrow}|0,0\rangle_k$,  $\hat{c}^\dagger_{k1,\downarrow}\hat{c}^\dagger_{k2,\uparrow}|0,0\rangle_k$, $\hat{c}^\dagger_{k1,\uparrow}\hat{c}^\dagger_{k1,\downarrow}|0,0\rangle_k$,  $\hat{c}^\dagger_{k2,\uparrow}\hat{c}^\dagger_{k2,\downarrow}|0,0\rangle_k$, with the opposite orientation of two mobile electrons leads to the following block Hamiltonian  ${\cal \widetilde{H}}_k$ for the particular case with $S^z_k=0$
\begin{eqnarray}
{\cal \widetilde{H}}_k(2,0)=
\left( 
\resizebox{5.4cm}{!}{$\displaystyle 
\begin{array}{cccc}
-J(\sigma_{k1}-\sigma_{k2}) & 0 & -t & -t\\
 0 & J(\sigma_{k1}-\sigma_{k2}) & t & t \\
 -t & t & 0 & 0\\
 -t & t & 0 & 0
\end{array}
$}
\right)\;,
\label{eq10}
\end{eqnarray}
which gives the following four eigenvalues
\begin{eqnarray}
E_{k8,k9}\!\!\!&=&\!\!\!-J'\sigma_{k1}\sigma_{k2}\;.
\label{eq11}\\
E_{k10,k11}\!\!\!&=&\!\!\!\pm\sqrt{J^2(\sigma_{k1}-\sigma_{k2})^2+4t^2}-J'\sigma_{k1}\sigma_{k2}\;.
\label{eq12}
\end{eqnarray}
\underline{\it (d) The subspace with $n_k=3$.}\\
There are four different basis states\- $\hat{c}^\dagger_{k1,\uparrow}\hat{c}^\dagger_{k1,\downarrow}\hat{c}^\dagger_{k2,\uparrow}|0,0\rangle_k$, $\hat{c}^\dagger_{k1,\uparrow}\hat{c}^\dagger_{k1,\downarrow}\hat{c}^\dagger_{k2,\downarrow}|0,0\rangle_k$, $\hat{c}^\dagger_{k1,\uparrow}\hat{c}^\dagger_{k2,\uparrow}\hat{c}^\dagger_{k2,\downarrow}|0,0\rangle_k$ and $\hat{c}^\dagger_{k1,\downarrow}\hat{c}^\dagger_{k2,\uparrow}\hat{c}^\dagger_{k2,\downarrow}|0,0\rangle_k$, which form two different $2\times2$ block Hamiltonians
\begin{eqnarray}
{\cal \widetilde{H}}_k(3,\pm1)=
\left( 
\begin{array}{cc}
\mp J\sigma_{k2}  & t\\
t & \mp J\sigma_{k1} 
\end{array}
\right)\;.
\label{eq13}
\end{eqnarray}
After the direct diagonalization of (\ref{eq13}) for $S^z_k=1$, one obtains the eigenvalues
\begin{eqnarray}
E_{k12,k13}\!\!\!&=&\!\!\!-\frac{J(\sigma_{k1}+\sigma_{k2})}{2}\pm\frac{\sqrt{J^2(\sigma_{k1}-\sigma_{k2})^2+4t^2}}{2}
\nonumber\\
\!\!\!&-&\!\!\!J'\sigma_{k1}\sigma_{k2}
\;,
\label{eq14}
\end{eqnarray}
and for $S^z_k=-1$
\begin{eqnarray}
E_{k14,k15}\!\!\!&=&\!\!\!\frac{J(\sigma_{k1}+\sigma_{k2})}{2}\pm\frac{\sqrt{J^2(\sigma_{k1}-\sigma_{k2})^2+4t^2}}{2}
\nonumber\\
\!\!\!&-&\!\!\!J'\sigma_{k1}\sigma_{k2}
\;.
\label{eq15}
\end{eqnarray}
It is evident that the energy spectrum for the particular case with three mobile electrons per decorating dimer is identical  to the particular case with one electron per decorating dimer. This reflects the particle-hole symmetry of the present model. 
\underline{\it (e) The subspace with $n_k=4$.}\\
Owing to the particle-hole symmetry, the system with $n_k=4$ is equivalent to the system without any electron. For this reason, the last eigenvalue for $n_k=4$ reads $E_{k16}=-J'\sigma_{k1}\sigma_{k2}$.

The sixteen eigenvalues can be straightforwardly used for the calculation of the bond grand-partition function $\Xi_k$. After tracing out the degrees of freedom of mobile electrons, the bond grand-partition function $\Xi_k$ depends only on the spin states of localized Ising spins, and one may employ the  generalized decoration-iteration transformation~\cite{Fisher,Syozi,Rojas} 
\begin{eqnarray}
\Xi_k\!\!\!&=&\!\!\!\sum_{i=1}^{16}\exp(-\beta E_{ki})\exp\left[\beta\mu n_k(E_{ki})\right]
\nonumber\\
\!\!\!&=&\!\!\!\exp(\beta J'\sigma_{k1}\sigma_{k2})\bigg\{1+4\cosh\left[\frac{\beta J}{2}(\sigma_{k1}+\sigma_{k2})\right]
\nonumber\\
\!\!\!&\times &\!\!\!
\cosh\left[\frac{\beta}{2}\sqrt{J^2(\sigma_{k1}-\sigma_{k2})^2+4t^2}\right](z+z^3)
\nonumber\\
\!\!\!&+&\!\!\! 2z^2\left\{ 1+\cosh\left[\beta J(\sigma_{k1}+\sigma_{k2})\right]\right.
\nonumber\\
\!\!\!&+ &\!\!\!
\left.
\cosh\left[\beta\sqrt{J^2(\sigma_{k1}-\sigma_{k2})^2+4t^2}\right]\right\} + z^4  \bigg\}
\nonumber\\
\!\!\!&=&\!\!\! A\exp(\beta R\sigma_{k1}\sigma_{k2})\;,
\label{eq16}
\end{eqnarray}
where $z=\exp(\beta\mu)$ is the fugacity  of the mobile electrons. The mapping parameters $A$ and $R$ are given by the "self-consistency" condition of the decoration-iteration transformation (\ref{eq16}), which must hold for all four combinations of two Ising spins $\sigma_{k1}$ and $\sigma_{k2}$. Using standard mathematical operations, one can obtain the following expressions
\begin{eqnarray}
A=(V_1V_2)^{1/2},\hspace*{1,5cm} \beta R=\frac{1}{2}\ln\left(\frac{V_1}{V_2}\right)+\beta J'\;,
\label{eq17}
\end{eqnarray}
where 
\begin{eqnarray}
V_1\!\!\!&=&\!\!\!1+4(z+z^3)\cosh(\beta J)\cosh(\beta t)
\nonumber\\
\!\!\!&+&\!\!\! 2z^2\left[ 1+\cosh(2\beta J)+
\cosh(2\beta t)\right]+z^4\; ,
\nonumber\\
V_2\!\!\!&=&\!\!\!1+4(z+z^3)\cosh\left(\beta\sqrt{J^2+t^2}\right)
\nonumber\\
\!\!\!&+&\!\!\! 2z^2\left[ 2+\cosh\left(2\beta\sqrt{J^2+t^2}\right)\right]+z^4\;.
\label{eq18}
\end{eqnarray}
By a straightforward substitution of the generalized decoration-iteration transformation (\ref{eq16}) into the expression (\ref{eq3}),  one  obtains a simple mapping relation between the grand-canonical partition function $\Xi$ of the interacting spin-electron system on doubly decorated 2D lattices and, respectively, the canonical partition function $Z_{IM}$ of a simple spin-1/2 Ising model on the corresponding undecorated lattice with an effective nearest-neighbour interaction $R$
\begin{eqnarray}
\Xi(\beta,J,J',t)=A^{Nq/2}Z_{IM}(\beta,R)\;.
\label{eq19}
\end{eqnarray}
The mapping parameter $A$ cannot cause a non-analytic behaviour of the grand-canonical partition function $\Xi$ and thus, the investigated spin-electron system becomes critical if and only if the corresponding Ising model becomes critical as well. The average number of mobile electrons per decorating dimer is given by 
\begin{eqnarray}
\rho\!\!\!&=&\!\!\!\langle n_k\rangle=\frac{z}{N}\frac{\partial}{\partial z}\ln\Xi
=z\frac{\partial}{\partial z}\ln A+z\varepsilon\frac{\partial}{\partial z}\beta R
\nonumber\\
\!\!\!&=&\!\!\!\frac{z}{2}\left(\frac{V'_1}{V_1}+\frac{V'_2}{V_2}\right)+\frac{z}{2}\varepsilon\left(\frac{V'_1}{V_1}-\frac{V'_2}{V_2}\right)
\;,
\label{eq20}
\end{eqnarray}
where
\begin{eqnarray}
V'_1\!\!\!&=&\!\!\!\frac{\partial V_1}{\partial z}=4(1+3z^2)\cosh\left(\beta J\right)\cosh(\beta t)
\nonumber\\
\!\!\!&+&\!\!\!4z\left[1+\cosh\left(2\beta J\right)+\cosh\left(2\beta t\right)\right]+4z^3\;,
\nonumber\\
V'_2\!\!\!&=&\!\!\!\frac{\partial V_2}{\partial z}=4(1+3z^2)\cosh\left(\beta\sqrt{J^2+t^2}\right)
\nonumber\\
\!\!\!&+&\!\!\!4z\left[2+\cosh\left(2\beta\sqrt{J^2+t^2}\right)\right]+4z^3\;.
\label{eq21}
\end{eqnarray}
The critical points of the coupled spin-electron system on doubly decorated 2D lattices can now be obtained from the expression (\ref{eq20}) for the average number of mobile electrons after taking into account the critical value of the nearest-neighbour pair correlation function $\varepsilon=\langle \sigma_{k1}\sigma_{k2}\rangle$ of the effective spin-1/2 Ising model along with the critical value of the effective coupling $\beta_cR$. Critical values of the effective coupling $\beta_cR$ and nearest-neighbour pair correlation functions $\varepsilon_c$  of the spin-1/2 Ising model on a few different 2D lattices  are listed in Tab.~\ref{tab1}.
\begin{table}[h!]
\begin{center}
\resizebox{0.48\textwidth}{!} {
\begin{tabular}{l||c|c}
lattice type & $\beta_cR$ & $\varepsilon_c^F$\\
\hline\hline
honeycomb & $\pm\frac{1}{2}\ln(2+\sqrt{3})$ & $\pm 4\sqrt{3}/9$\\
square & $\pm\frac{1}{2}\ln(1+\sqrt{2})$ & $\pm 1/\sqrt{2}$\\
kagome & $\frac{1}{4}\ln(3+2\sqrt{3})$& $(1+2\sqrt{3})/6$\\
triangular & $\frac{1}{4}\ln{3}$ & 2/3\\
diced & $\pm\frac{1}{2}\ln\left[\frac{1}{2}\left(1+\sqrt{3}+\sqrt[4]{12}\right)\right]$ & $\pm\frac{1}{3^{3/4}}\left(\sqrt{2/3}+1/\sqrt{2}\right)$\\
\hline
\end{tabular}}
\caption{\small Critical parameters for a few different planar Ising lattices with spin $\pm1$. 
The $\pm$ sign corresponds to the $F$(+)/$AF$(-) model.}
\label{tab1}
\end{center}
\end{table}
It is generally known that the critical values for  the $AF$ Ising model are the same (but of opposite signs) as for the $F$ ones on loose-packed lattices (e.g. honeycomb and square), while  the $AF$ Ising models on the close-packed lattices (like the triangular and kagome lattice) cannot exhibit criticality at non-zero temperatures.

The mapping relation (\ref{eq19}) between the partition functions allows us to study also  thermodynamic quantities, like the grand potential $\Omega$, the internal energy $U$, the entropy $S$ and the specific heat $C$ using the basic relations 
\begin{eqnarray}
\begin{array}{llll}
\Omega=-k_BT\ln \Xi\;, & &&U=-\displaystyle\frac{\partial \ln \Xi}{\partial \beta}\;,\\
S=-\displaystyle\left(\frac{\partial \Omega}{\partial T}\right)_z\;,& && C=\displaystyle\frac{\partial U}{\partial T}\;.
\end{array}
\label{eq22}
\end{eqnarray} 
In addition, we will concentrate our attention on a detailed analysis of the uniform and staggered magnetizations of the localized Ising spins  and the  mobile electrons, which can serve as order parameters for the $F$ and $AF$ states, respectively. Applying exact mapping theorems developed by Barry {\it et al}.~\cite{Barry1,Barry2,Barry3,Barry4}, the magnetization of nodal Ising spins equals to the magnetization of the corresponding spin-1/2  Ising model on the undecorated lattice
\begin{eqnarray}
m_i\equiv\frac{1}{2}\langle\hat{\sigma}^z_{k1}+\hat{\sigma}^z_{k2}\rangle=\frac{1}{2}\langle \hat{\sigma}^z_{k1}+\hat{\sigma}^z_{k2}\rangle_{IM}  \equiv m_{IM}\;.
\label{eq23}
\end{eqnarray}
The symbols $\langle\cdots\rangle$ and $\langle\cdots\rangle_{IM}$ denote the standard  ensemble average within the original spin-electron model and the effective Ising model, respectively. The magnetization of the Ising spins can be calculated from the following expressions
\begin{eqnarray}
m_{IM}=(1-P)^{1/8}\;,
\nonumber
\end{eqnarray}
\begin{eqnarray}
P=\left\{
\begin{array}{lll}
\displaystyle\frac{16y^3(1+y^3)}{(1-y)^3(1-y^2)^3}&$~\cite{Naya}$ & \mbox{(honeycomb)}\\\\
\displaystyle\frac{16y^4}{(1-y^2)^4}&$~\cite{Yang}$ & \mbox{(square)}\\\\
\displaystyle\frac{16y^6}{(1+3y^2)(1-y^2)^3}&$~\cite{Potts}$ & \mbox{(triangular),}
\end{array}
\right.
\label{eq24}
\end{eqnarray} 
where $y=\exp(-2\beta R)$. The total magnetization of mobile electrons per decorating dimer can be derived from the generalized Callen-Suzuki identity~\cite{Callen,Suzuki}
\begin{eqnarray}
\resizebox{0.4\textwidth}{!}{$\displaystyle 
\langle f(\hat{c}^\dagger_{k\alpha,\gamma}\hat{c}_{k\alpha,\gamma})\rangle=\left\langle 
\frac{\mbox{Tr}_kf(\hat{c}^\dagger_{k\alpha,\gamma}\hat{c}_{k\alpha,\gamma})\exp(-\beta\hat{\cal H}_k)\exp(\beta\mu\hat{n}_k)}{\mbox{Tr}_k\exp(-\beta\hat{\cal H}_k)\exp(\beta\mu\hat{n}_k)}\right\rangle\;,
\label{eq25}$}
\end{eqnarray}
where $\alpha$=1,2, $\gamma=\uparrow,\downarrow$  and $f$ is an arbitrary function of creation and annihilation operators from the $k$-th bond Hamiltonian $\hat{\cal H}_k$. As a result, one obtains the following formula for the magnetization of mobile electrons
\begin{eqnarray}
m_e\!\!\!&=&\!\!\!\left\langle \sum_{i=1}^2(n_{ki,\uparrow}-n_{ki,\downarrow})\right\rangle 
=\left\langle \frac{1}{\Xi_k}\left[\frac{\partial \Xi_k}{\partial (\beta J\sigma_{k1})}+\frac{\partial \Xi_k}{\partial (\beta J\sigma_{k2})}\right]\right\rangle
\nonumber
\\
\!\!\!&=&\!\!\!
\frac{4m_i}{V_1}\left[(z+z^3)\cosh(\beta t)\sinh(\beta J)+z^2\sinh(2\beta J)\right].
\nonumber\\
\label{eq26}
\end{eqnarray} 
It should be noted that the expressions for the spontaneous magnetizations $m_i$ and $m_e$ hold just for the case with the $F$ effective interaction $\beta R>0$, which supports the existence of the $F$ phase. For the $AF$ effective interaction $\beta R<0$, it is necessary to calculate new order parameters known as the staggered magnetization of localized Ising spins ($m^s_i$) and the staggered  magnetization of mobile electrons ($m^s_e$) per decorating dimer. As in the previous case, the staggered  magnetization $m^s_i$ can be obtained from the exact mapping theorem
\begin{eqnarray}
m^s_{i}\equiv\frac{1}{2}\langle \hat{\sigma}^z_{k1}-\hat{\sigma}^z_{k2}\rangle=
\frac{1}{2}\langle \hat{\sigma}^z_{k1}-\hat{\sigma}^z_{k2}\rangle_{IM}
 \equiv m_{IM}^{s}\;.
\label{eq27}
\end{eqnarray}
Owing to this fact, the staggered magnetization of the localized Ising spins $m^s_i$ has the following explicit form on the honeycomb~\cite{Naya} and square~\cite{Yang} lattices
\begin{eqnarray}
\begin{array}{ll}
m_{i}^s=\displaystyle\frac{1}{2}\left[1-\frac{16x^3(1+x^3)}{(1-x^2)^3(1-x)^3}\right]^{1/8}, & \mbox{(honeycomb)} \\
m_{i}^s=\displaystyle\frac{1}{2}\left[1-\frac{16x^4}{(1-x^2)^4}\right]^{1/8}, & \mbox{(square)}
\end{array}
\label{eq28}
\end{eqnarray}
where $x=\exp(-2\beta |R|)$. The staggered magnetization $m_i^s$ follows  the same formula (\ref{eq28}) on the loose-packed lattices as does the uniform magnetization $m_i$ given by Eq.~(\ref{eq24}), while it becomes identically zero on close-packed lattices due to a lack of spontaneous $AF$ long-range order caused by a geometric spin frustration. The staggered magnetization of mobile electrons  can be derived by the use of the Callen-Suzuki identity, which provides for $m_e^s$ the following expression depending on the  bond grand-partition function $\Xi_k$ and its derivatives
\begin{eqnarray}
m_{e}^s\!\!\!&=&\!\!\!\left\langle (n_{k1,\uparrow}-n_{k1,\downarrow})-(n_{k2,\uparrow}-n_{k2,\downarrow})\right\rangle 
\nonumber\\
\!\!\!&=&\!\!\!\left\langle \frac{1}{\Xi_k}\left[\frac{\partial \Xi_k}{\partial (\beta J\sigma_{k1})}-\frac{\partial \Xi_k}{\partial (\beta J\sigma_{k2})}\right]\right\rangle\;.
\label{eq29}
\end{eqnarray} 
After straightforward but cumbersome calculations, one obtains an explicit formula for the staggered magnetization of mobile electrons $m_e^s$ on the 2D doubly decorated  lattices
\begin{eqnarray}
m_{e}^s\!\!\!&=&\!\!\!\frac{(z+z^3)\sinh(\beta\sqrt{J^2+t^2})+z^2\sinh(2\beta\sqrt{J^2+t^2})}{V_2\sqrt{J^2+t^2}}
\nonumber\\
&\times &4Jm_{i}^s\;,
\label{eq30}
\end{eqnarray} 
which is expressed in terms of the formerly derived staggered magnetization of the Ising spins $m_i^s$. Finally, the calculation of the  electron compressibility follows from the knowledge of the average electron density $\rho$ according to~\cite{Reichl} 
\begin{eqnarray}
\kappa=\frac{1}{\rho^2}\left(\frac{\partial \rho}{\partial  \mu}\right)_{T}\;.
\end{eqnarray}
The exact expression (\ref{eq20}) for the average number of mobile electrons can be now utilized in order to find the following expression for the electron compressibility
\begin{eqnarray}
\kappa\!\!\!&=&\!\!\!\frac{z}{2}\left\{\beta\Gamma+\frac{
(1+\varepsilon)}{V_1^2}\left[\frac{\partial V_1'}{\partial \mu}V_1-V_1'\frac{\partial V_1}{\partial \mu}\right]\right.
\nonumber\\
\!\!\!&+&\!\!\!\left.\frac{
(1-\varepsilon)}{V_2^2}\left[\frac{\partial V_2'}{\partial \mu}V_2-V_2'\frac{\partial V_2}{\partial \mu}\right]
\right\}\;,
\end{eqnarray}
where 
\begin{eqnarray}
\Gamma=\left[
\frac{V_1'}{V_1}(1+\varepsilon)+\frac{V_2'}{V_2}(1-\varepsilon)
\right]\;.
\end{eqnarray}
\section{Results and discussion}
\label{Results}
In the following  we will provide a detailed discussion of the most interesting results obtained for  the extended spin-electron  model on doubly decorated planar lattices. First of all, it is worth mentioning that all  exact results derived in the previous section remain unchanged under the transformation $J\to-J$. A change of the $F$ Ising interaction $J>0$ to the $AF$ one $J<0$ results in a rather trivial change of the mutual spin orientation of the mobile electrons with respect to their nearest Ising neighbours. Consequently,  the critical temperature as well as other thermodynamic quantities (except the sign of order parameters) remain unchanged under the transformation $J\to-J$ and  one may further consider the $F$ interaction $J>0$ without  loss of generality.
On the other hand, it should be expected that different types of further-neighbour Ising interaction  $J'$ between the localized Ising spins may basically influence the physical properties of the model under investigation. For this reason, we will investigate the model with the  $F$ ($J'>0$) as well as  $AF$ ($J'<0$) further-neighbour Ising interaction in addition to the  zero interaction limit ($J'=0$).  Our further discussion of thermodynamic characteristics will be restricted  only to the case  $\rho\leq 2$ due to the validity of particle-hole symmetry. For simplicity, we will use the magnitude of the nearest-neighbour Ising interaction $J$ between localized spins and mobile electrons as the energy unit when normalizing all other parameters with respect to this coupling.

\subsection{Ground state}
\begin{figure}[h!]
\begin{center}
{\includegraphics[scale=0.3,trim=0 0 1.3cm 0.5cm, clip]{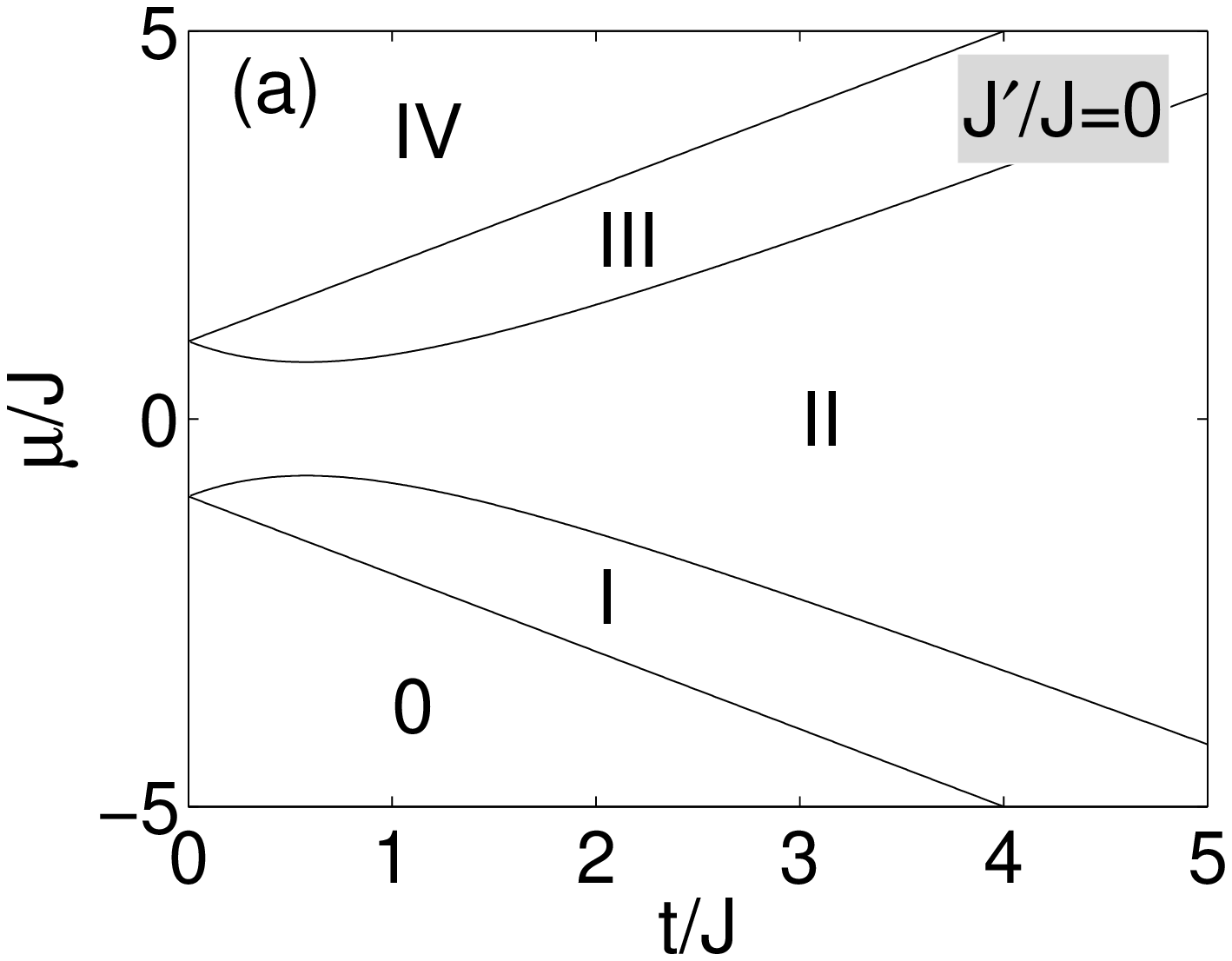}}
\includegraphics[scale=0.3,trim=0 0 1.3cm 0.5cm, clip]{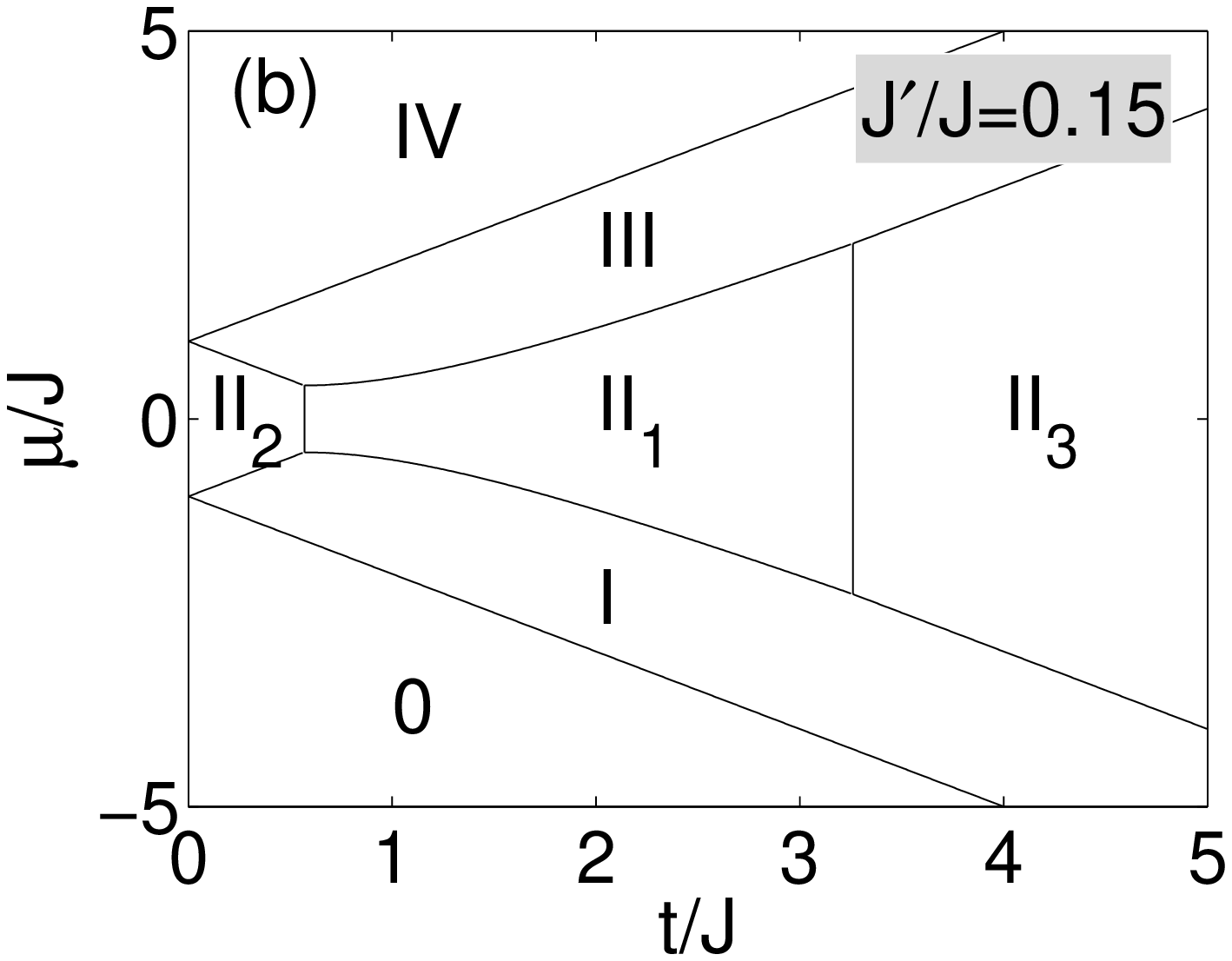}\\
\includegraphics[scale=0.3,trim=0 0 1.3cm 0.5cm, clip]{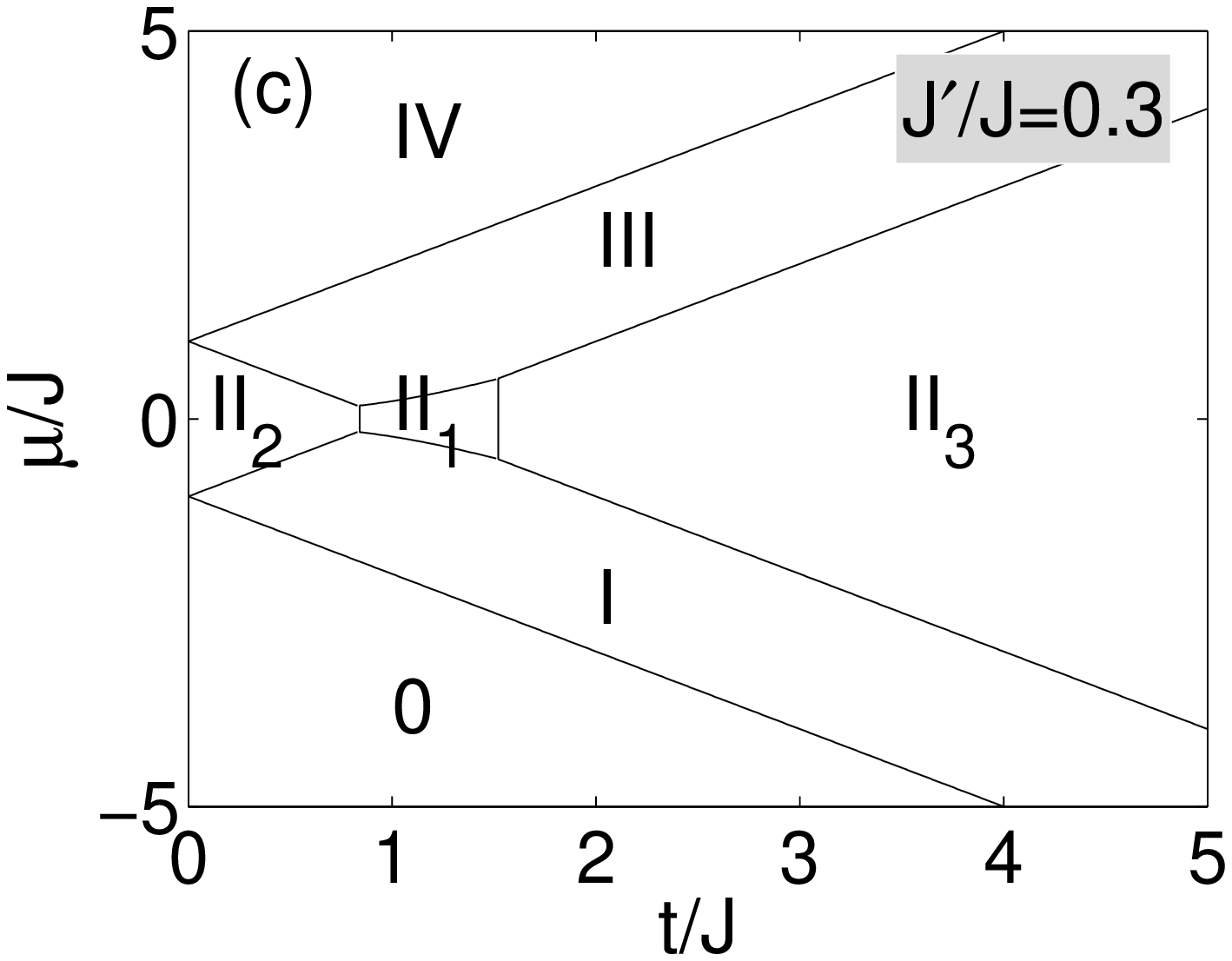}
\includegraphics[scale=0.3,trim=0 0 1.3cm 0.5cm, clip]{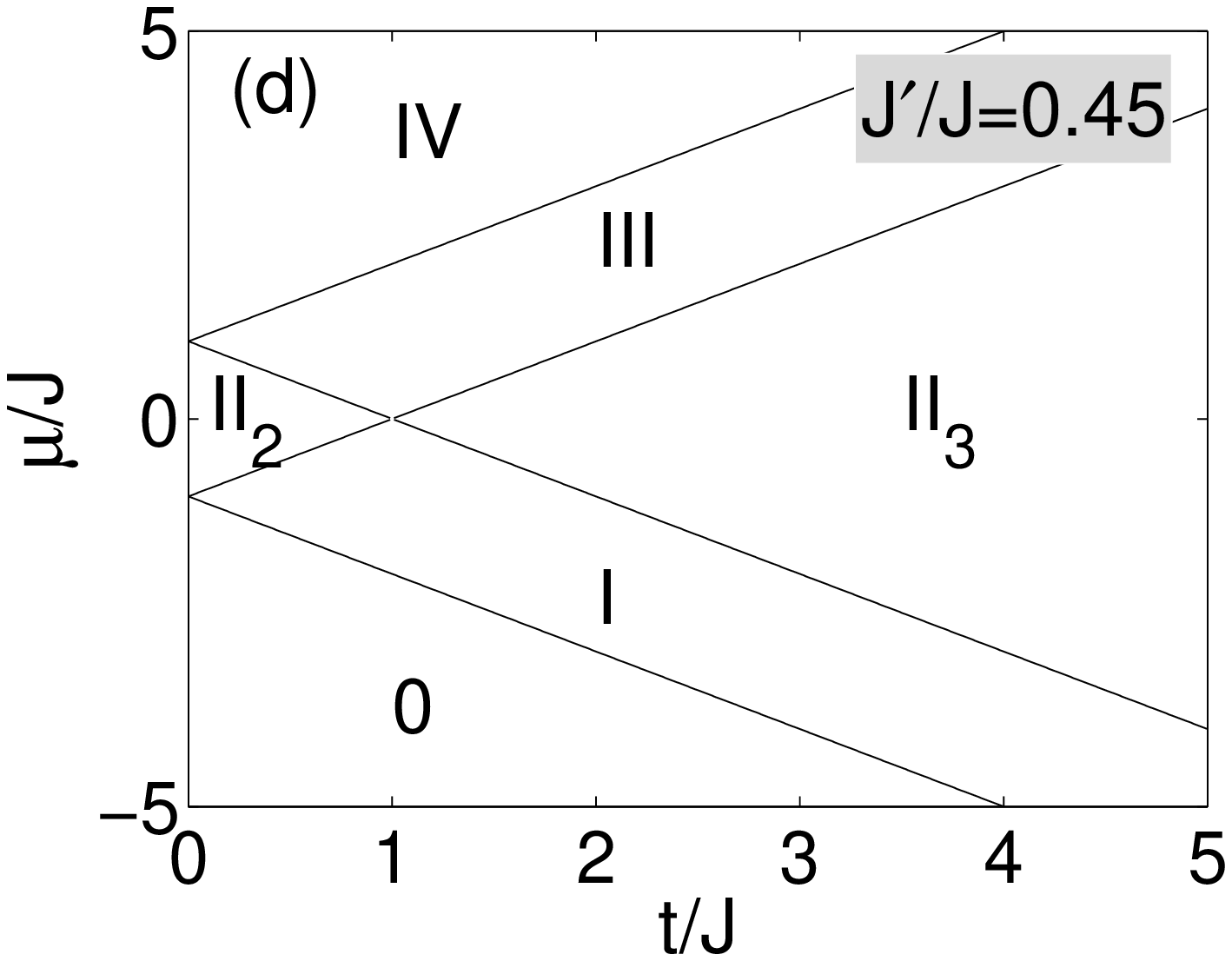}
\caption{\small Ground-state phase diagrams of the coupled spin-electron model (\ref{eq1}) in the $t/J-\mu/J$ plane for $J'/J=0$ and  the $F$ further-neighbour coupling $J'/J>0$.}
\label{fig2aa}
\end{center}
\end{figure}

At first, let us  perform  an exhaustive study of the ground state of correlated spin-electron system for different values of the relative strength of the further-neighbour interaction $J'/J$. Our analysis has shown that the competition between the model parameters ($\mu, J, J'$ and $t$) leads to several  ground states with different number of mobile electrons per decorating dimer, which may result in qualitatively different ground-state phase diagrams. It turns out that even small non-zero values of the further-neighbour coupling $J'/J$ strongly influences the ground-state phase diagram by generating new ground states, which are totally absent in the ground-state phase diagram for $J'/J=0$ (Fig.~\ref{fig2aa}(a)). For better clarity, we have collected the ground-state eigenvectors together with the respective eigenenergies and phase boundaries in Tab.~2. The microscopic nature of the ground states for $J'/J=0$ has been examined in detail in our preceding work~\cite{cenci1} to which the
  interested readers are referred to for further details. Herewith we will concentrate our attention to the effect of further-neighbour Ising interaction $J'/J$.

\begin{figure}[h!]
\begin{center}
\includegraphics[scale=0.3,trim=0 0 1.3cm 0.5cm, clip]{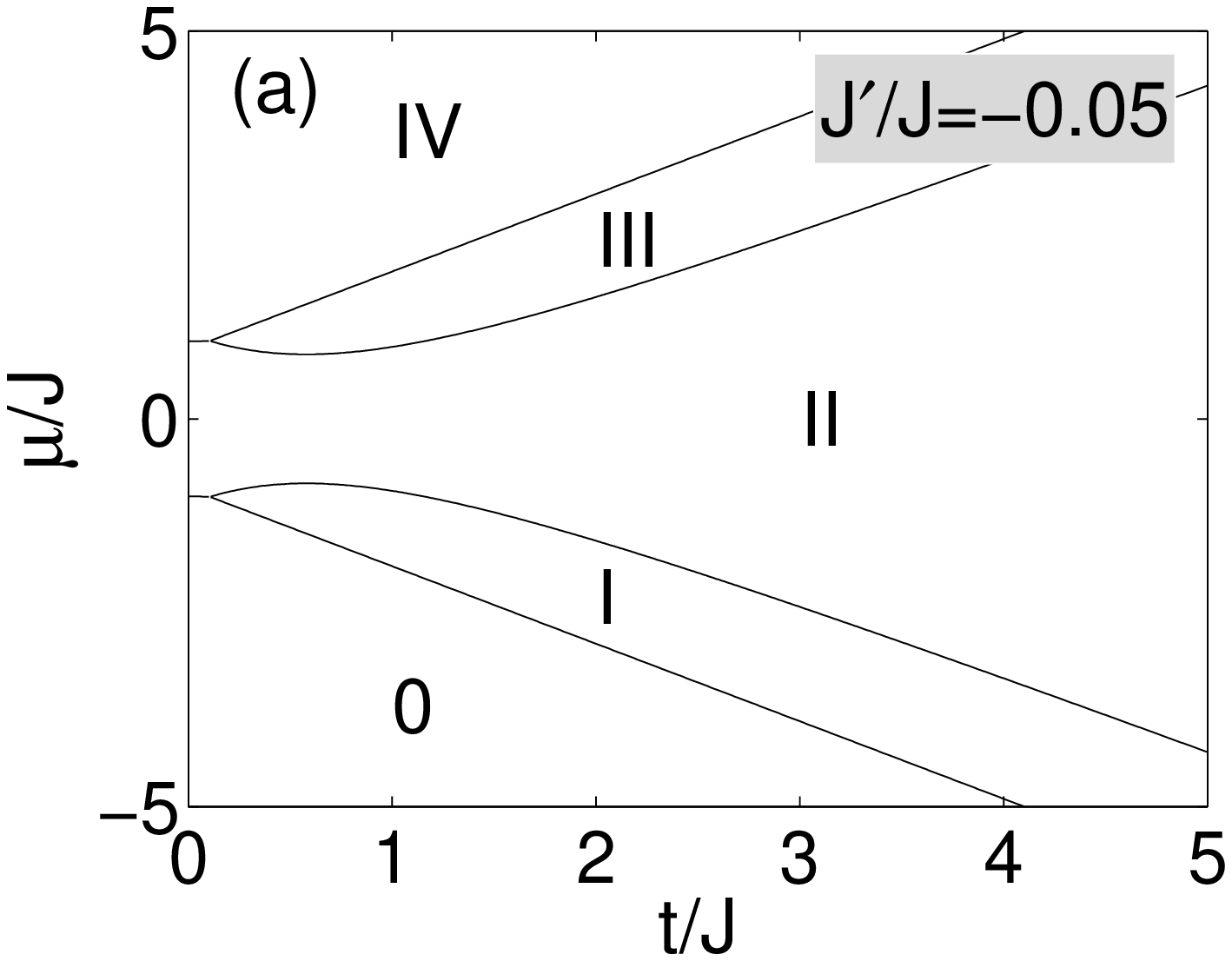}
\includegraphics[scale=0.3,trim=0 0 1.3cm 0.5cm, clip]{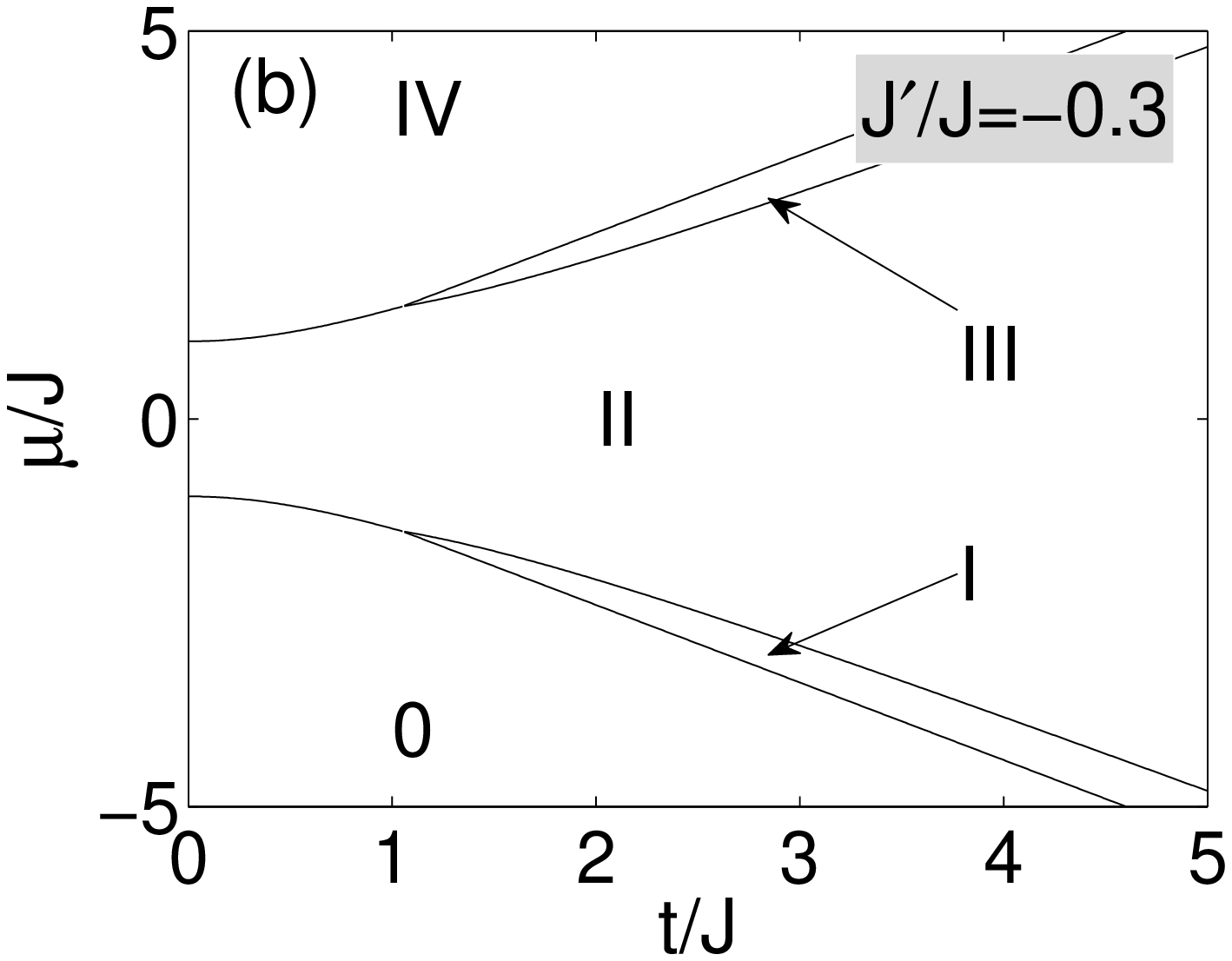}\\
\includegraphics[scale=0.3,trim=0 0 1.3cm 0.5cm, clip]{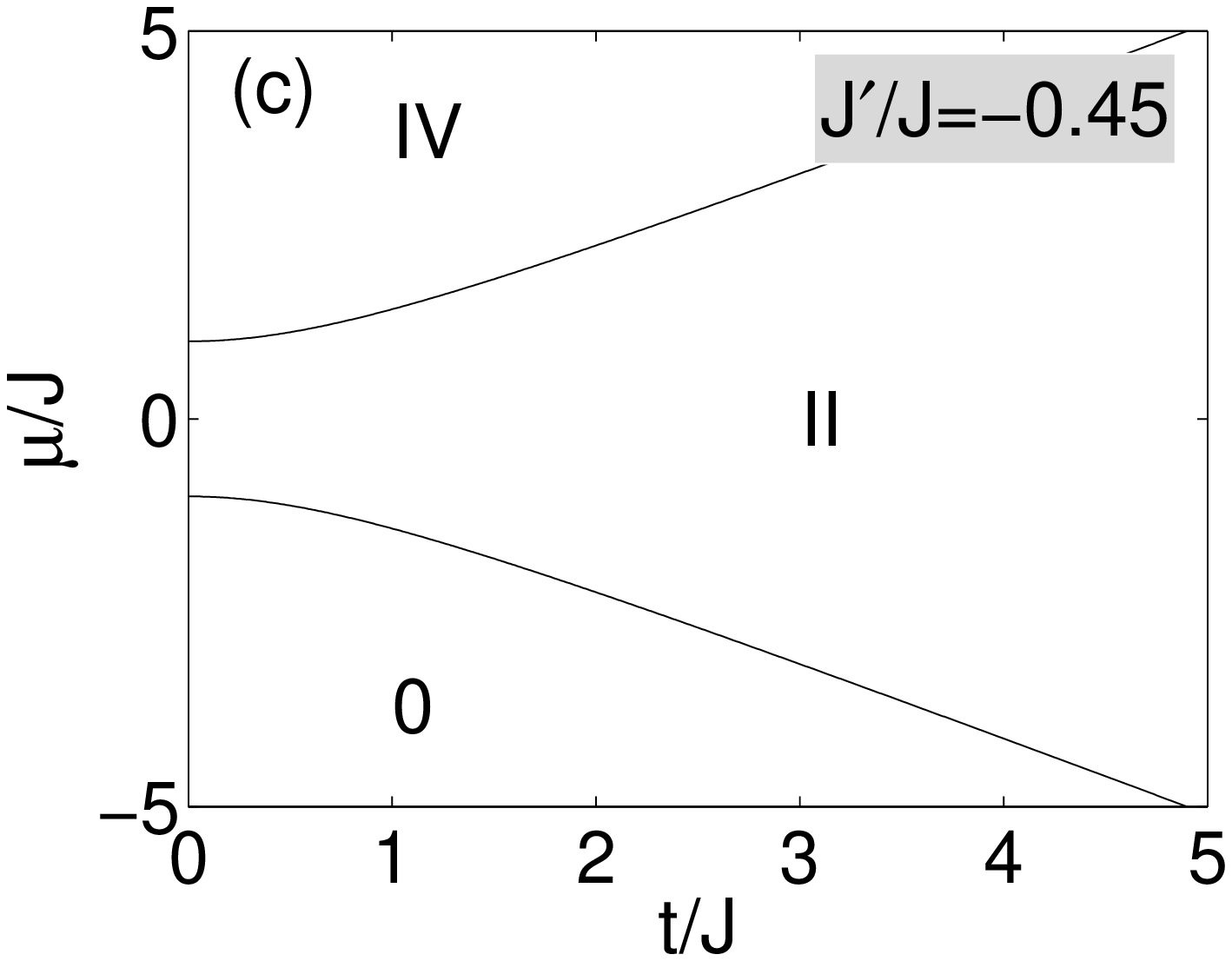}
\includegraphics[scale=0.3,trim=0 0 1.3cm 0.5cm, clip]{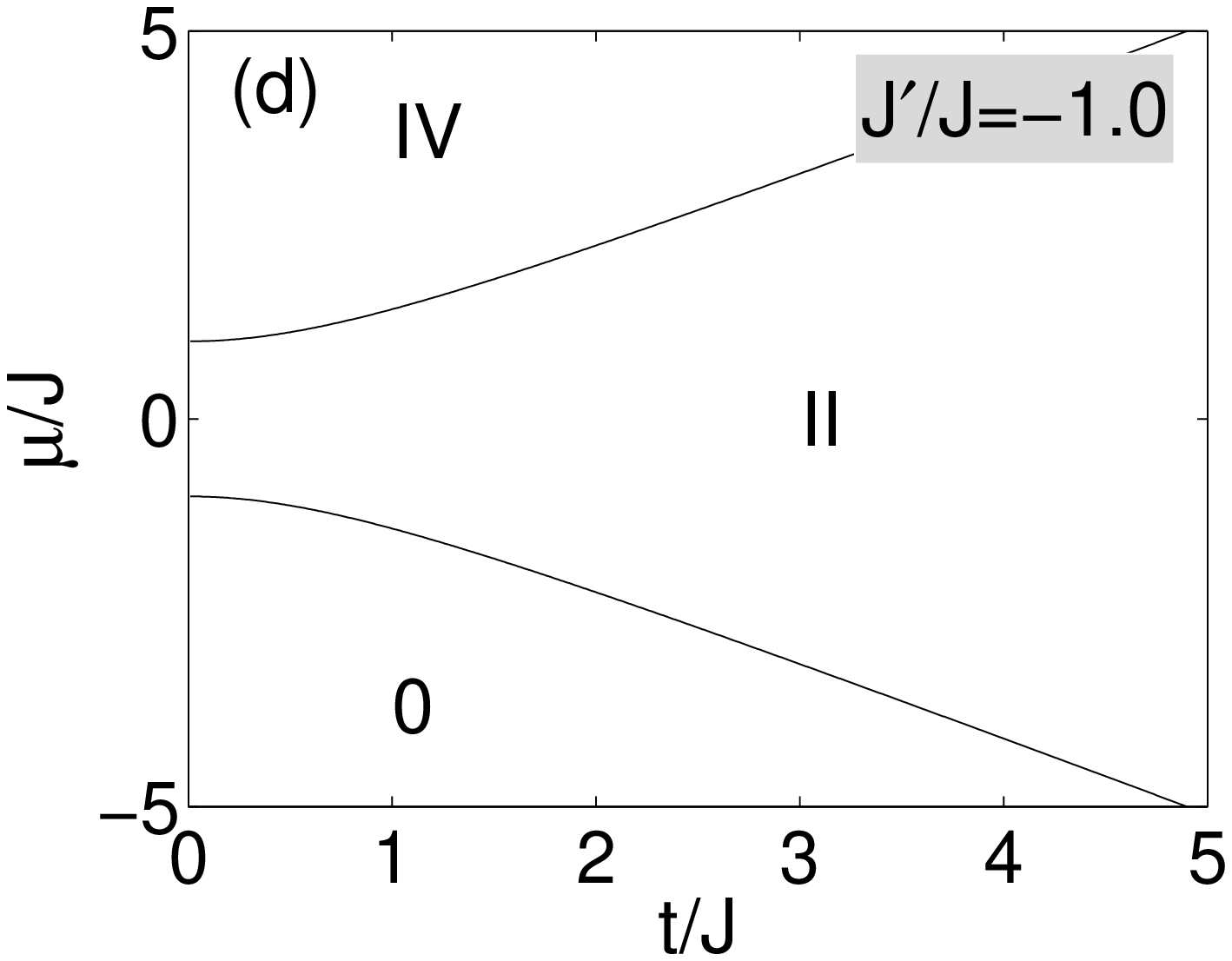}
\caption{\small Ground-state phase diagrams of the extended spin-electron model (\ref{eq1}) in the $t/J-\mu/J$ plane for the $AF$ further-neighbour coupling $J'/J<0$.}
\label{fig2bb}
\end{center}
\end{figure}

We start our discussion with the particular case with $F$ further-neighbour  interaction $J'/J>0$. It could be expected that the $F$ further-neighbour interaction will stabilize the $F$ phases (the phase I and III) at the expense of the remaining phases (0, II and IV) that may additionally undergo qualitatively changes. Indeed, we have found that the eigenvectors of the phases I and III remain qualitative unchanged, whereas the corresponding eigenenergies are only shifted by the constant $-J'$ (see Tab.~2). The  ground-state phase diagrams are presented in Fig.~\ref{fig2aa}(b)-(d) for a few selected values of the $F$ further-neighbour interaction $J'/J>0$. 
For the ground states  0 and IV, in which the electron subsystem is empty or fully occupied, the additional $F$ interaction $J'/J$ influences only the spin subsystem. As a result, 
the $F$ spin alignment of the localized Ising spins is strictly preferred instead of a random spin orientation observed in the ground states 0 and IV on assumption that $J'/J=0$. The corresponding eigenenergies are therefore shifted only by the constant $-J'$. In the spirit of these facts, it is not surprising that the phase boundaries between phases 0-I and III-IV are identical with the corresponding ones for $J'/J=0$ (Tab.~2). The situation is much more involved for the ground states with two mobile electrons per decorating dimer, which cause the $AF$ Ne\'el order II in the limit $J'/J=0$. Namely, the $F$ interaction $J'/J$ competes with the $AF$ arrangement of the localized Ising spins and it thus strongly influences both subsystems. While the classical $F$ spin arrangement between the localized Ising spins and the mobile electrons is preferred within the phase II$_2$ emerging for sufficiently small values of the hopping term $t<\sqrt{2JJ'+J'^2}$,  the quantum superposition of two $AF$ and two
  non-magnetic ionic states of the mobile electrons accompanies a perfect $F$ alignment of the localized Ising spins within the ground state II$_3$ for strong enough hopping amplitudes for $t>(J^2-J'^2)/2J'$ (see Tab.~2). These two novel ground states become dominant with increasing $J'/J$ until the $AF$ Ne\'el ground state II$_1$ completely disappears from the ground-state phase diagram. Nevertheless,  all three phases coexist together at the appropriate small values of $J'/J$.  Thus, we can conclude that the variation of the kinetic term may lead to a magnetic phase transition, while the number of mobile electrons is kept constant. On the other hand, the $AF$ further-neighbour interaction $J'/J<0$ should stabilize the $AF$ phase (the phase II) at the expense of the  remaining ground states (0, I, III and IV), as it is illustrated in Fig.~\ref{fig2bb}. 
It actually turns out that the ground state II with two mobile electrons per decorating dimer cannot be accompanied with the $F$ alignment of localized Ising spins, which originate in the phases II$_2$ and II$_3$ exclusively from the $F$ further-neighbour interaction (the hopping process of two mobile electrons transmits an effective $AF$ interaction between the localized Ising spins). Bearing this in mind, the $AF$ further-neighbour interaction does not alter the ground state II with two mobile electrons per decorating dimer, which still exhibits the quantum N\'eel ordering with a perfect $AF$  arrangement of localized Ising spins and  the quantum $AF$ arrangement of the mobile electrons underlying a quantum superposition of two $AF$ and two non-magnetic states. The corresponding eigenenergy of the phase II has the similar form with a rather trivial extension by the further-neighbour interaction $+|J'|$ (see Tab.~2). Similarly as for $J'/J>0$, the  $AF$ further-neighbour interaction
  $J'/J$ influences only the spin subsystem of the phase 0 and IV with an empty or fully occupied electron subsystem, where the classical Ne\'el long-range order of the localized Ising spins is preferred and the corresponding eigenenergies are only shifted by the constant $-J'$. Surprisingly, the $AF$ further-neighbour interaction still favors the same arrangement of the electron as well as spin subsystems as for the case $J'/J=0$ for the ground states I and III with  one and three electrons per decorating dimer. When the hopping term is sufficiently weak $t<-2J'[(J+J')/(J+2J')]$ with respect to the $AF$ further-neighbour interaction $J'$,  the ground states  with odd number of mobile electrons per dimer are suppressed by the quantum $AF$ ground state II with two electrons per decorating dimer. The ground-state phase diagram  consists only of phases with an even number of mobile electrons per  dimer if $\left[-\displaystyle(1+t/J)-\sqrt{(t/J)^2+1}\right]/2<J'/J<\left[-(1+t/J)+\sqrt{(t/J)^2+1}\right]/2$, as is illustrated in Fig.~\ref{fig2bb}(d). 
%
%
\subsection{Finite-temperature phase diagrams}
\begin{figure}[h!]
\begin{center}
{\includegraphics[scale=0.3,trim=0 0 1.3cm 0.5cm, clip]{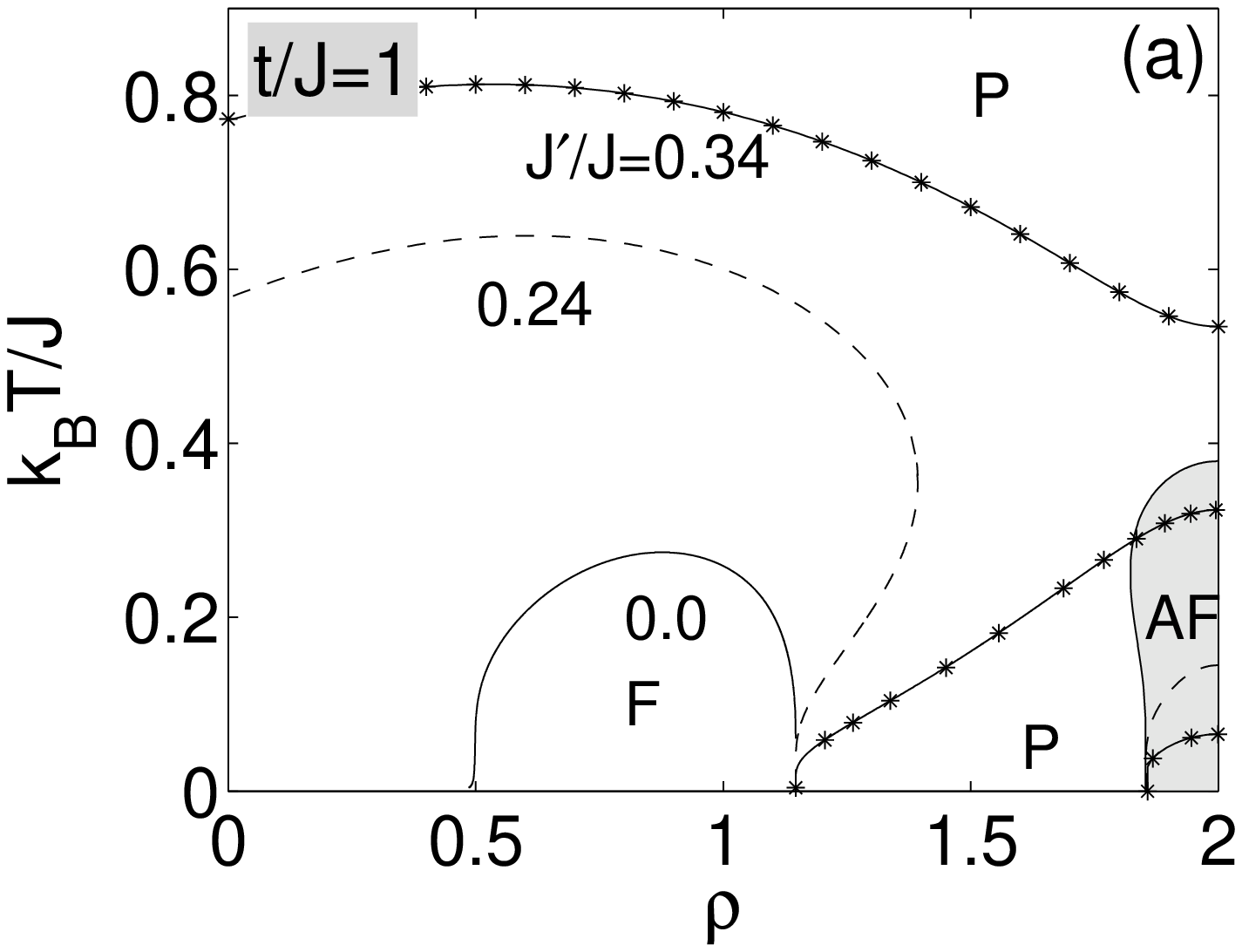}}
\includegraphics[scale=0.3,trim=0 0 1.3cm 0.5cm, clip]{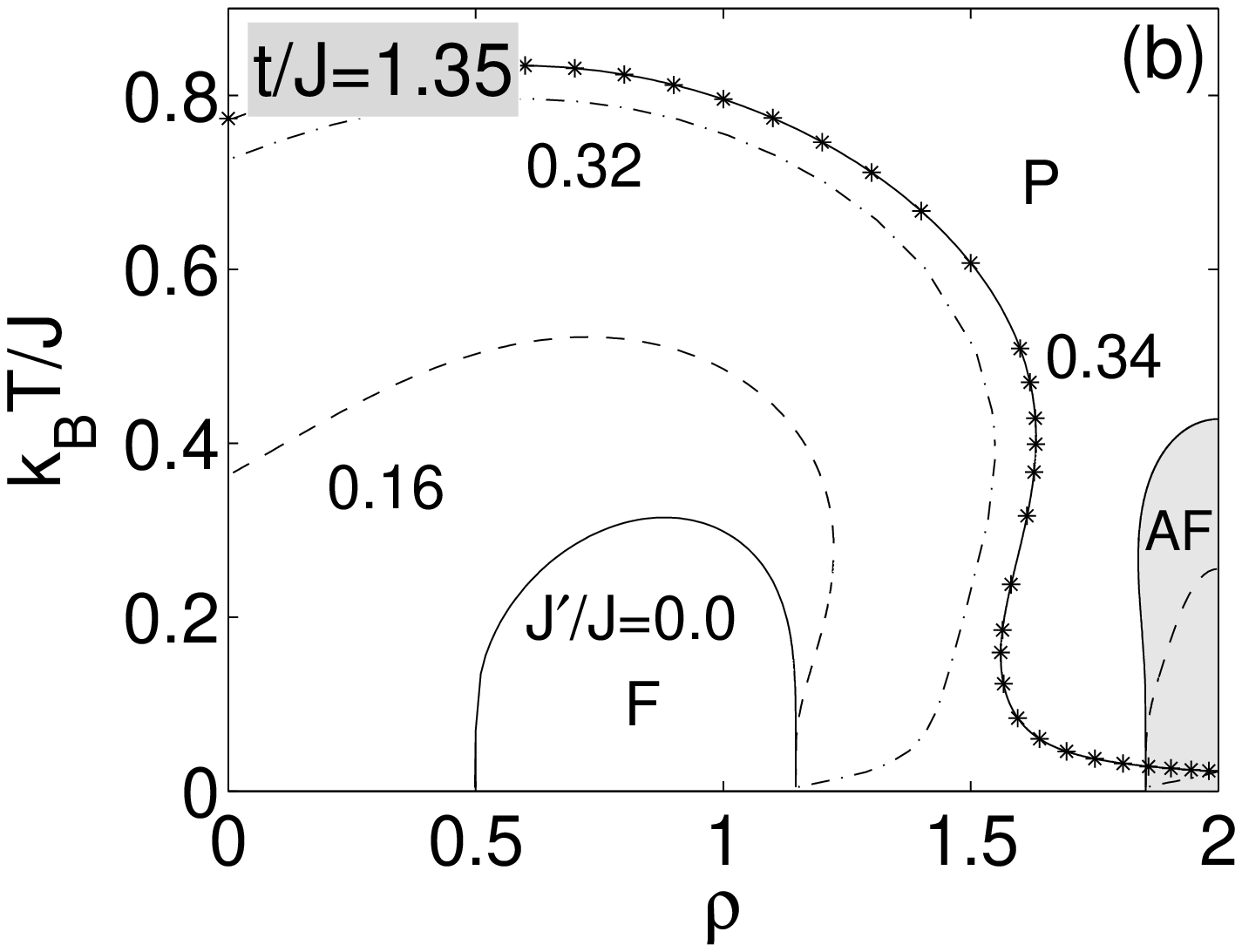}
\caption{\small Phase diagrams in the $\rho-k_BT/J$ plane for two representative values of $t/J$ ($t/J=$1 and 1.35) and distinct $F$ further-neighbour interaction $J'/J$.  Different  lines illustrate the borders between the $AF$-$P$ and $F$-$P$ phases.  The shaded area has been used for the better visualization of the $AF$ phase.}
\label{fig4}
\end{center}
\end{figure}

It has been demonstrated in our previous works~\cite{Doria,cenci1} that the finite-temperature phase diagram of the coupled spin-electron model on doubly decorated 2D lattices displays similar features for several lattice topologies. Therefore our further discussion will be restricted only to the representative case of the double decorated  square lattice. Fig.~\ref{fig4} illustrates a few typical finite-temperature phase diagrams in the form of critical temperature versus electron concentration plots for several values of the $F$ further-neighbour coupling $J'/J>0$.
\\

\begin{figure}[h!]
\begin{center}
\includegraphics[scale=0.3,trim=0 0 1.3cm 0.5cm, clip]{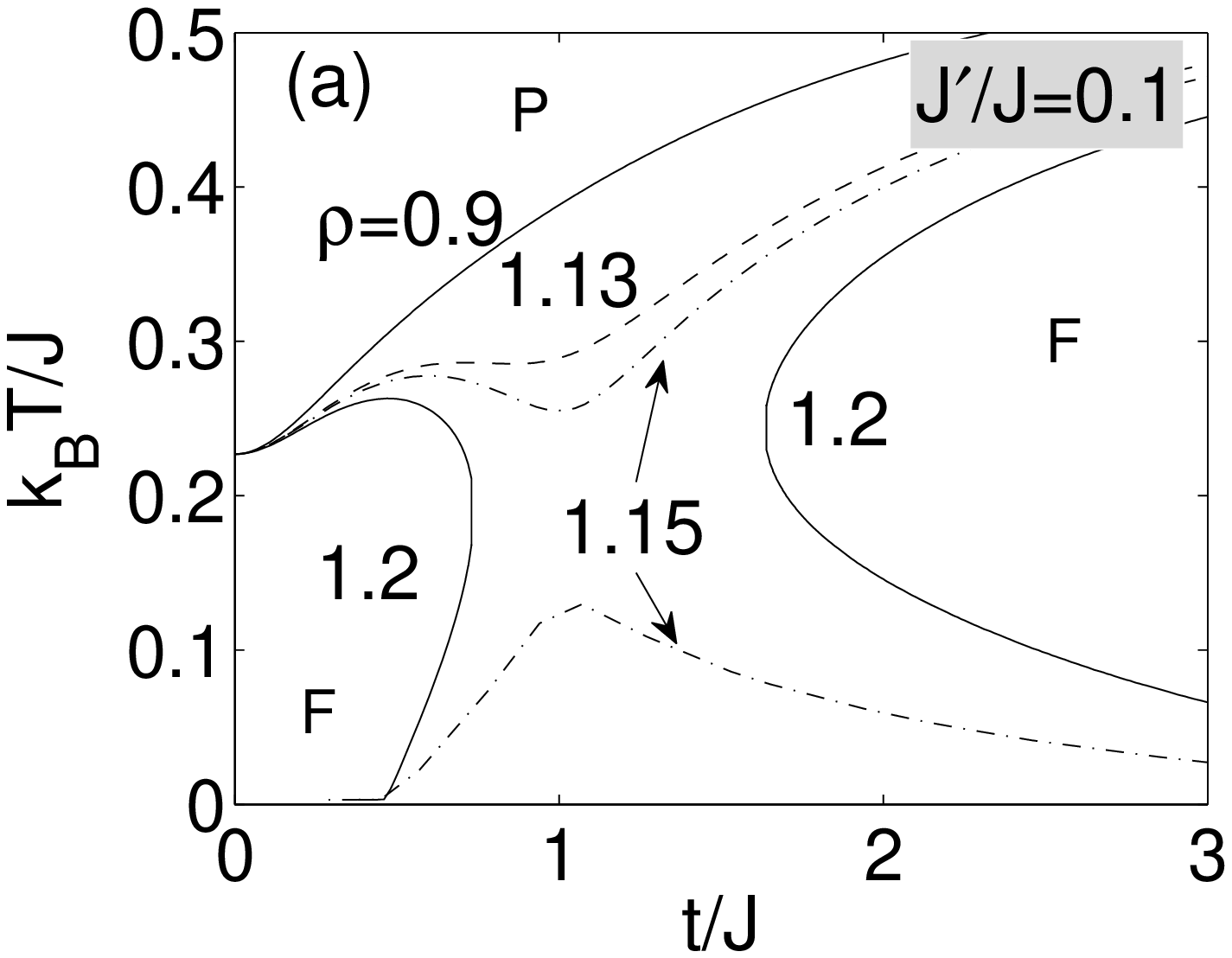}
\includegraphics[scale=0.3,trim=0 0 1.3cm 0.5cm, clip]{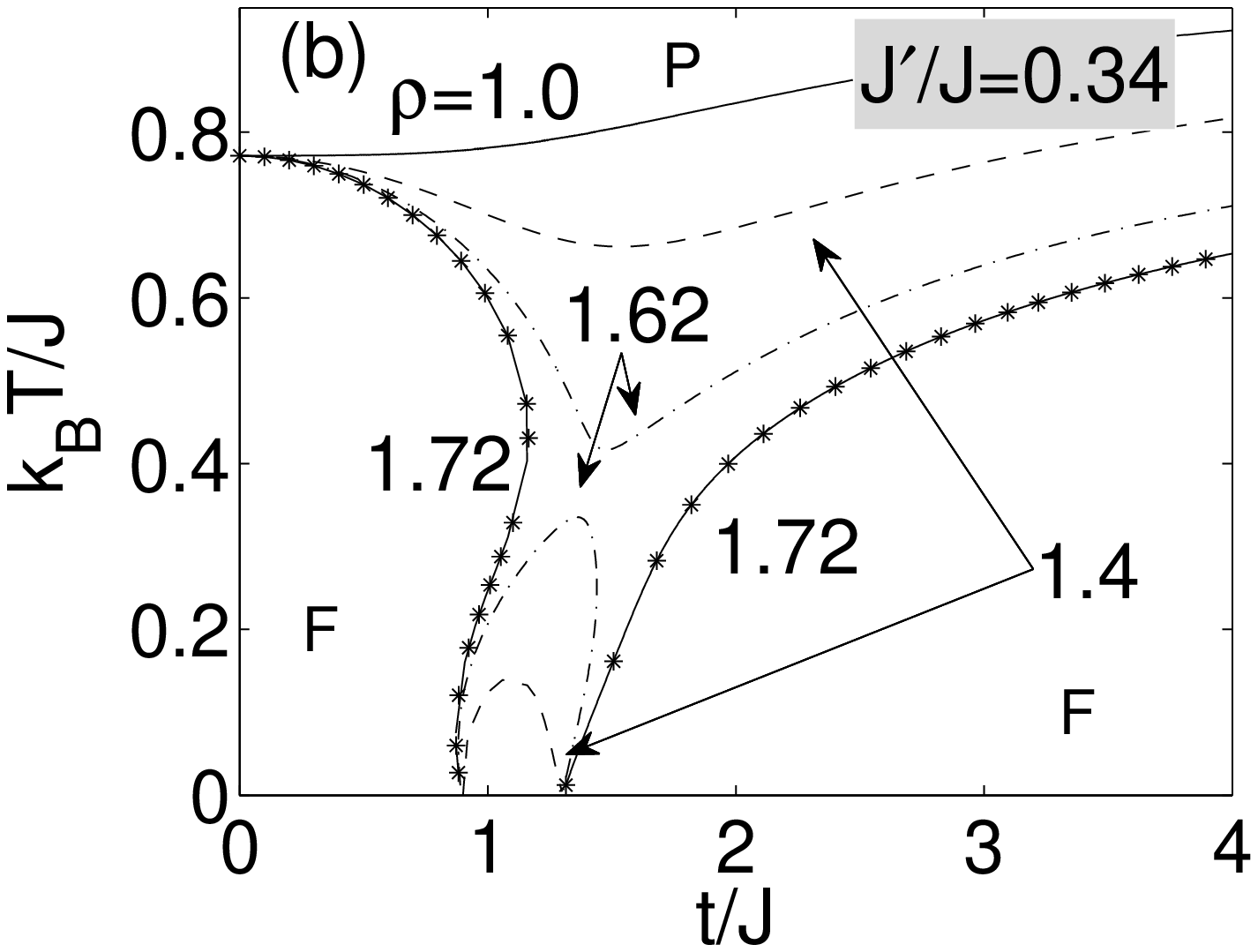}
\caption{\small Phase diagrams in the $t/J-k_BT/J$ plane for a few selected values of $J'/J$ and $\rho$ corresponding to the $F$ phase. Different  lines illustrate the borders between the $F$ and $P$ phase. For $\rho=1.15$ in the panel (a) and $\rho=1.4$ and 1.62 in the panel (b) the $F$ phase is delimited by border lines, above and below which the $P$ phase occurs.}
\label{fig5}
\end{center}
\end{figure}

It is evident that the area corresponding to the $F$ long-range order (in the vicinity of $\rho\approx 1$) generally  increases with increasing $J'/J$, whereas the area corresponding to the $AF$ state (in the vicinity of $\rho\approx 2$) gradually diminishes. The most pronounced changes in the phase diagram can be primarily observed at low electron concentrations, where the spontaneous $F$ long-range order appears due to the non-zero further-neighbour interaction $J'/J\neq0$ also below the bond percolation threshold $\rho=0.5$, in contrast with the particular case of $J'/J=0$. In addition, there exists a critical value of $J'/J$ above which  only the $F$ state can be detected for all possible electron concentrations $\rho$ (e.g., see the curve $J'/J$=0.34 in Fig.~\ref{fig4}(b)). As far as the reentrant transitions are concerned, we have found that the $F$ further-neighbour interaction $J'/J>0$ may produce many types of magnetic reentrant phenomena, some of which are absent in the model with $J'/J=0$. If the further-neighbour coupling is weaker 
 than  $J'/J\approx 0.05$, the system exhibits a reentrant phase transitions with two consecutive critical points similar to the ones observed for $J'/J=0$ for both the $F$ and $AF$ phases. The reentrant phase transitions with two consecutive critical points can be also observed for $J'/J\geq 0.05$, but only for the $F$ case. Moreover for the higher further-neighbour interaction ($J'/J\gtrsim 0.08$), a different mechanism underlines their formation, as illustrated in Fig.~\ref{fig5}. 
Surprisingly, the temperature fluctuations support an appearance of the $F$ long-range order above the $P$ ground state when the hopping integral $t$ is smaller or greater  than the Ising coupling and the electron filling $\rho\gtrsim 1$, see Fig.~\ref{fig5}(a). The $F$ state then  persists  in a rather narrow temperature interval. A further temperature increase destroys this state. It was found that the $P$-$F$-$P$ reentrance observed for $t/J>1$ is markedly reduced for a stronger further-neighbour interaction ($J'/J\gtrsim 0.2$) and the system exhibits a single transition from the $F$ to the $P$ state, as seen Fig.~\ref{fig5}(b). As one can see from Fig.~\ref{fig5}(b), a sufficiently strong further-neighbour $F$ coupling is responsible for the existence of other reentrant phase transitions with three consecutive critical points, namely $F$-$P$-$F$-$P$, near the electron filling $\rho\approx 1.5$, e.g., for $\rho=1.62$.
Last but not least, new reentrant phenomena produced by the non-zero $F$ further-neighbour interaction are the  mixed reentrant phase transitions with three consecutive critical points of the $AF$-$P$-$F$-$P$ type.
 The term mixed reentrant phase transition is used to denote the situation when the investigated model system re-enters at higher temperatures to a spontaneous ordering of another type than at lower temperatures.
 Such behaviour exists only near the half-filled band case ($\rho\approx 2$) and appropriate model parameters  $t/J\approx 1$ and $J'/J\approx 1/3$ (see border lines for $J'/J$ = 0.34 in Fig.~\ref{fig4}(a)). It is clear that this behaviour is a direct effect of the additional further-neighbour Ising interaction $J'/J$, because its existence cannot be observed for the $J'/J=0$.

\begin{figure}[h!]
\begin{center}
\includegraphics[scale=0.3,trim=0 0 1.3cm 0.5cm, clip]{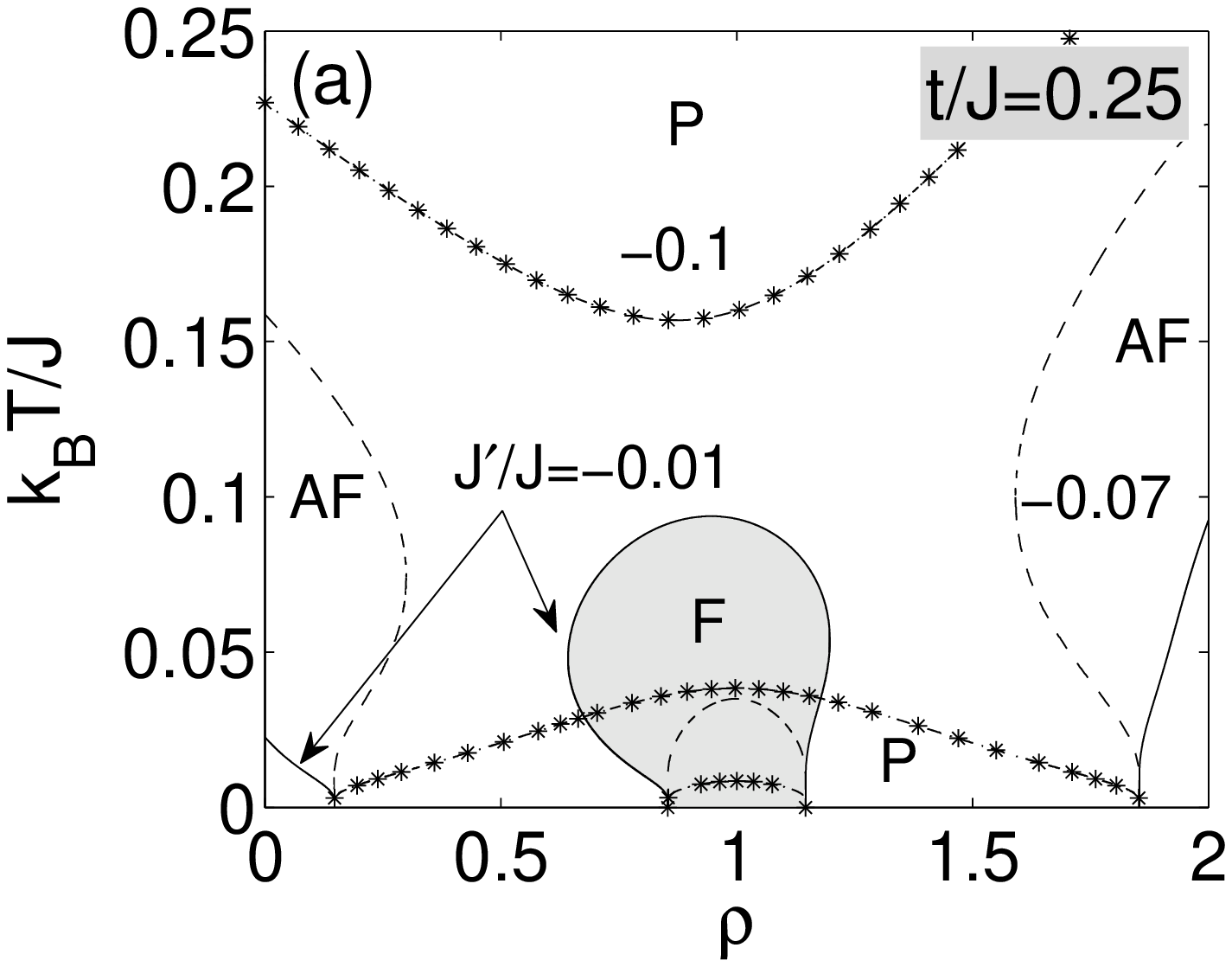}
\includegraphics[scale=0.3,trim=0 0 1.3cm 0.5cm, clip]{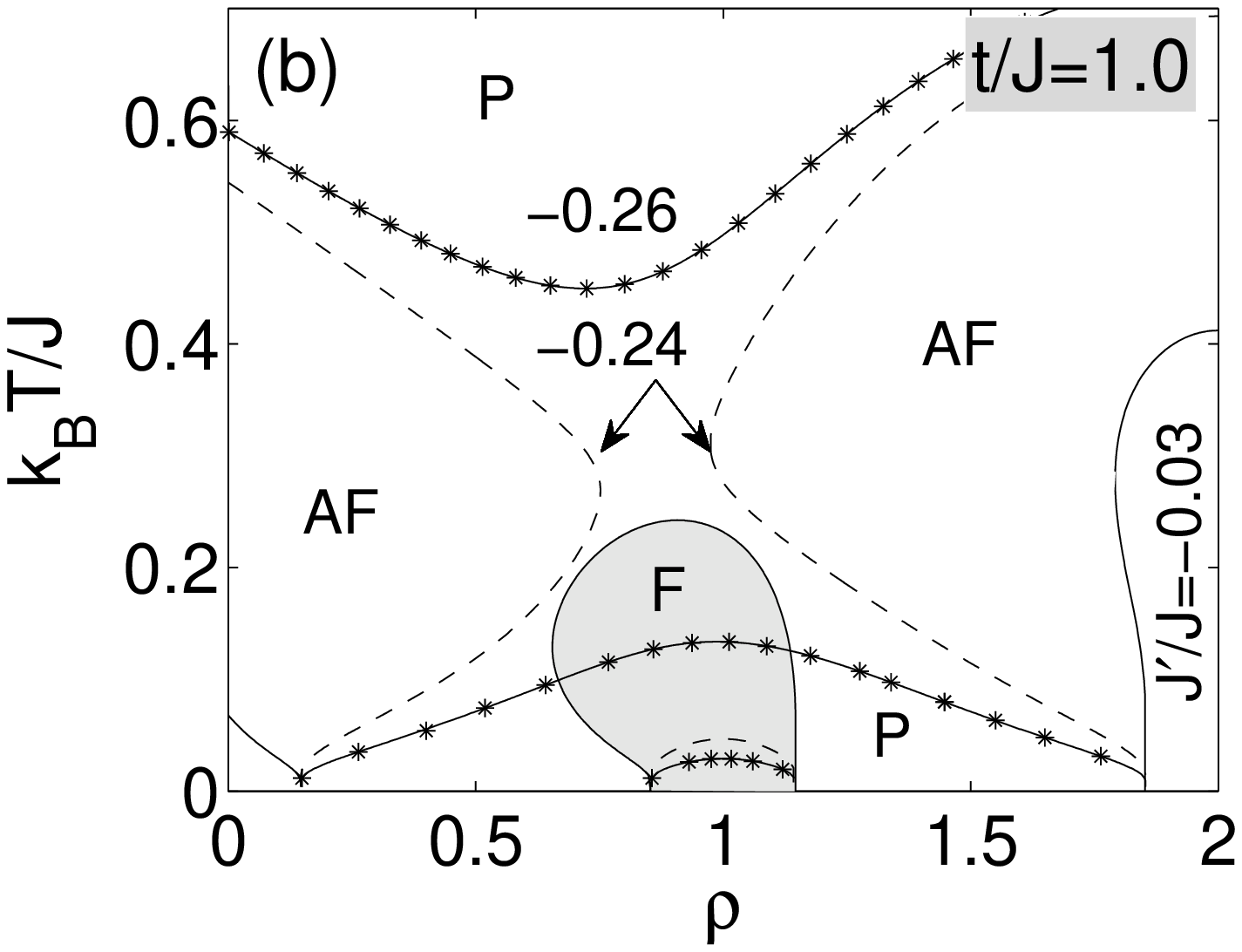}
\caption{\small Phase diagrams in the $\rho-k_BT/J$ plane for two representative values of $t/J$  calculated for $J'/J<0$.  Different  lines illustrate the borders between the $AF$-$P$ and $F$-$P$ phases.  The shaded area has been used for the better visualization of the $F$ phase.}
\label{fig6}
\end{center}
\end{figure}

The obtained phase diagrams are more complex for the $AF$ further-neighbour interaction $J'/J<0$. In Fig.~\ref{fig6} we present a few typical finite-temperature phase diagrams for two representative values of the hopping term $t/J=0.25$ and 1. It is evident that the $AF$ further-neighbour  interaction between the localized Ising spins reduces the critical temperature of the $F$ phase (in the vicinity of $\rho\approx 1$) and increases (in comparison with the $J'/J=0$) the lower critical concentration (up to $\rho_c=0.854$) of the mobile electrons needed for the onset of the $F$ long-range order at relatively low temperatures. Moreover, the small additional $AF$ further-neighbour interaction generates a new $P-F-P$ reentrant transition for smaller electron concentrations $\rho<1$.
Even though the $F$ long-range order is generally reduced with the $AF$ further-neighbour interaction $J'/J$, the $F$ long-range order still ends up at the same upper critical concentration $\rho_c^U=1.146$ as for $J'/J=0$. Contrary to this,  the $AF$ phase gradually fills up the whole region of the phase diagram, because the new $AF$ phase emerges at small electron concentrations $\rho\to 0$. Both  $AF$ phases are stabilized through the $AF$ further-neighbour interaction $J'/J<0$ and are subsequently connected into one large common $AF$ area above a certain threshold value. Also for the $AF$ further-neighbour interactions $J'/J<0$, the interesting  mixed reentrant sequence of transitions $F$-$P$-$AF$-$P$ has been detected. Contrary to the former case with $J'/J>0$, where  similar reentrant transitions $AF$-$P$-$F$-$P$  occurs just for the electron concentrations close to a half-filling $\rho\approx 2$, the reentrant transitions $F$-$P$-$AF$-$P$ can be observed only for electron concentrations in the vicinity of  $\rho\approx 1$. 
This effect is also a direct consequence of the additional further-neighbor Ising interaction $J'/J$, because its existence cannot  be observed for $J'/J=0$.
\subsection{The magnetization, specific heat and compressibility}
\begin{figure}[h!]
\begin{center}
{\includegraphics[width=4.3cm,height=3.85cm,trim=0.25cm 0 1.05cm 0.5cm, clip]{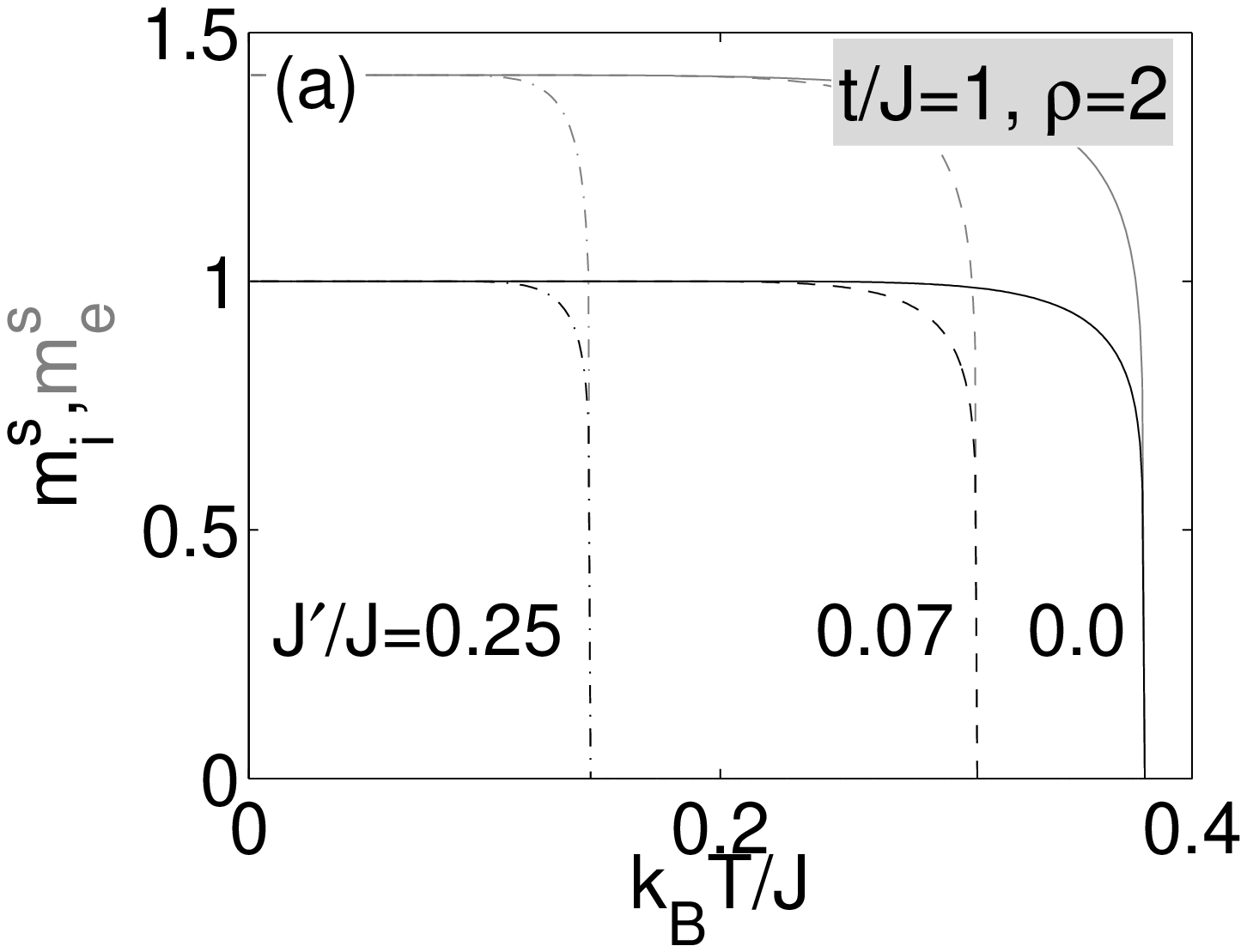}}
{\includegraphics[width=4.3cm,height=3.85cm,trim=0.2cm 0 1.1cm 0.5cm, clip]{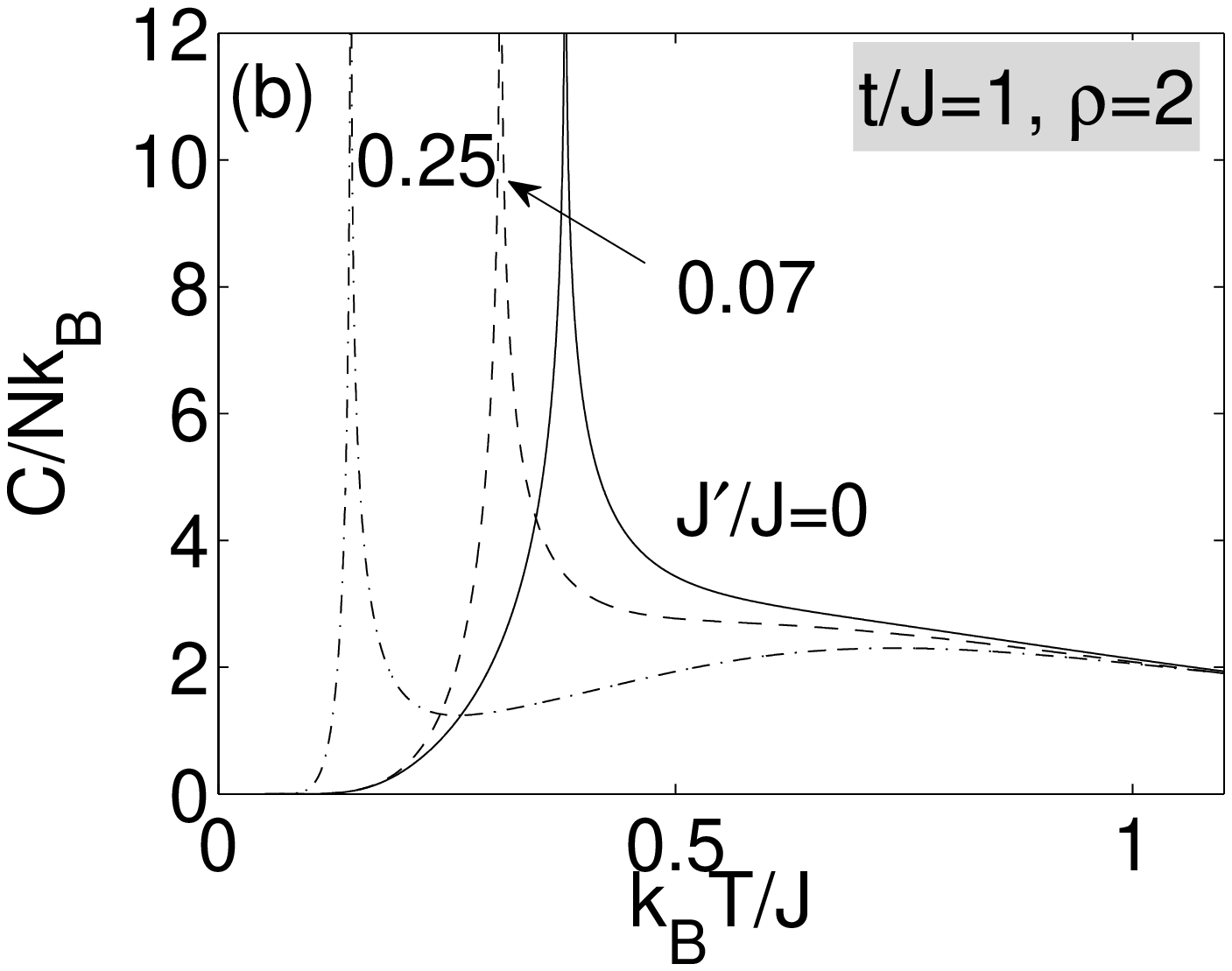}}
\caption{\small The staggered magnetizations $m_i^s$ (black lines) and $m_e^s$ (gray lines) and corresponding specific heat as a function of temperature for $t/J=1$, $\rho=2$ and $J'/J\geq0$ generating the $AF$ state.}
\label{fig9}
\end{center}
\end{figure}

\begin{figure}[h!]
\begin{center}
{\includegraphics[scale=0.4,trim=0 0 0.9cm 0.5cm, clip]{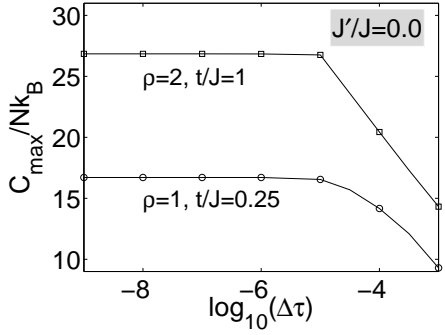}}
\caption{\small The dependencies of the cusp maximum on the temperature step for integer electron concentrations ($\rho=1$ and $\rho=2$) and two representative values of $t/J$. The lines are the guide for eyes.}
\label{fig9a}
\end{center}
\end{figure}

\begin{figure}[h!]
\begin{center}
\includegraphics[width=4.3cm,height=3.85cm,trim=0.2cm 0 1.cm 0.5cm, clip]{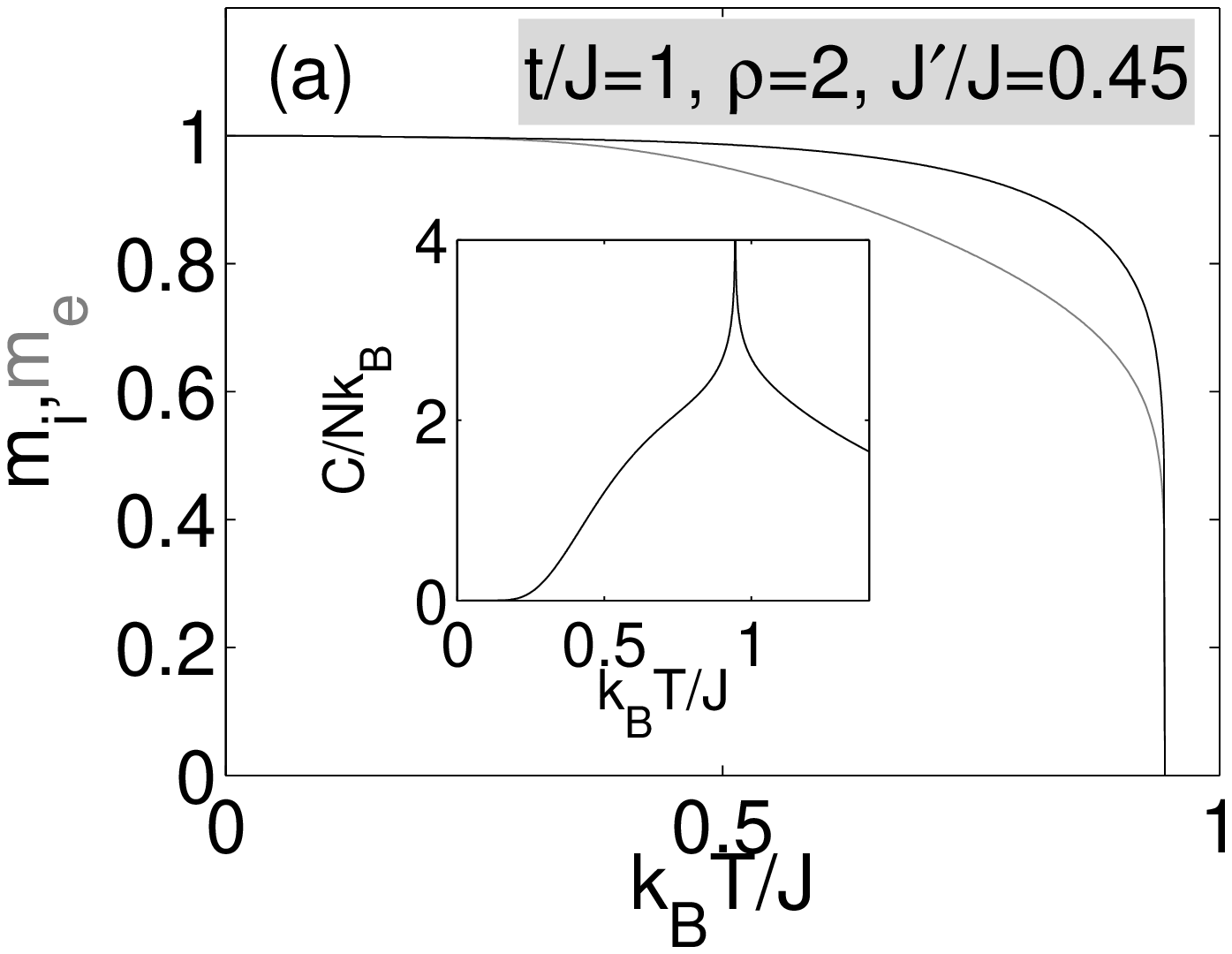}
\includegraphics[width=4.3cm,height=3.85cm,trim=0.2cm 0 1.cm 0.5cm, clip]{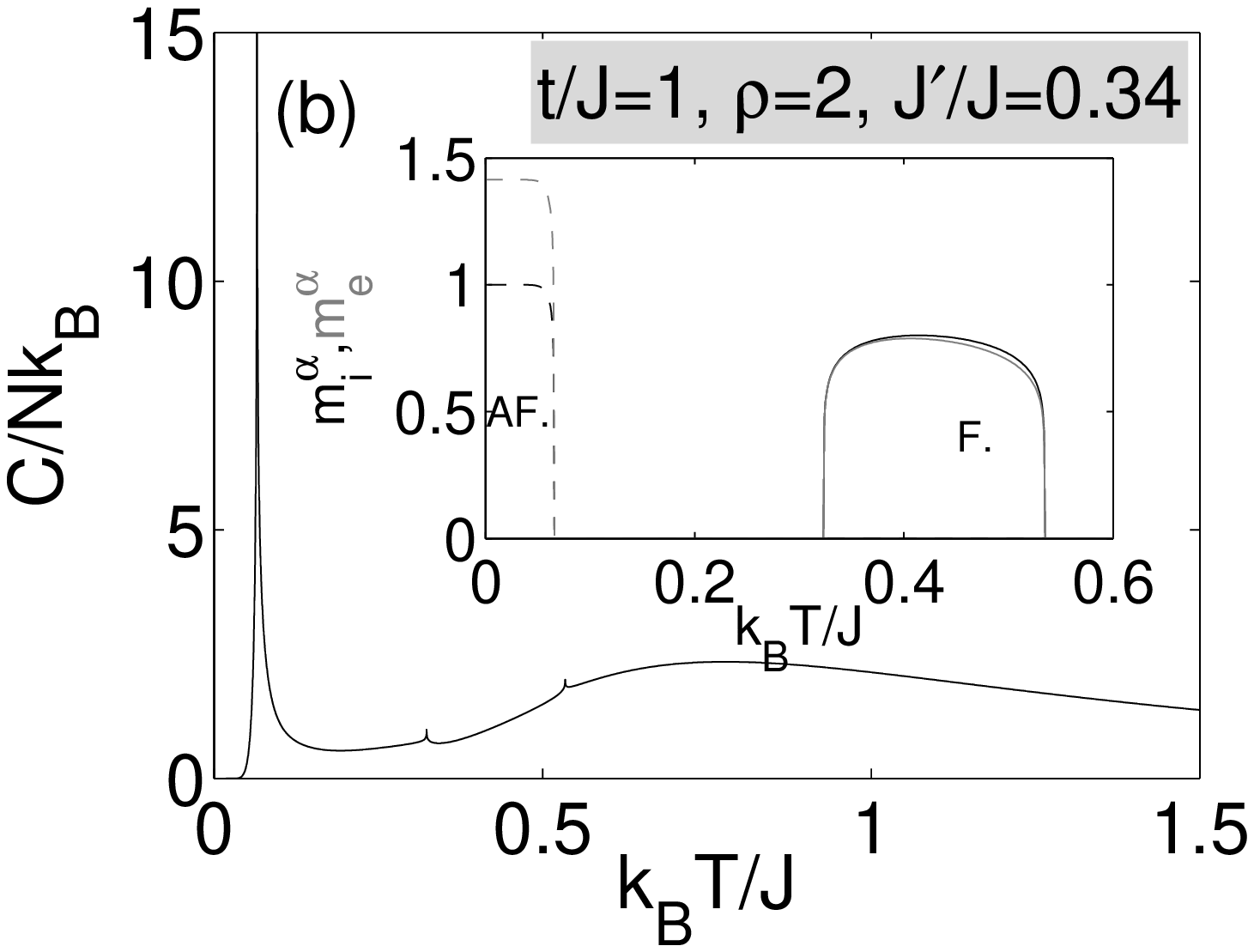}
\caption{\small (a) The magnetizations $m_i$ (black lines) and $m_e$ (gray lines) and corresponding specific heat (inset) as a function of temperature for $t/J=1$, $\rho=2$ and $J'/J=0.45$ generating just the $F$ state. (b) The thermal dependence of the specific heat  in the case with  mixed reentrant transitions $AF$-$P$-$F$-$P$ for $J'/J=0.34$. Inset: uniform (solid lines) and staggered (dashed lines) magnetizations as a function of temperature for the same model parameters.}
\label{fig10}
\end{center}
\end{figure}

The existence of  mixed reentrant transitions motivated us to  investigate also the behaviour of selected physical quantities, e.g., the magnetizations, specific heat  and compressibility  with the goal to provide a more  complete understanding of the considered coupled spin-electron system. We start our discussion with the particular case with a $F$ further-neighbour interaction $J'/J>0$ at  half-filling  $\rho=2$. If the  additional further-neighbour interaction is relatively small, the system should exhibit the $AF$ long-range order at low enough temperatures and the disordered $P$ phase at higher temperatures. The spontaneous staggered magnetizations $m^s_i$ and $m^s_e$ of the localized Ising spins and mobile electrons shown in Fig.~\ref{fig9}(a) indeed confirm the $AF$ nature of both spin as well as electron subsystems.  
The magnetic moment of localized Ising spins exhibits a perfect Ne\'el long-range order characterized by the maximal value of $m_i^s=1$ at zero temperature. On the other hand, the quantum fluctuations present in the electron subsystem lead to a quantum reduction of the staggered magnetization of mobile electrons $m_e^s$. For this reason, the saturation value of $m_e^s$ is not equal to its maximal value,  but  reaches the value $2/\sqrt{1+(t/J)^2}$. Both staggered magnetizations remain nearly constant as temperature increases up to moderate temperatures. Then they  rapidly vanish in the vicinity of the critical temperature  with the identical critical exponent $\beta_m = 1/8$ from the standard Ising universality class. It is evident from Fig.~\ref{fig9}(a) that  the critical temperature declines upon strengthening of the $F$ further-neighbour  coupling $J'/J$. The significant changes in the magnetization curves are also reflected in the specific heat, where a relatively  narrow sharp but still finite maximum (cusp) is observed at the critical temperature  for  both non-integer and integer electron concentrations. The previously conjectured possibility of a logarithmic divergence of the specific heat for integer average electron concentrations~\cite{cenci1} has been ruled out by  more accurate numerical calculations with the temperature step up to 10$^{-9}$. The present results definitively confirm the finite character of the narrow sharp maximum of the specific heat also for integer values of electron concentration, as illustrated in Fig.~\ref{fig9a}. 
The specific-heat curves may also exhibit an additional broad maximum located at higher temperatures with the dominant contribution from the electron subsystem.  The situation is very different for a sufficiently large $F$ further-neighbour coupling $J'/J$, where only the  $F$ long-range order is present, e.g., $J'/J=0.45$ (see Fig.~\ref{fig10}(a)).
In this case both uniform magnetizations start from the identical value $m_i=m_e=1$ and commonly vanish at the critical temperature  keeping the critical exponent $\beta_m = 1/8$ identical with the standard Ising universality class. Nevertheless, there is an evident difference in magnetizations at moderate temperatures, where  the electron subsystem is more susceptible with respect to  temperature fluctuations than the spin subsystem. Also the character of the specific heat is slightly different with a narrow finite cusp, whose position (contrary to the small $J'/J$) shifts to a higher temperature and its magnitude becomes smaller. However, the most attractive type of reentrant phase transitions relates to the thermally driven  magnetic transition from the $AF$ to the $F$ state through the intermediate $P$ state. Under this circumstance the $AF$ order vanishes at relatively low critical temperature with an identical behaviour of magnetic moments for spin and electron subsystems.  Afterwards the $F$ phase becomes favorable at moderate temperatures, which is manifested by a loop character
  of uniform magnetizations  with $m_i\gtrsim m_e$. The specific heat consequently displays an interesting temperature dependence, in which the development or disappearance of spontaneous (uniform or staggered) magnetizations is reflected by finite cusps (Fig.~\ref{fig10}(b)).
\begin{figure}[h!]
\begin{center}
\includegraphics[scale=0.4,trim=0 0 0.9cm 0.5cm, clip]{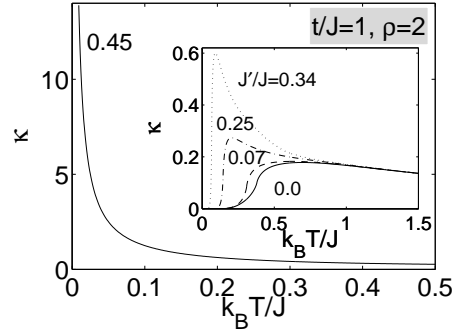}
\caption{\small Thermal dependencies of the electron compressibility for a few different values of the $F$ further-neighbour interactions, the electron concentration $\rho=2$ and the hopping parameter $t/J=1$. }
\label{fig11}
\end{center}
\end{figure}
It is found from the analysis of the electron compressibility $\kappa$ that the system in the $AF$ state  exhibits a huge rigidity ($\kappa(0)=0$), which is  generally weakened as the ratio  $J'/J$ increases (inset in Fig.~\ref{fig11}). An increase in the further-neighbour Ising interaction of the $F$ type leads to the formation of a visible kink connected to the fluctuation of particles, which become more free and destroy the stability of the system. Thus, the magnitude of the kink determines the degree of its compressibility.  On the other hand, the transition to the $F$ state ($J'/J=0.45$ in Fig.~\ref{fig11}) due to the further-neighbour $J'/J$ interaction is accompanied with a rapid divergence of the compressibility, which indicates a reduction of the system's rigidity. 
 
\begin{figure}[h!]
\begin{center}
\includegraphics[width=4.3cm,height=3.85cm,trim=0.15cm 0 0.7cm 0.5cm, clip]{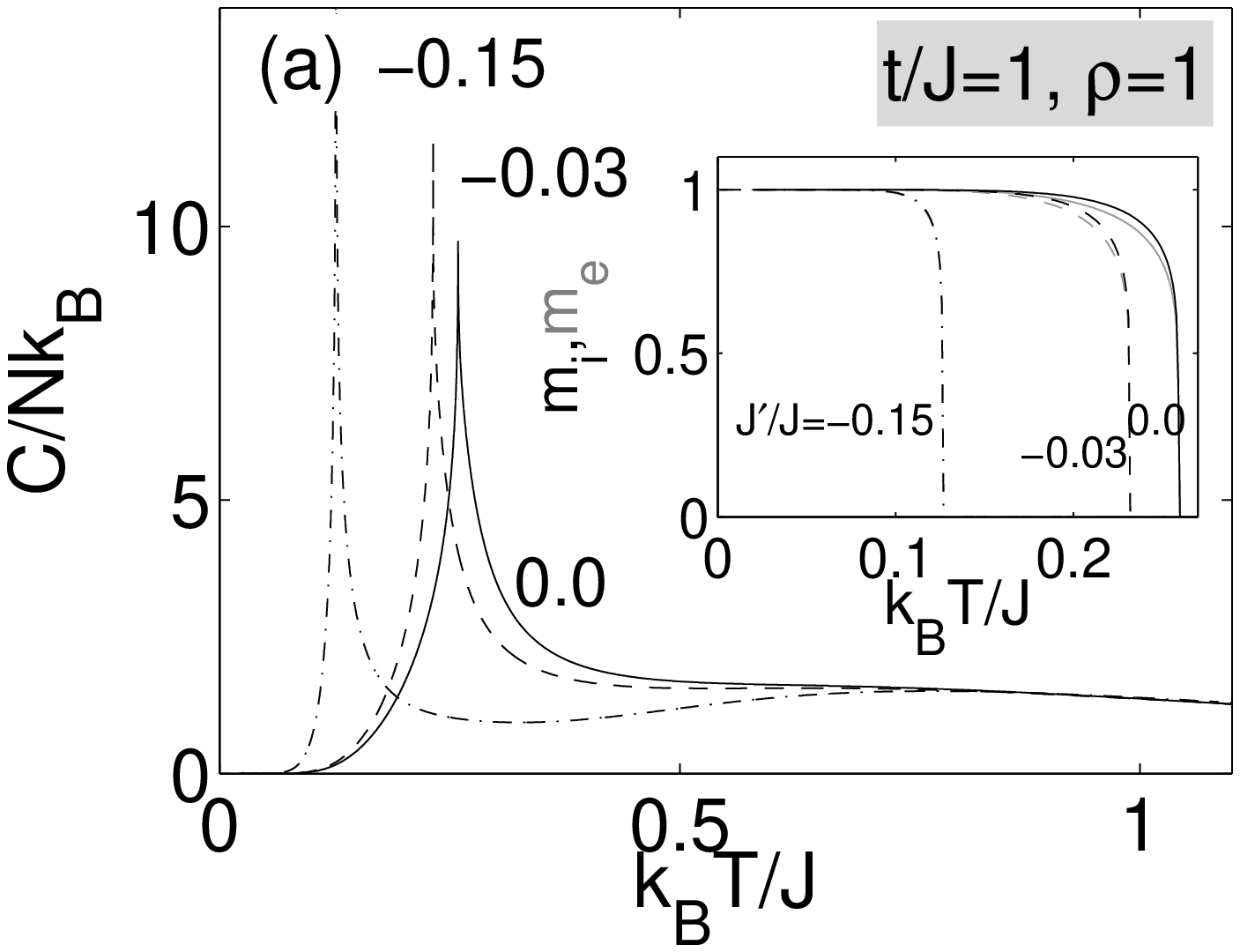}
\includegraphics[width=4.3cm,height=3.85cm,trim=0.2cm 0 0.7cm 0.5cm, clip]{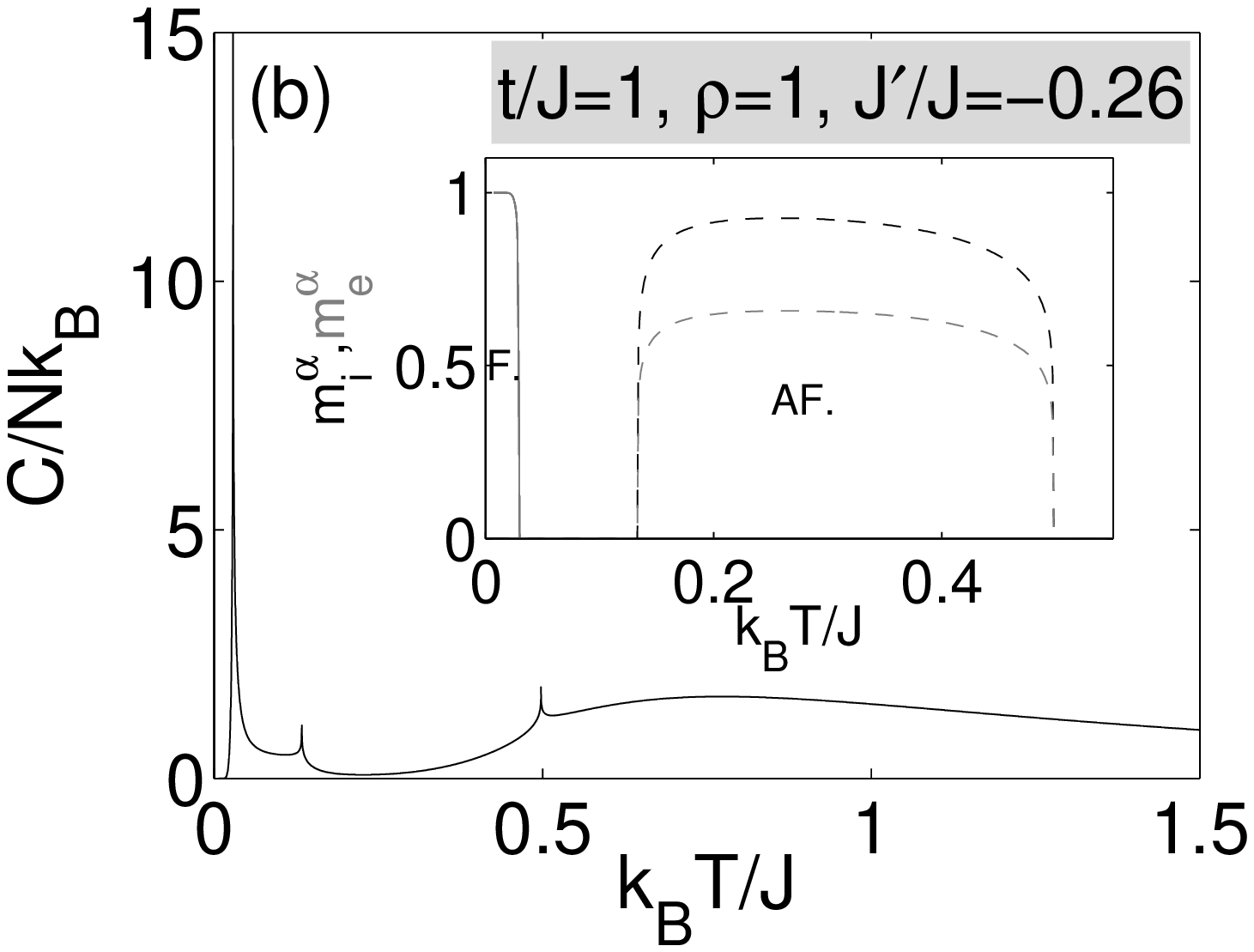}
\caption{\small (a) The specific heat and corresponding magnetizations $m_i$ (black lines) or $m_e$ (gray lines) as a function of temperature for $t/J=1$, $\rho=1$ and $J'/J<0$. (b) The thermal dependence of the specific heat  in a case with  mixed reentrant transitions for $J'/J<0$. Inset: uniform (solid lines) and staggered (dashed lines) magnetizations as a function of temperature for the same model parameters.}
\label{fig12}
\end{center}
\end{figure}

Contrary to the previous case, the $AF$ further-neighbour interaction $J'/J$ reduces the $F$ phase. Owing to this fact, the mechanism of thermally-induced changes in the spontaneous magnetization is different. For a relatively small $|J'|/J$, which is not strong enough to destroy the $F$ ground state, the uniform  magnetization of both subsystems starts from the zero-temperature asymptotic value equal to its saturation value provided that the electron concentration is equals to $\rho$=1, as shown in the inset of Fig.~\ref{fig12}(a).
The specific heat curves show just two separate maxima, one significant finite cusp connected to the order-disorder phase transition and  one more or less visible broad maximum whose origin lies predominantly in thermal excitations of the  electron subsystem. 
For the case with  mixed reentrant phase transitions $F$-$P$-$AF$-$P$, the uniform spontaneous magnetizations within the $F$ phase decline until they both vanish at the lowest critical temperature. A further temperature increase is responsible for the up rise of the staggered magnetizations within the $AF$ phase with a more interesting loop thermal dependencies with $m_i^s\gtrsim m_e^s$.  In accordance with this phenomenology,  the corresponding temperature dependence of the specific heat shows a finite cusp at each critical temperature, where either spontaneous uniform or staggered magnetization disappears, as illustrated in Fig.~\ref{fig12}(b). 
\begin{figure}[h!]
\begin{center}
\includegraphics[width=4.3cm,height=3.85cm,trim=0.15cm 0 1.cm 0.5cm, clip]{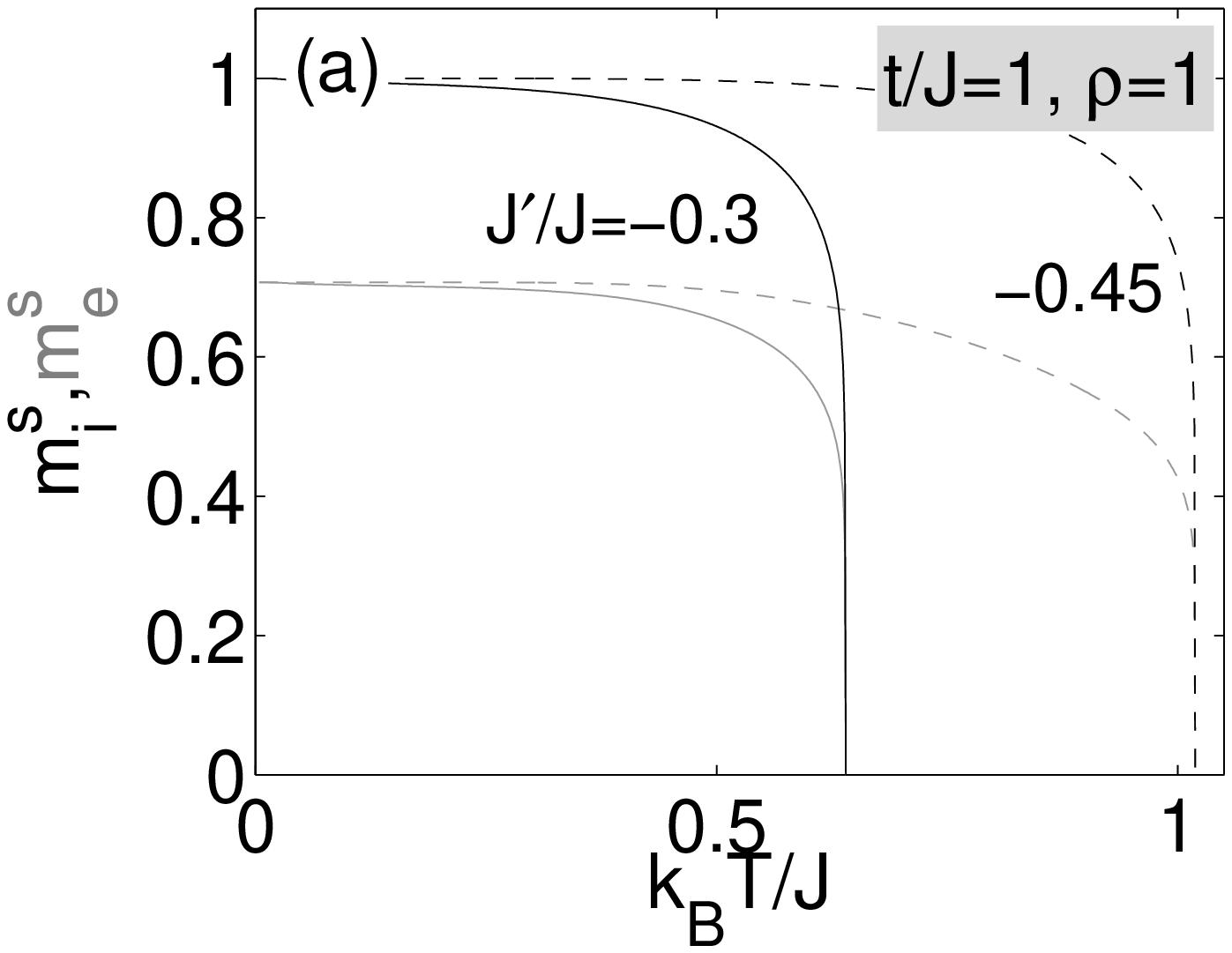}
\includegraphics[width=4.3cm,height=3.85cm,trim=0.2cm 0 1.cm 0.5cm, clip]{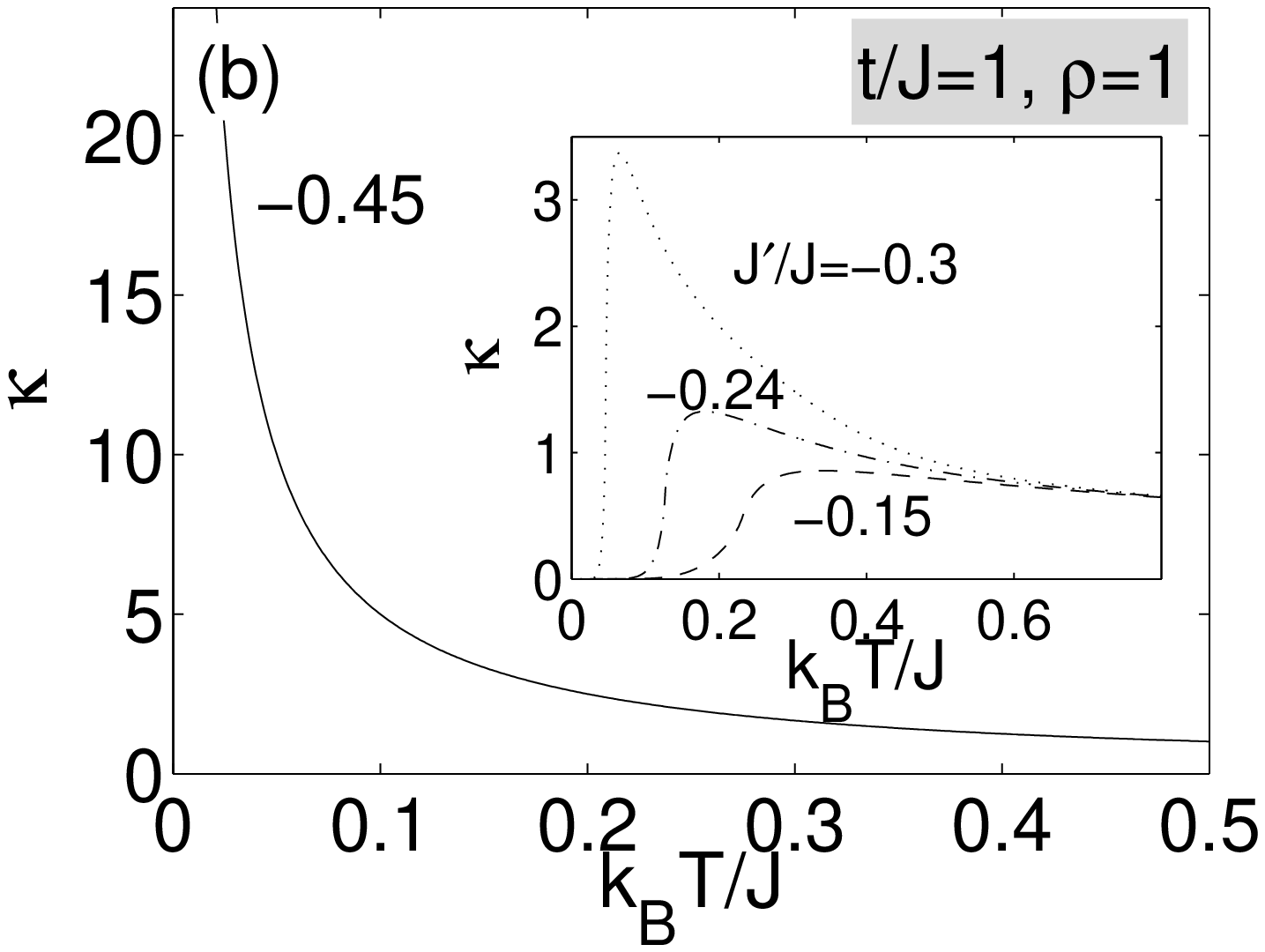}
\caption{\small (a) The staggered magnetizations $m^s_i$ (black lines) or $m^s_e$ (gray lines) as a function of temperature for $t/J=1$, $\rho=1$ and $J'/J<0$ producing the $AF$ phase. (b) The thermal dependence of the electron compressibility for a few different values of the $AF$ further-neighbour interactions, electron concentration $\rho=1$ and hopping parameter $t/J=1$.}
\label{fig12a}
\end{center}
\end{figure}
If the $AF$ further-neighbour  Ising interaction is sufficiently strong to enforce a perfect Ne\'el arrangement of the localized Ising spins,  the staggered magnetizations  follow  similar trends except  that the staggered magnetization of mobile electrons $m_e^s$ does not start from its saturation value in contrast to the staggered magnetization of the localized Ising spins (Fig.~\ref{fig12a}(a)). The corresponding specific heat has a simple thermal dependence  with a single sharp cusp located at the critical point. The electron compressibility of the system (Fig.~\ref{fig12a}(b)) with  $AF$ further-neighbour interaction indicates a huge rigidity for the $F$ phase with $\kappa=0$, while large $\kappa$ values points to a large instability in the $AF$ phase.
\\\\
We have also analyzed other basic thermodynamic characteristics out of  mixed reentrant phase transitions.  In the following, we will present just the most interesting results to demonstrate the richness of the present model. We start our discussion with the case $J'/J\gtrsim 0$ and high electron concentrations leading to the presence of a spontaneous $AF$ arrangement. It will be demonstrated that the increasing temperature influences the behaviour of sublattice magnetizations in a different way.  Out of the reentrant regime,  the staggered magnetizations  of localized Ising spins as well as mobile electrons gradually fall down with increasing temperature until they completely vanish at a critical temperature, as illustrated in Fig.~\ref{fig13}(a). 
\begin{figure}[h!]
\begin{center}
\includegraphics[width=4.3cm,height=3.85cm,trim=0.0cm 0 1.cm 0.5cm, clip]{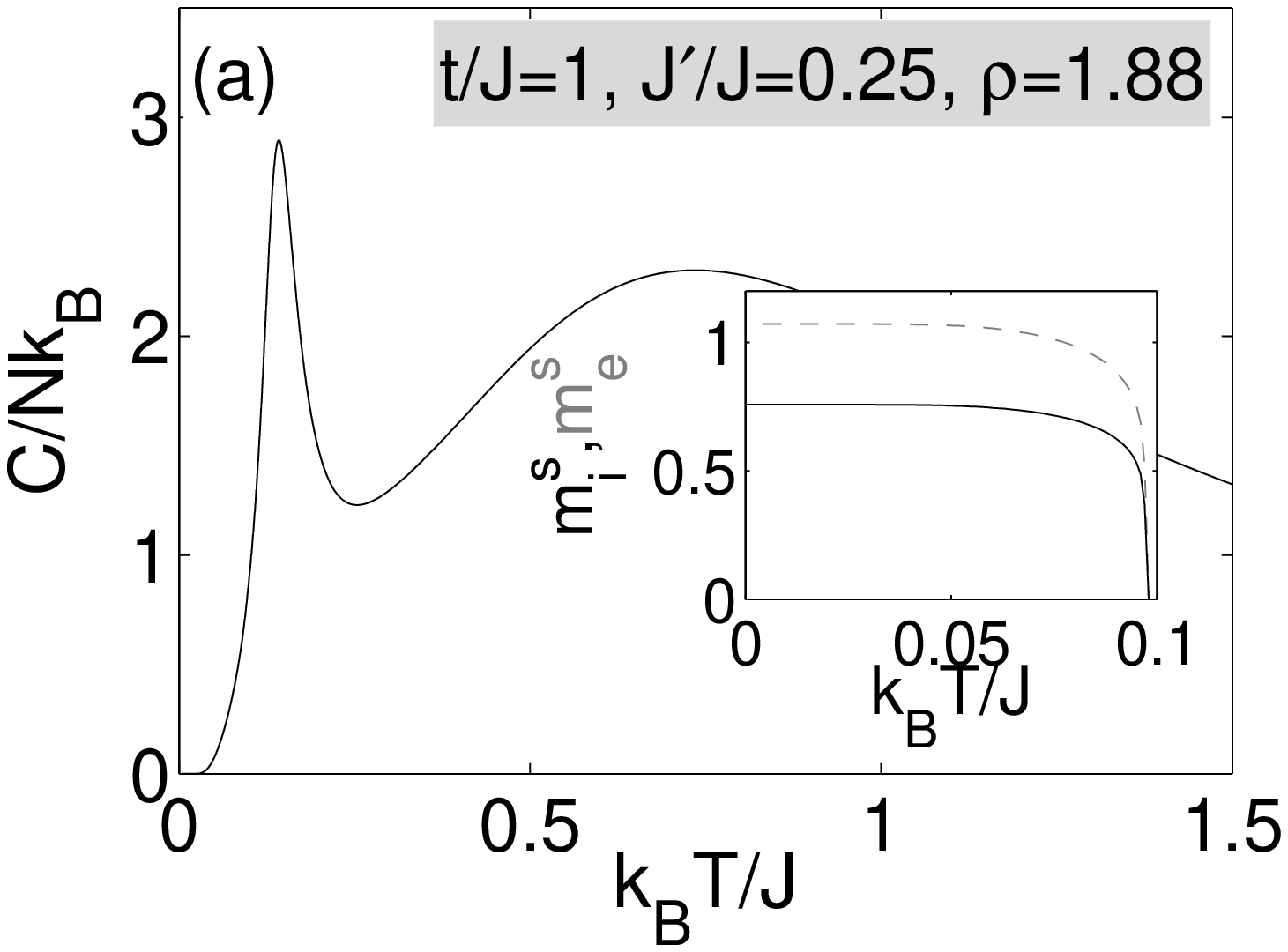}
\includegraphics[width=4.3cm,height=3.85cm,trim=0.0cm 0 1.cm 0.5cm, clip]{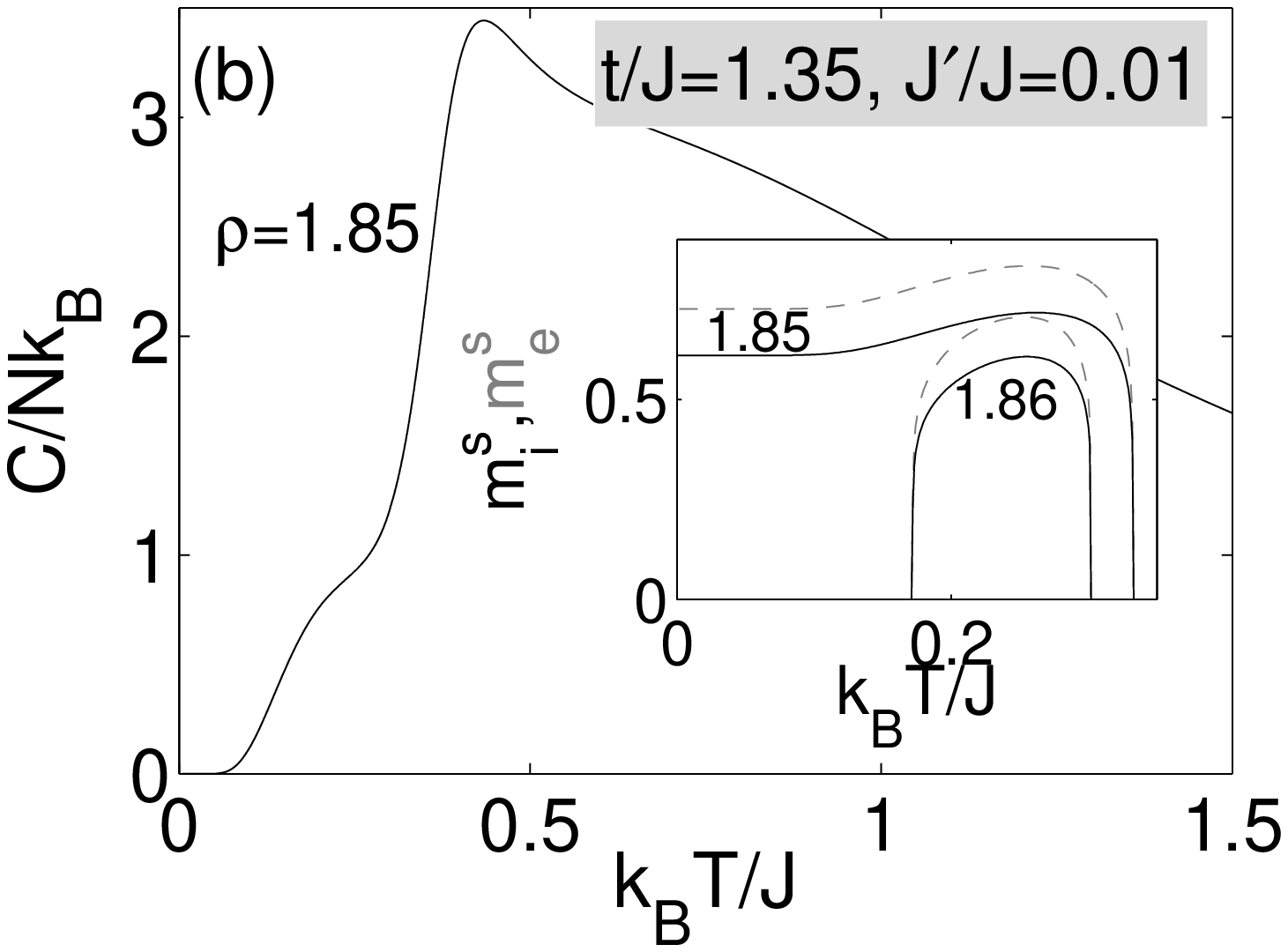}
\caption{\small Thermal dependencies of the specific heat for the $AF$ phase for $J'/J>0$, $\rho\to 2$ and (a) $t/J=1$ or (b) $t/J=1.35$. Insets: the respective thermal variations of staggered magnetizations.}
\label{fig13}
\end{center}
\end{figure}
However, it should be pointed out that the staggered magnetization  $m_e^s$ is  higher than the staggered magnetization $m_i^s$. While the staggered magnetization $m_e^s$ undergoes a quantum reduction, the staggered magnetization $m_i^s$ is substantially reduced by the annealed bond disorder. Also the specific heat displays a more diverse temperature dependence  including a sharp finite cusp along with  a  broad high-temperature maximum sometimes accompanied with another smaller broad maximum. Other types of magnetization curves are connected to the $P-AF-P$ reentrant phase transition (with a loop character) or dependences in their close neighbourhood [see upper curves in the inset of Fig.\ref{fig13}(b)]. In this parameter space region,  the specific heat has a simple behaviour, as shown in Fig.~\ref{fig13}(b). The influence of the $F$ further-neighbour  Ising interaction $J'/J$  is very significant, especially on the opposite parameter space with low electron concentrations ($\rho<1$), where the $F$ further-neighbour Ising interaction stabilize the $F$ phase. For $J'/J\gtrsim 0$, the localized Ising spins display a perfect spontaneous $F$ ordering  with $m_i\equiv 1$, quite similarly as the electron subsystem does  ($m_e\equiv \rho$). However, temperature fluctuations cause a rather steep decrease of the spontaneous magnetizations of localized Ising spins (see Fig.~\ref{fig15}(a)).
\begin{figure}[h!]
\begin{center}
\includegraphics[width=4.3cm,height=3.85cm,trim=0.0cm 0 1.1cm 0.5cm, clip]{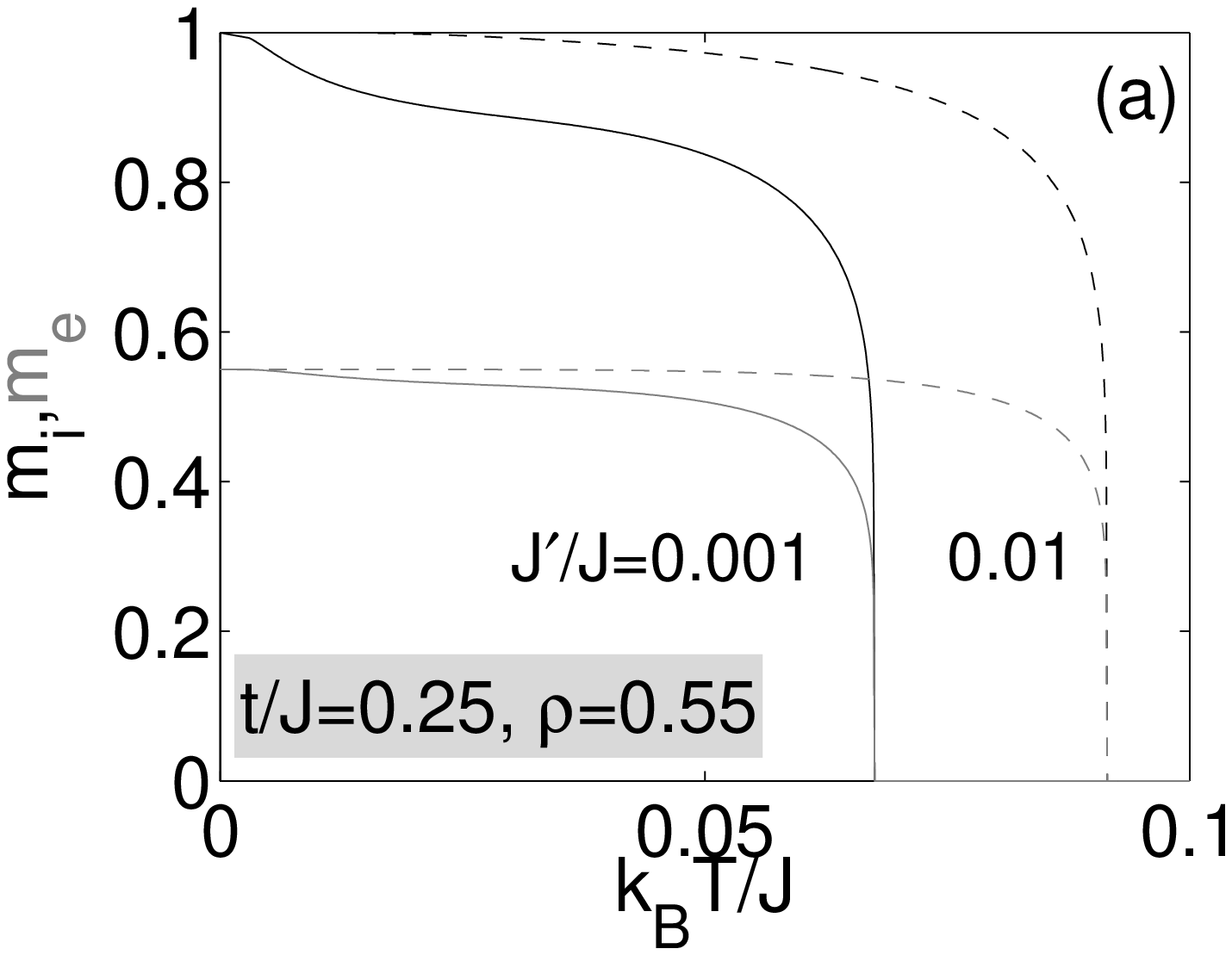}
\includegraphics[width=4.3cm,height=3.85cm,trim=0.0cm 0 1.1cm 0.5cm, clip]{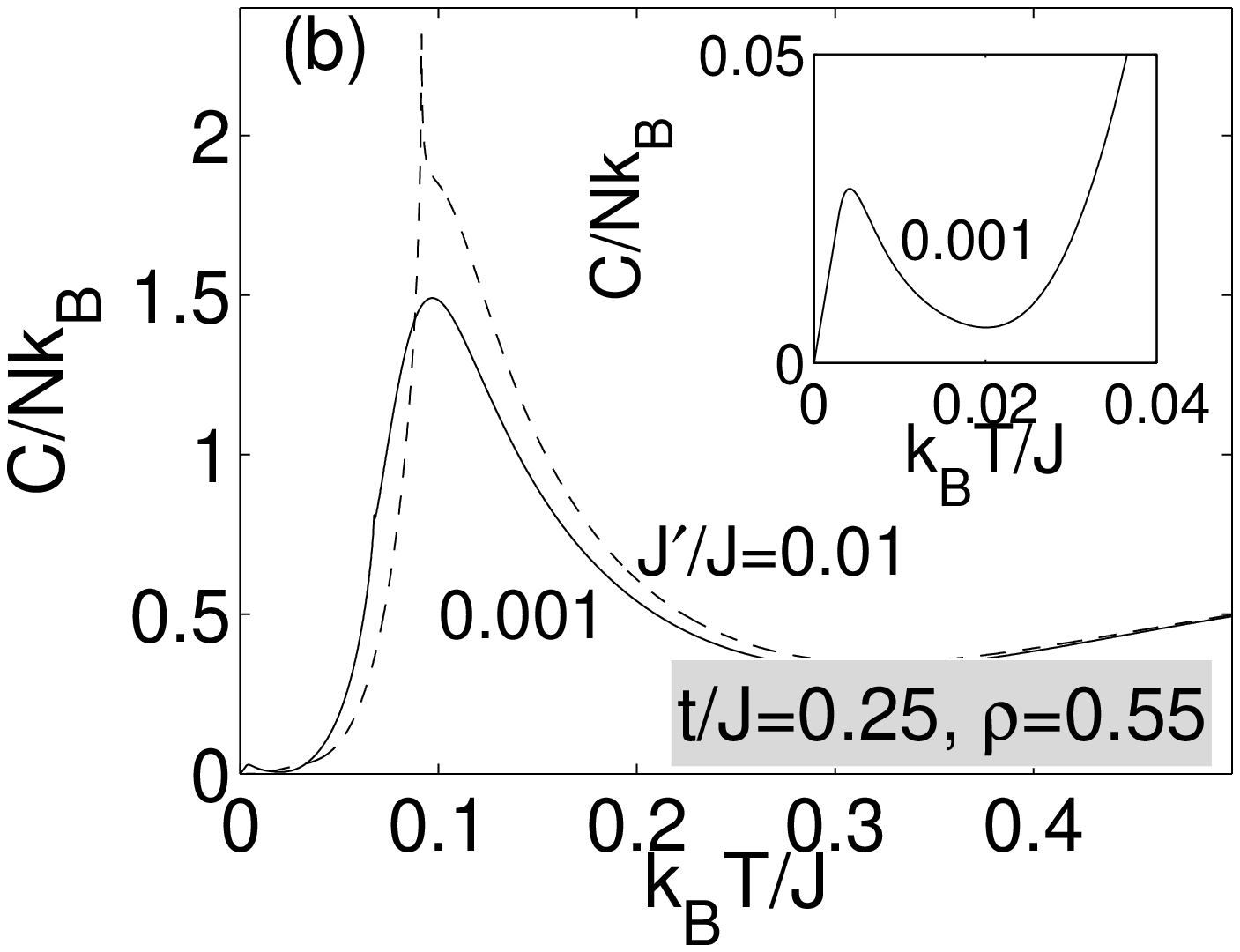}
\caption{\small (a) Thermal dependencies of spontaneous magnetizations $m_i$ (black lines) and $m_e$ (gray lines); (b) Thermal dependencies of the specific heat for the $F$ phase.}
\label{fig15}
\end{center}
\end{figure}
The increasing strength of the further-neighbour interaction $J'/J$ reduces the effect of thermal fluctuations and the spontaneous magnetizations persist at their maximum values up  to higher temperatures. The specific heat displays for this particular case a broad high-temperature maximum, originating predominantly from thermal excitations of the  electron subsystem, and an additional sharp cusp singularity, arising out from  both spin and electron subsystems, Fig.~\ref{fig15}(b). Note, furthermore, that the specific heat exhibits another low-temperature maximum (inset in Fig.~\ref{fig15}(b)) for $J'/J\to 0$ with a very small height. Such low-temperature maximum is missing in the model without the further-neighbour coupling $J'/J=0$, so its presence can be attributed to the Ising subsystem. For electron concentrations above a quarter-filling, the thermal variation of the spontaneous magnetization can be divided into two groups, namely,  with or without reentrant phase transitions. The 
  reentrant transitions manifest themselves in the corresponding magnetizations as loop dependencies. Depending on the hopping amplitude, the spontaneous magnetization of mobile electrons $m_e$ can reach  higher ($t/J< 1$) or smaller ($t/J\geq 1$) values than the spontaneous magnetization of localized spins $m_i$, as it is shown in Fig.~\ref{fig16}. Comparing this observation with the $J'/J=0$ case, it is plausible to suppose  that the change of the relationship between $m_i$ and $m_e$ is a direct consequence of the further-neighbour interaction $J'/J$.
\begin{figure}[h!]
\begin{center}
\includegraphics[width=4.3cm,height=3.85cm,trim=0.0cm 0 0.5cm 0.5cm, clip]{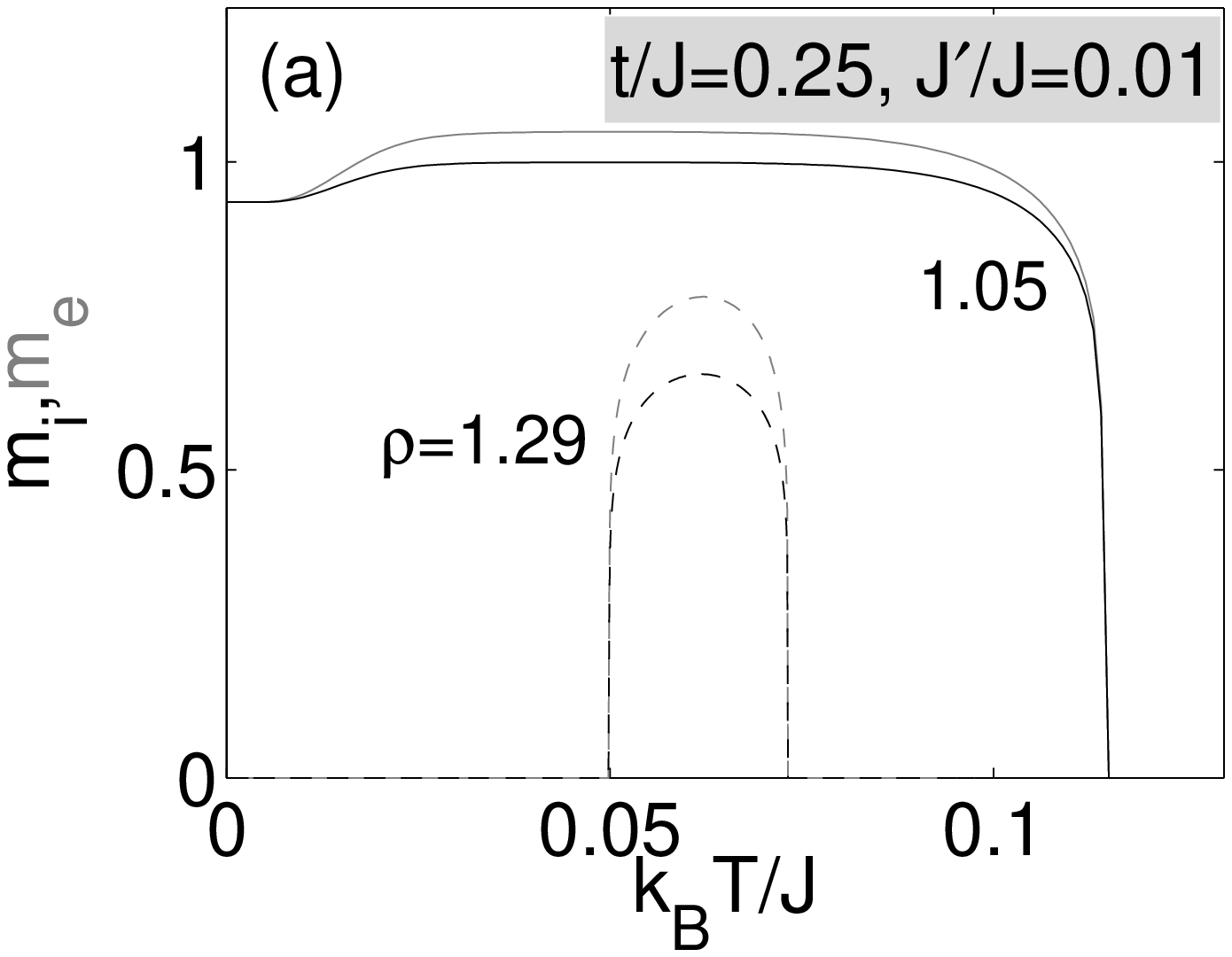}
\includegraphics[width=4.3cm,height=3.85cm,trim=0.0cm 0 1.cm 0.5cm, clip]{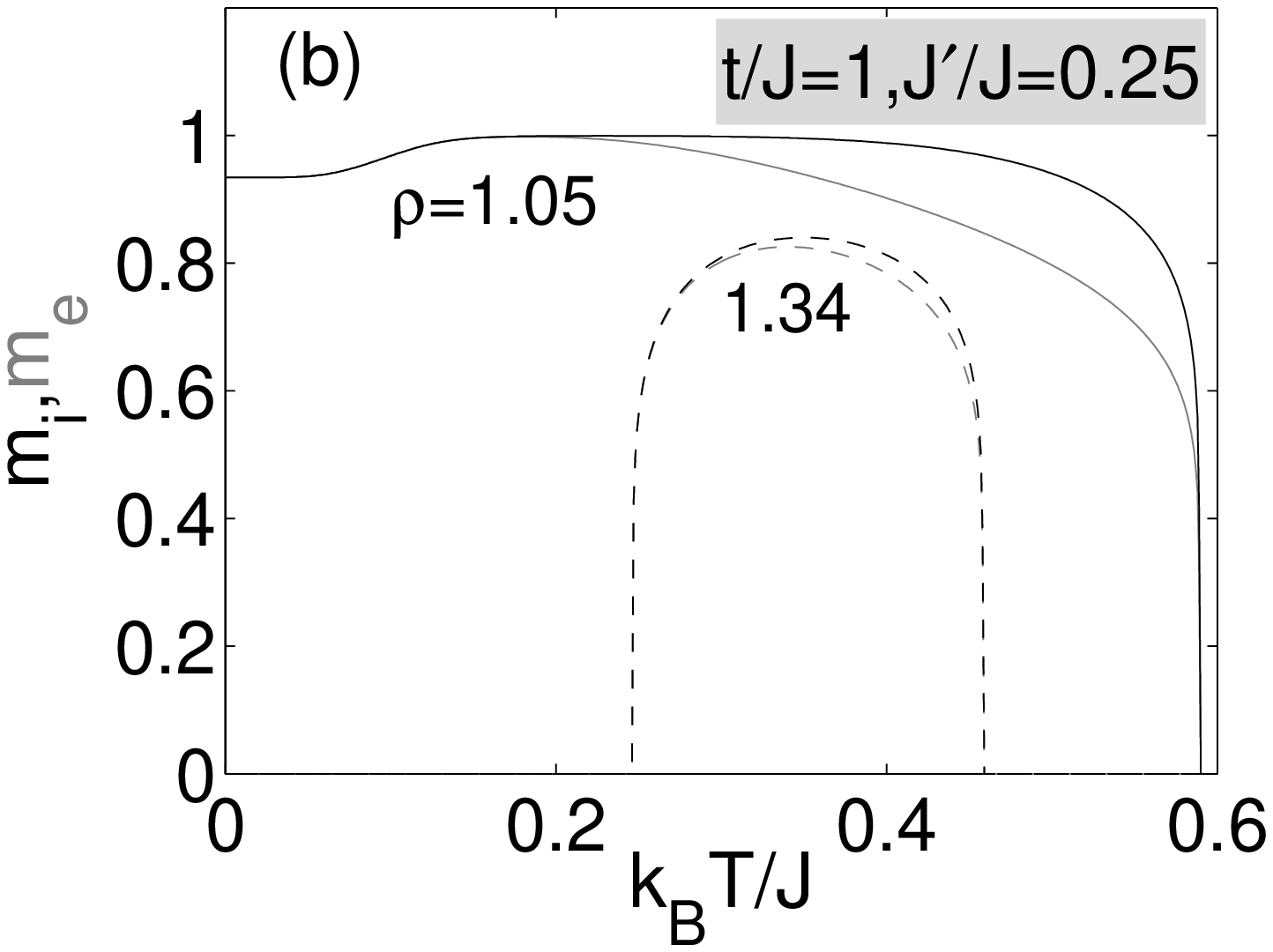}
\caption{\small Thermal dependencies of spontaneous magnetizations $m_i$ (black lines) and $m_e$ (gray lines) in the $F$ phase for $\rho>1$ and (a) $t/J< 1$, (b) $t/J\geq1$. }
\label{fig16}
\end{center}
\end{figure}\\
Out of the reentrant regime, it is evident that the spontaneous magnetizations are reduced by the annealed bond disorder of the decorating dimers caused by fractional electron concentrations. The disorder can be partially lifted by a  non-zero temperature whose increase leads to an almost ideal ferromagnetic alignment of magnetic moments. Of course, both magnetizations vanish together  at the critical point, where thermal fluctuations become too large. 
The region with $t/J\approx 1.35$ exhibits a very interesting diversity of temperature variations of the spontaneous magnetizations, as reported in  Fig.~\ref{fig17}. 
\begin{figure}[h!]
\begin{center}
\includegraphics[width=4.3cm,height=3.85cm,trim=0.0cm 0 0.5cm 0.5cm, clip]{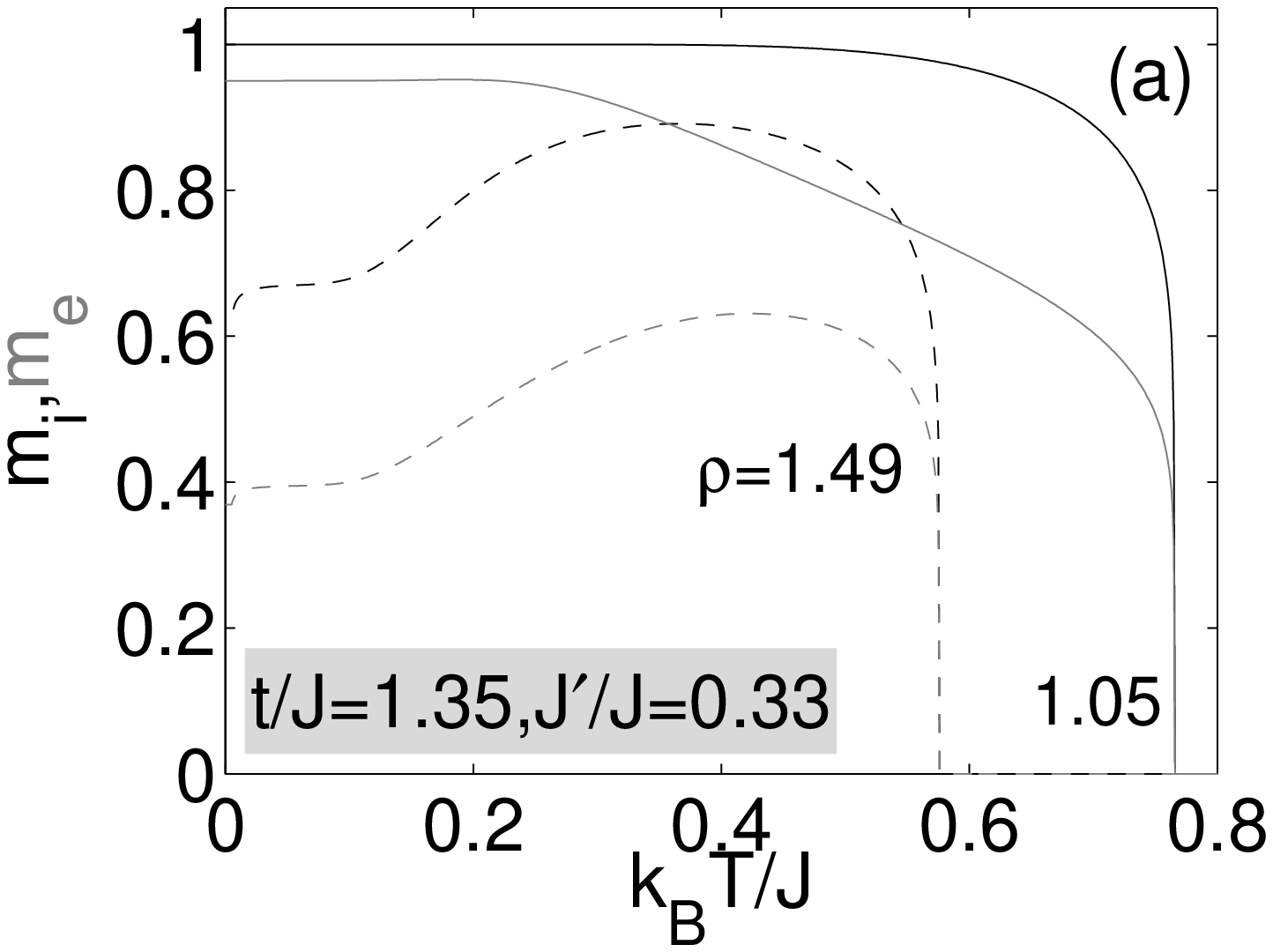}
\includegraphics[width=4.3cm,height=3.85cm,trim=0.0cm 0 0.5cm 0.5cm, clip]{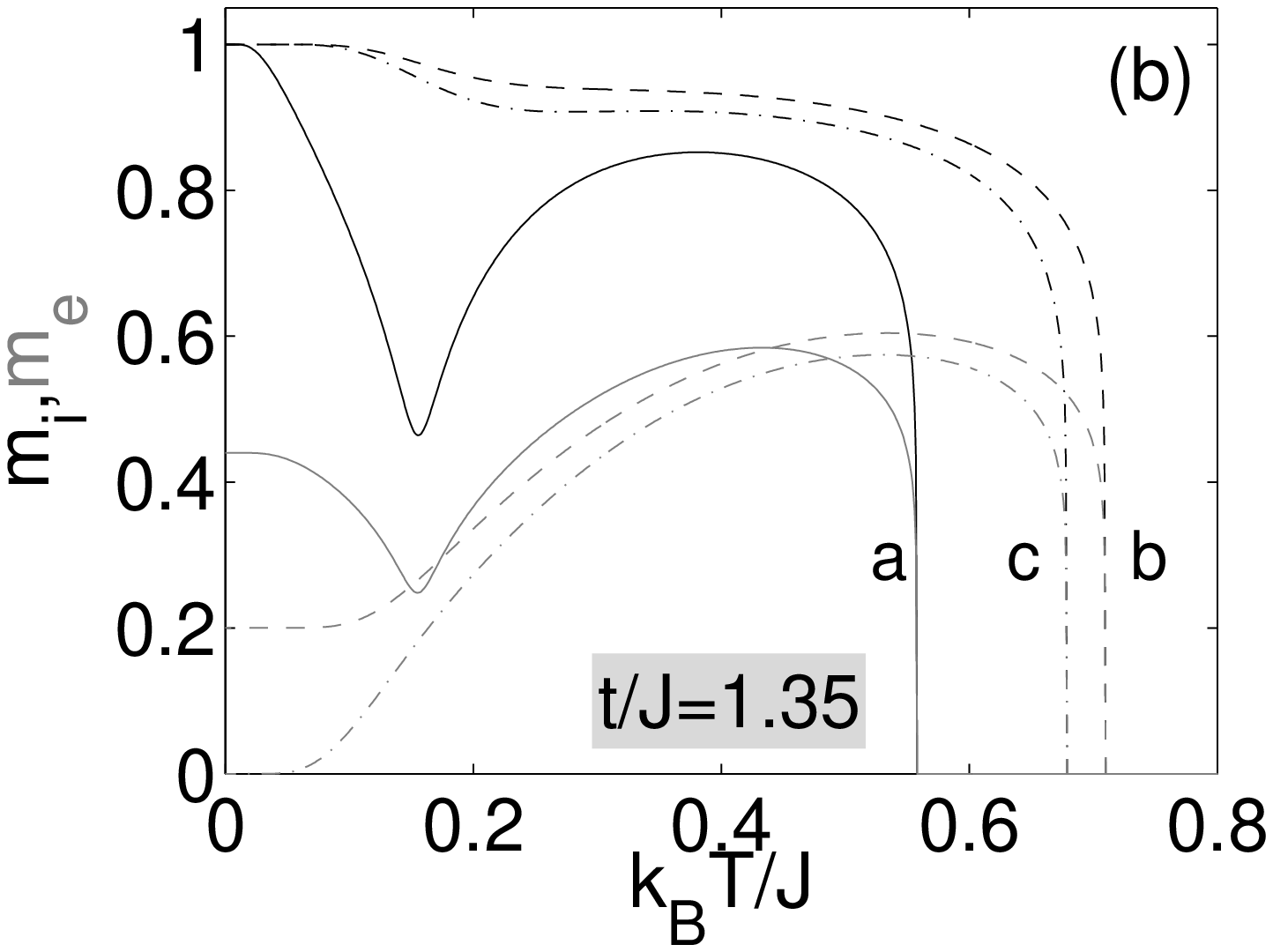}
\caption{\small Thermal dependencies of spontaneous magnetizations $m_i$ (black lines) and $m_e$ (gray lines) for $J'/J>0$, $t/J=1.35$  and various model parameters. The letters in figure (b) denote the following model parameters a:  $J'/J=0.34$, $\rho=1.56$, b: $J'/J=0.4$, $\rho=1.8$ and c: $J'/J=0.4$, $\rho=2$.}
\label{fig17}
\end{center}
\end{figure}
It should be stressed that the $F$ state is detected for all electron concentrations $\rho$ owing to a relatively strong $F$ further-neighbour coupling $J'/J$, which causes the $F$ alignment of the localized Ising spins even though the spontaneous alignment of mobile electrons is incomplete.
It can be seen from Fig.~\ref{fig17} that  increasing temperature diminishes the differences between both sublattice magnetizations, which merge together at the critical point. 

\begin{figure}[h!]
\begin{center}
\includegraphics[width=4.3cm,height=3.85cm,trim=0.0cm 0 0.5cm 0.5cm, clip]{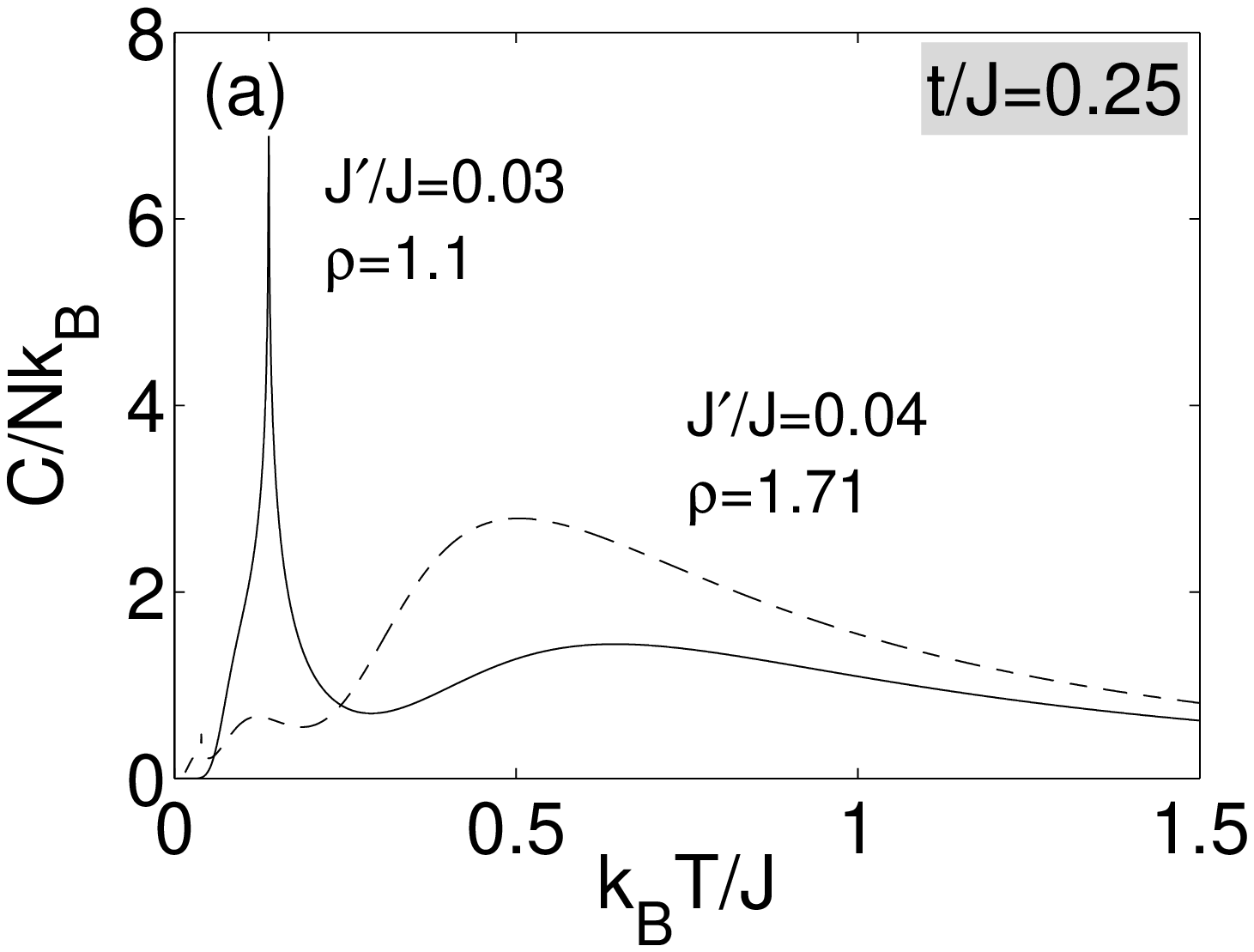}
\includegraphics[width=4.3cm,height=3.85cm,trim=0.0cm 0 0.5cm 0.5cm, clip]{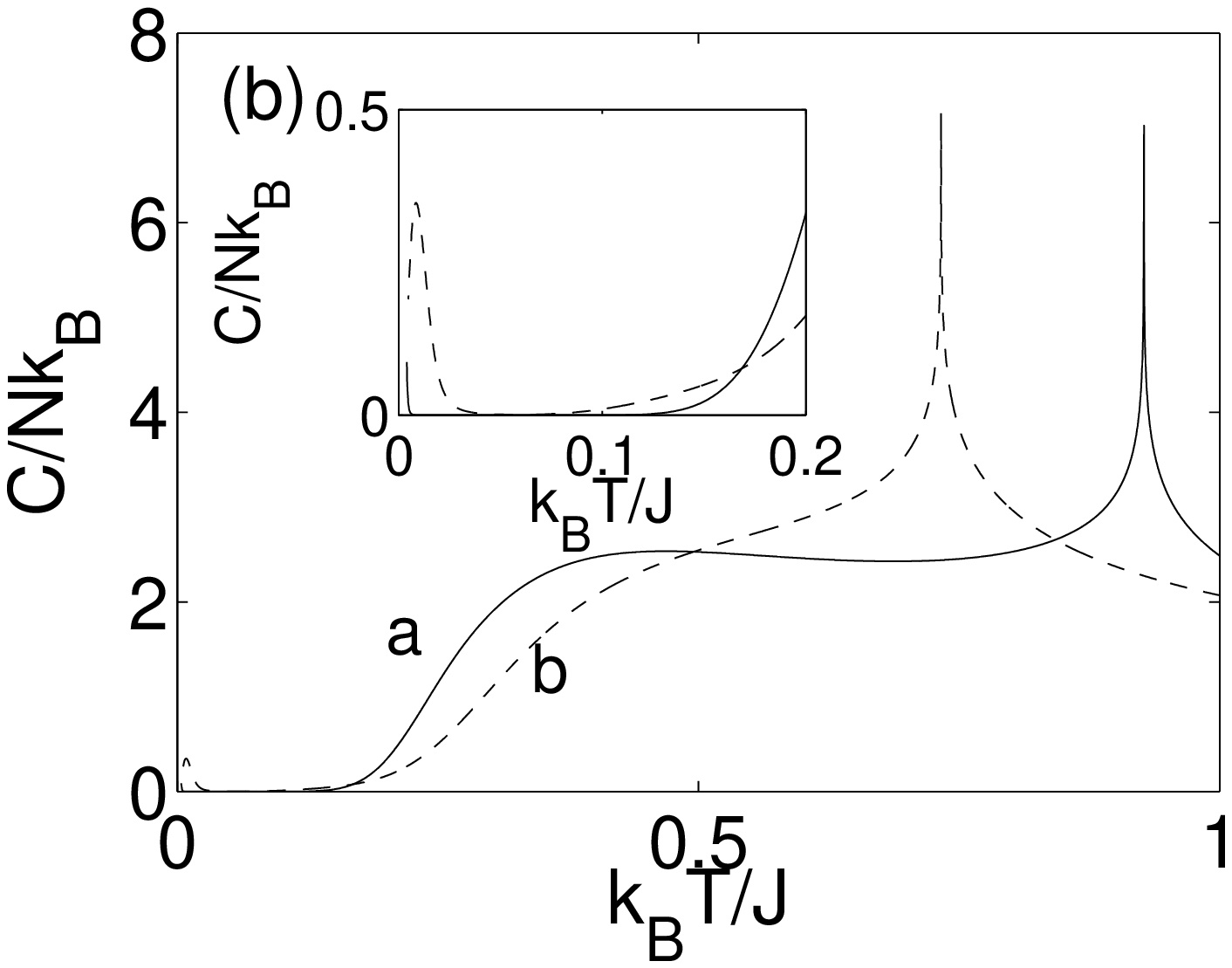}
\caption{\small Thermal dependencies of the specific heat for  $J'/J>0$, $\rho>1$ and selected values of model parameters. Different letters in figure (b) correspond to different model parameters.  a: $t/J=1$, $J'/J=0.41$, $\rho=1.05$ and b: $t/J=1.35$, $J'/J=0.32$, $\rho=1.1$. }
\label{fig18}
\end{center}
\end{figure}

\begin{figure}[h!]
\begin{center}
\includegraphics[width=4.35cm,height=3.85cm,trim=0.0cm 0 0.5cm 0.5cm, clip]{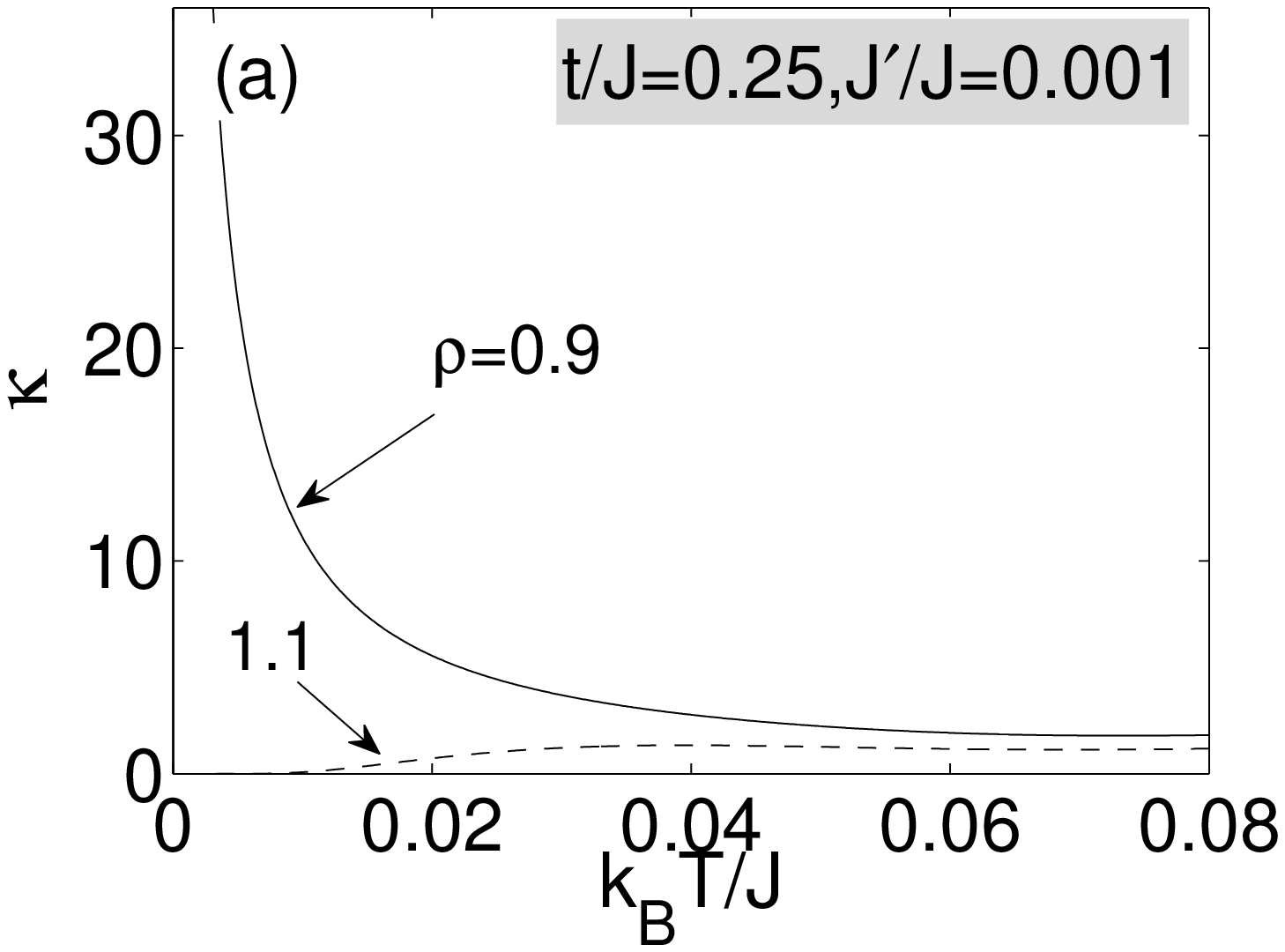}
\includegraphics[width=4.35cm,height=3.85cm,trim=0.0cm 0 0.5cm 0.5cm, clip]{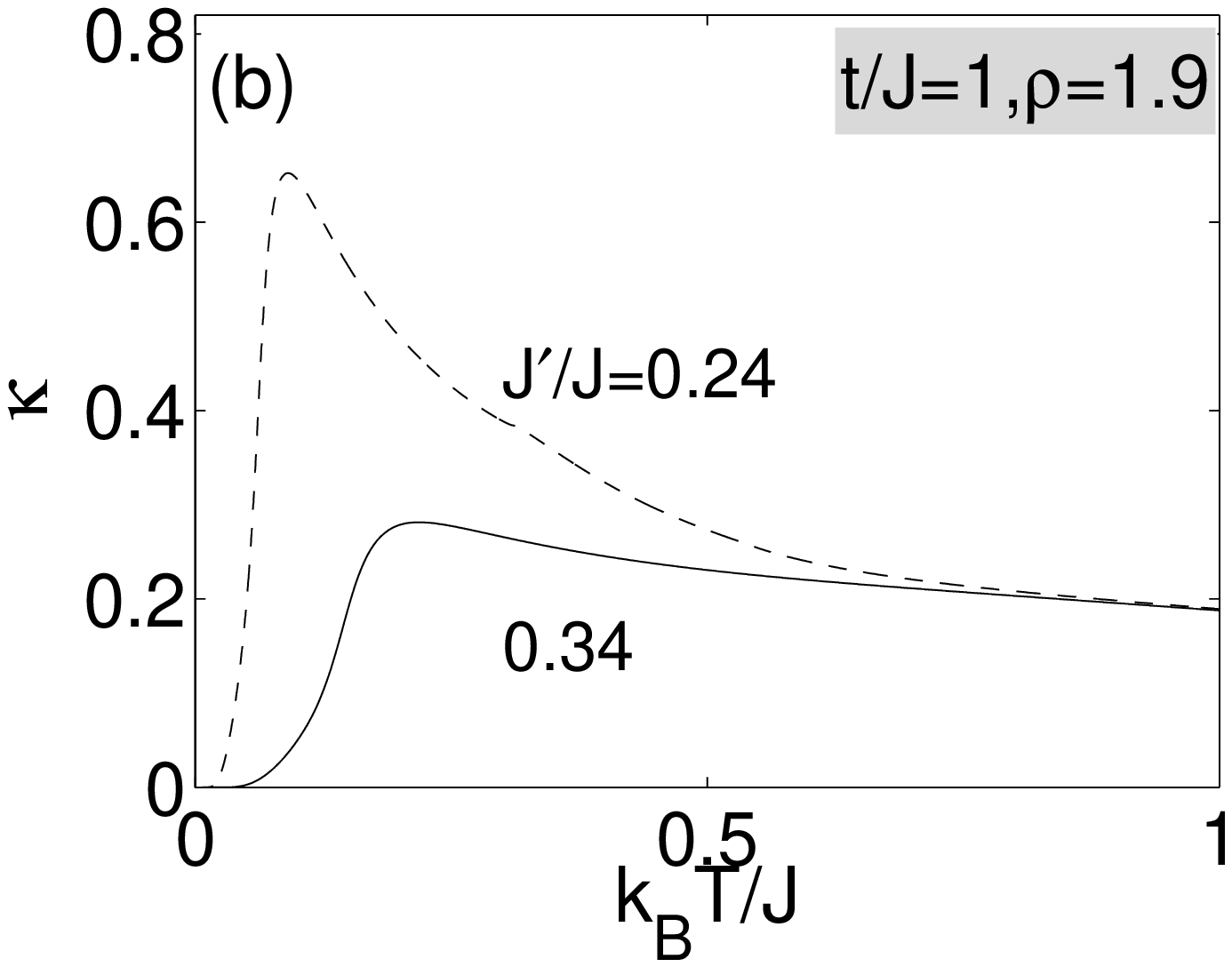}
\caption{\small The electron compressibility as a function of temperature for the $F$ further-neighbour interaction and different model parameters. }
\label{fig18b}
\end{center}
\end{figure}

The thermal variation of the specific heat in the parameter space with the electron concentration above  quarter-filling is also rich. In Fig.~\ref{fig18}, we present some typical curves obtained for different model parameters. In general, all specific heats have the multipeak structure with a more or less visible broad high-temperature maximum, which is predominantly formed by the contribution of the electron subsystem  accompanied with one or more narrow finite-cusp singularities connected with  continuous order-disorder phase transitions.
The most fascinating behaviour of the specific heat has been detected near quarter-filling provided that the hopping integral is sufficiently large ($t/J\geq 1$) and the further-neighbour Ising interaction $J'/J>0$ leads to almost perfect spontaneous $F$ order of all magnetic moments (e.g., for $t/J=1, J'/J=0.41$ or $t/J=1.35, J'/J=0.32$ in Fig.~\ref{fig18}(b)). Under this condition, the specific heat exhibits a small narrow maximum at very low temperature rapidly falling down almost to zero (inset in Fig.~\ref{fig18}(b)), with a further temperature rise. The specific heat do not reach zero value within this interval of moderate temperatures, but it is very small (of the order $10^{-4}-10^{-5}$). The analysis of the electron compressibility showed that all $F$ phases generated by the $F$ further-neighbour interaction below $\rho<1$ always exhibit a weak system stability characterized by a rapid divergence of $\kappa$ (e.g., $\rho=0.9$, $t/J=0.25$ and $J'/J=0.001$ in Fig.~
 \ref{fig18b}(a)).
On the other hand, the electron compressibility for the $F$ phase at $\rho\geq 1$ shows a huge system rigidity, which basically depends on all other parameters. The system rigidity in the $F$ phase for $\rho\geq 1$ is quite reduced at sufficiently large $J'/J$, for which the $F$ phase fills up the whole phase diagram. In the $AF$ phase, the electron compressibility always tends to zero  for very low temperatures,  indicating the huge system rigidity. However, the $F$ further-neighbour interaction rapidly reduces this phase.
\\ 
Let us turn our attention to the case $J'/J<0$. It is evident from the  phase diagrams shown in Fig.~\ref{fig6} that the $AF$ further-neighbour interaction $J'/J$ stabilizes the $AF$ phase, and constraints the $F$ one. For this reason, let us firstly analyze the $F$ phase. 
\begin{figure}[h!]
\begin{center}
\includegraphics[width=4.3cm,height=3.85cm,trim=0.0cm 0 0.5cm 0.cm, clip]{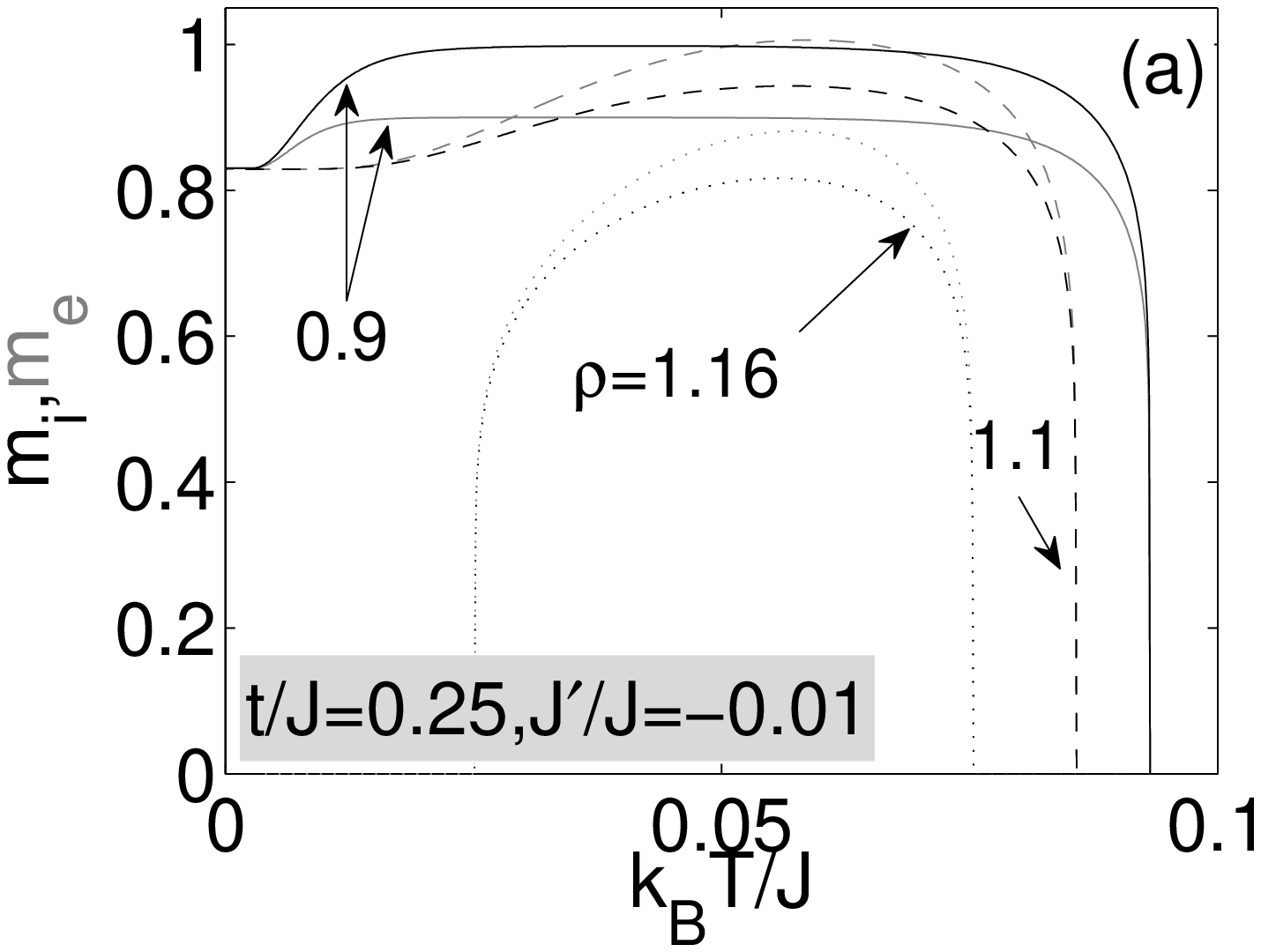}
\includegraphics[width=4.3cm,height=3.85cm,trim=0.0cm 0 0.5cm 0.cm, clip]{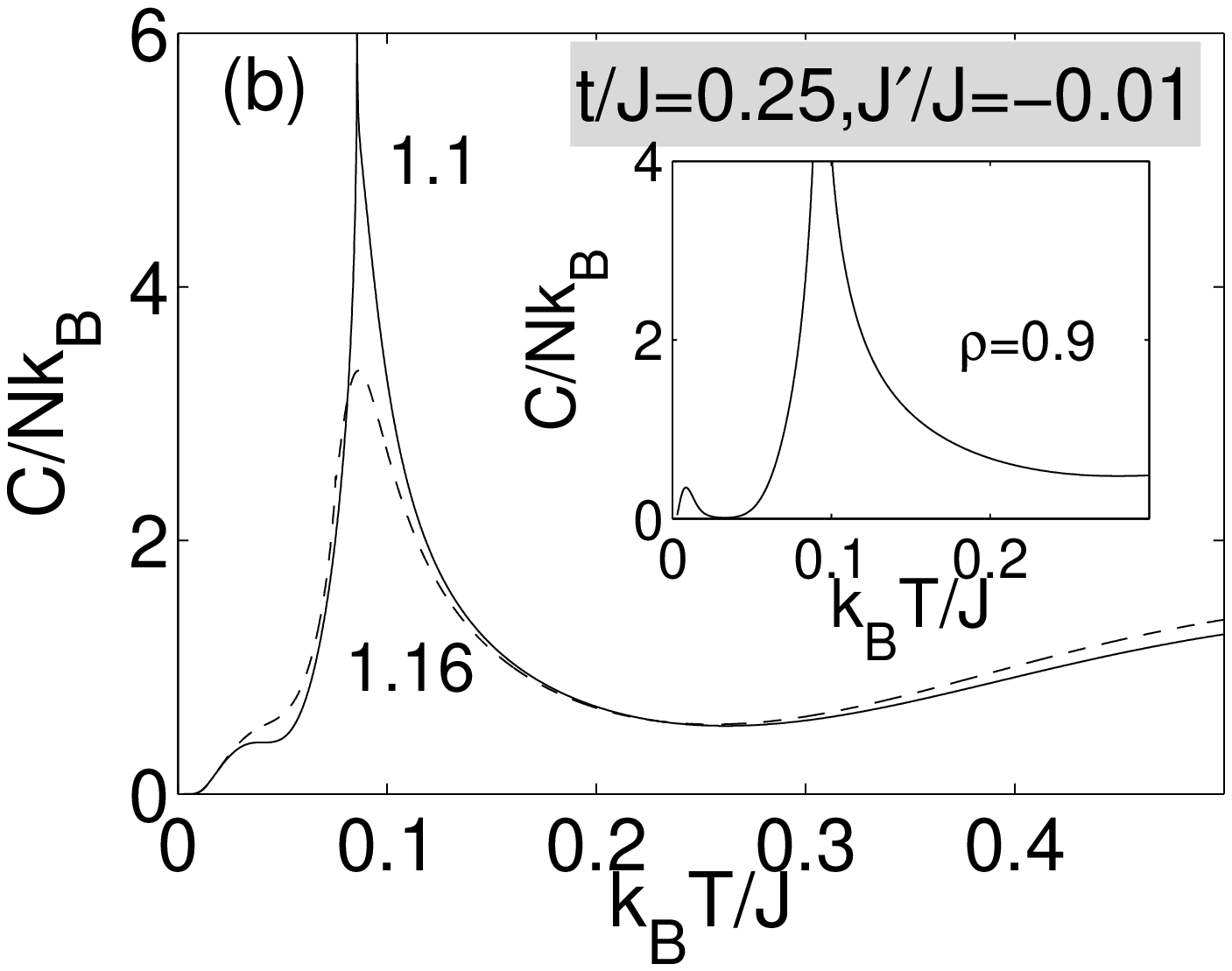}
\caption{\small (a) Thermal dependencies of spontaneous magnetizations $m_i$ (black lines) and $m_e$ (gray lines); (b) Thermal dependencies of the corresponding specific heat. }
\label{fig20}
\end{center}
\end{figure}
If the $AF$ further-neighbour interaction is sufficiently small and the system exhibits a reentrant phase transition in the $F$ phase, the character of the  spontaneous magnetizations $m_i$ and $m_e$ is influenced by the electron concentration $\rho$. Typical examples are presented in Fig.~\ref{fig20}(a). As one could expect, the loop behaviour of both magnetizations has been detected for electron concentrations near the lower and upper percolation thresholds connected to the reentrant phase transitions (e.g., $\rho=1.16$). Between these borders, both spontaneous magnetizations are reduced by the annealed bond disorder, presented at non-zero temperatures (e.g., $\rho=0.9$ or 1.1) and then commonly vanish at the critical point. It is interesting to note that this reduction is fully absent for  very small values of the further-neighbour interaction $|J'|/J\to 0$ (e.g., $J'/J=-0.001$) where an almost constant behaviour of the uniform magnetizations has been observed at low temperatures. For
  the higher value of $|J'|/J$, where  reentrant transitions in the $F$ phase are absent, both spontaneous magnetizations $m_i$ and $m_e$ are almost indistinguishable functions, whereas the electron concentration influences only the critical temperature as well as their saturation value. Moreover, our analysis shows that for $t/J >1$ the magnetization of mobile electrons is always smaller than the magnetization of localized spins. However, in the opposite limit $t/J<1$, such behaviour is observed only for $\rho\lesssim 1$. A slightly different picture has been detected for the special case $t/J=1$, where  reentrant transitions are present only for $\rho\leq 1$. For this parameter set, a  behaviour similar to the one described above for $t/J=0.25$ and $J'/J=-0.01$ has been observed. On the other hand,  both  magnetizations are almost indistinguishable for $\rho\geq 1$ and arbitrary $J'/J$.
The thermal variation of the specific heat is very similar to the ones observed for $J'/J>0$  (see Fig.~\ref{fig20}(b)). In general, the specific heat may exhibit a multipeak structure with a broad high-temperature maximum accompanied with relatively narrow finite cusps associated with the formation of spontaneous long-range order. 

The spontaneous staggered magnetizations in the $AF$ phase may also show some peculiar features.
\begin{figure}[h!]
\begin{center}
\includegraphics[width=4.4cm,height=3.85cm,trim=0.0cm 0 0.5cm 0.cm, clip]{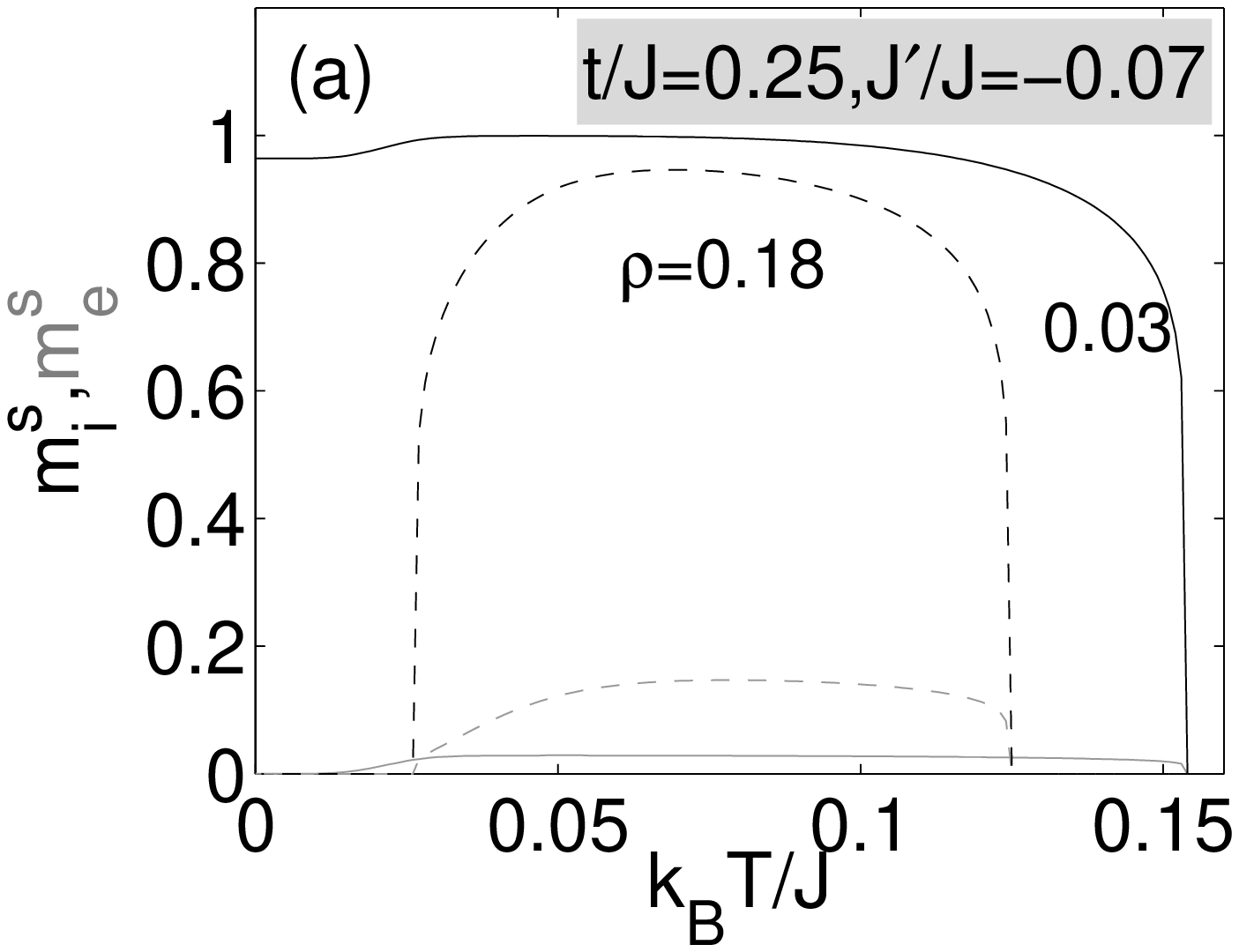}
\includegraphics[width=4.3cm,height=3.85cm,trim=0.0cm 0 0.5cm 0.cm, clip]{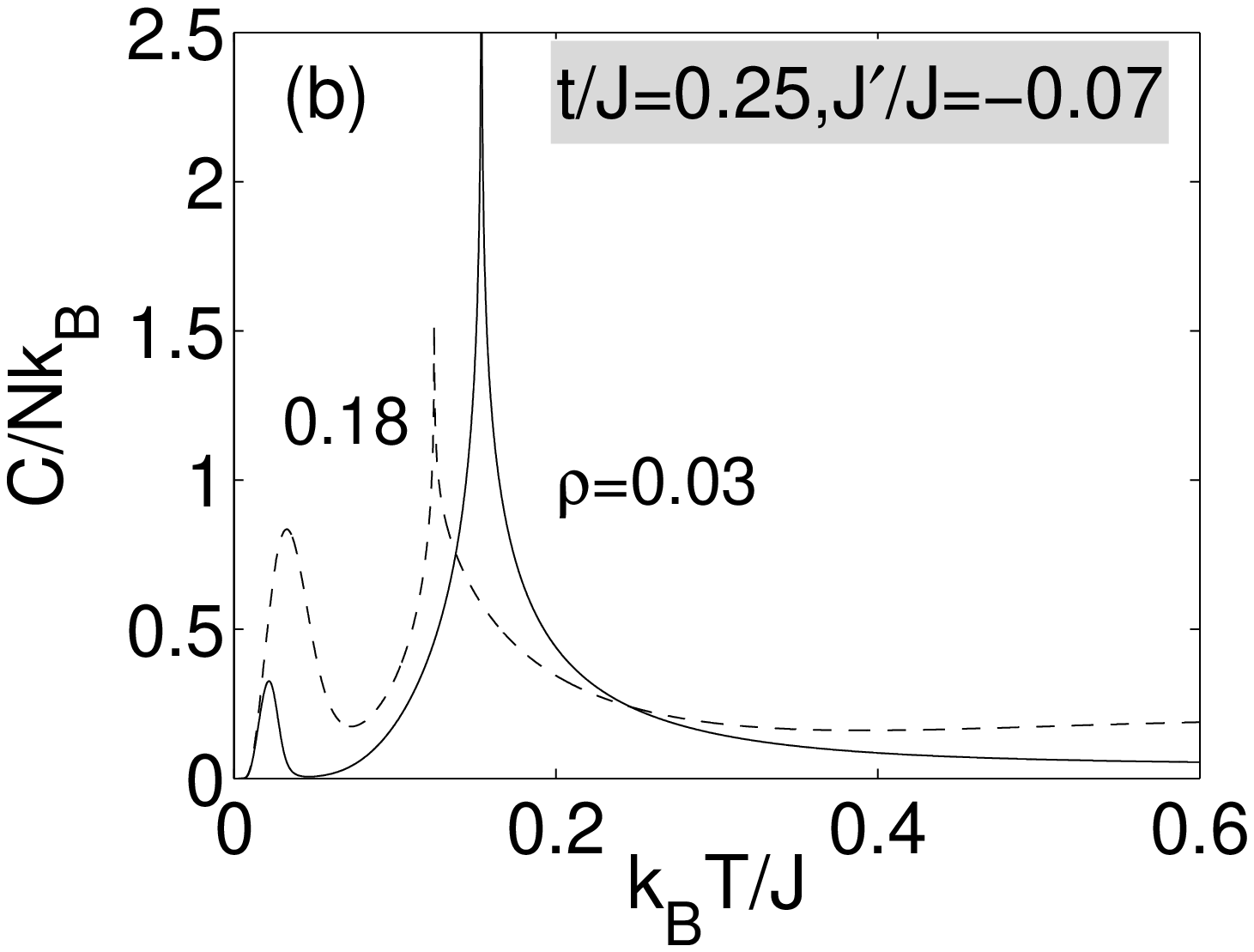}
\includegraphics[width=4.3cm,height=3.85cm,trim=0.0cm 0 0.5cm 0.cm, clip]{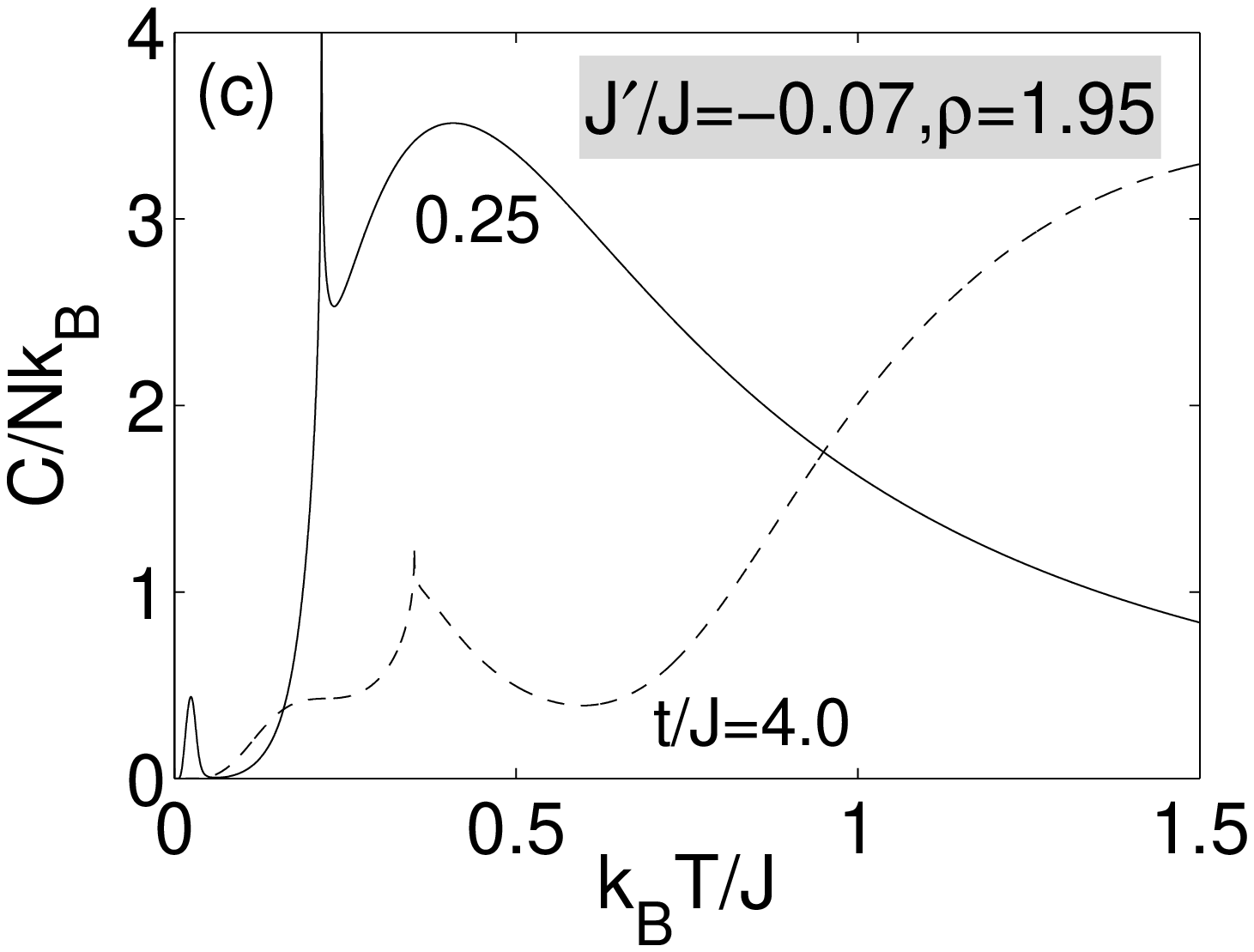}
\includegraphics[width=4.3cm,height=3.85cm,trim=0.0cm 0 0.5cm 0.cm, clip]{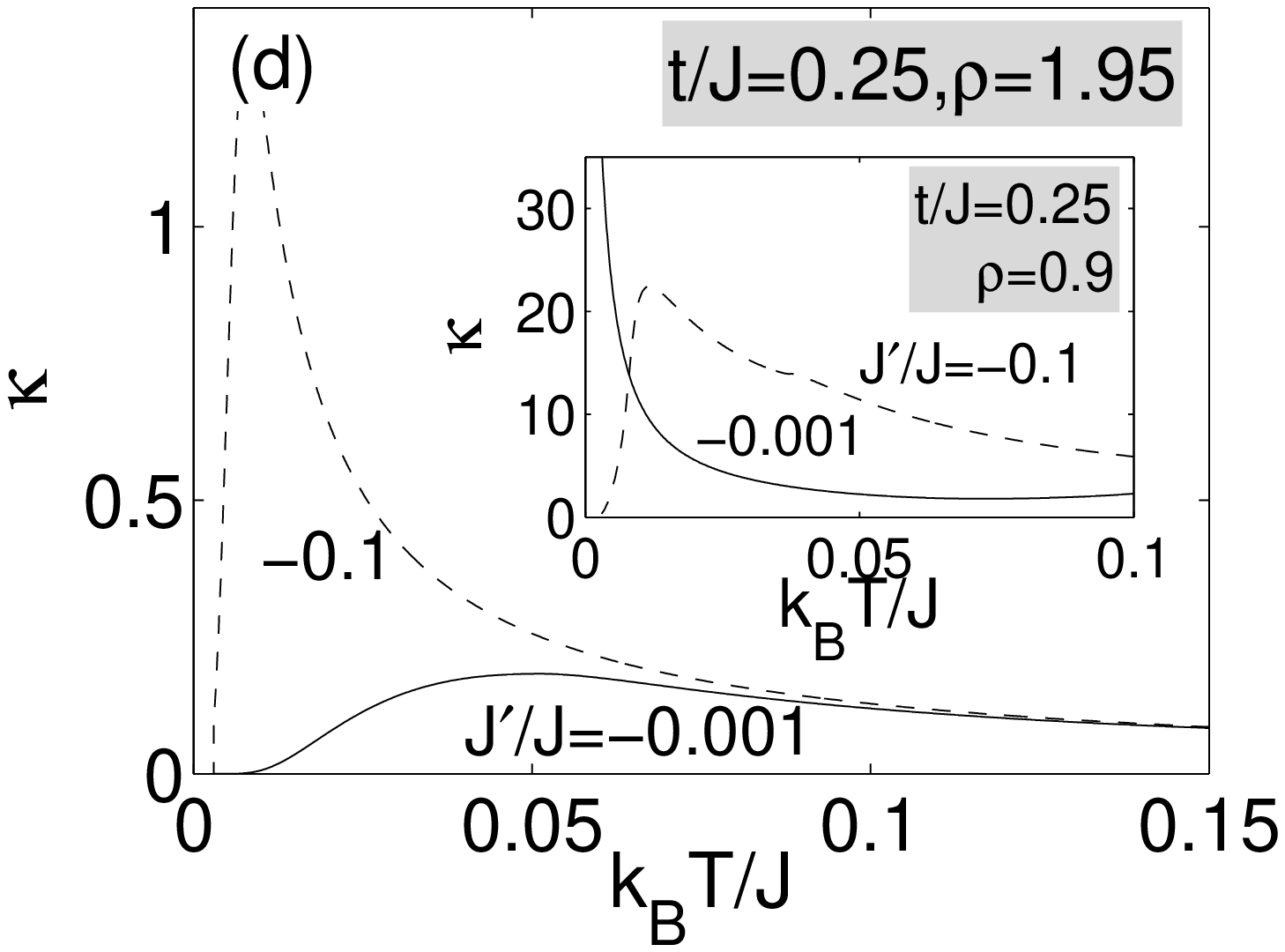}
\caption{\small Thermal dependencies of staggered magnetizations (a),  specific heat (b)-(c)  and electron compressibility (d) for $J'/J<0$  for different model parameters. Inset: The electron compressibility for the $F$ state and different $AF$ further-neighbour interaction. }
\label{fig22}
\end{center}
\end{figure}
Depending on the model parameters one can observe staggered magnetizations with or without a loop character, whereas the staggered magnetization of the mobile electrons  does not saturate neither at zero temperature. In general,  the electron concentration determines whether $m_i^s>m_e^s$ holds or vice versa. However, the $AF$ further-neighbour coupling  $J'/J<0$ is responsible for the existence of a spontaneous $AF$ long-range order also for  small electron concentrations. In this region, the spontaneous order due to the electron subsystem is marginal on account of  $m_e^s\to 0$ (Fig.~\ref{fig22}(a)). 
Contrary to this, the specific heat shows that the electron subsystem has a major influence on the low temperature behaviour. The specific heat exhibits two significant peaks, which arise from the superposition of both subsystem contributions (Fig.~\ref{fig22}(b)). Other types of magnetization dependencies were observed for sufficiently large $|J'|/J$ where the $AF$ ground state dominates  for all electron concentrations $\rho$. Namely, the localized Ising spins display a perfect  spontaneous Ne\'el order  with the saturated staggered magnetization $m_i^s\equiv 1$, while the staggered magnetization of the mobile electron obey a quantum reduction $m_e^s\equiv \rho/\sqrt{1+(t/J)^2}$. The increasing temperature gradually destroys the $AF$ long-range order due to the mutual interplay of thermal fluctuations and annealed bond disorder. Fig.~\ref{fig22}(c) illustrates a few typical thermal variations of specific heat for the special case without  mixed reentrant phase transitions. The strong influence of $J'/J<0$ has been also observed on the relative magnitude of the electron compressibility $\kappa$. While the electron compressibility of the $AF$ state  reaches zero at $T=0$ for an arbitrary $|J'|/J$ [as illustrated in Fig.~\ref{fig22}(d)], the electron compressibility of the $F$ state depends basically on the further-neighbour interaction $|J'|/J$ and electron concentration $\rho$. For $\rho\geq 1$ and arbitrary $|J'|/J$, its character is similar to the $AF$ one, while  the electron compressibility diverges below this value for small $|J'|/J$. A decrease in the  further-neighbour coupling changes this behaviour and the electron compressibility tends to zero at zero absolute temperature (see inset in Fig.~\ref{fig22}(d)). 
%
\section{Conclusion}
\label{Conclusion}
In the present work we analyzed the thermodynamic behavior of an interacting spin-electron system with a variable electron filling  on  decorating positions of doubly decorated  planar lattices by adapting the  exact solution based on a generalized decoration-iteration transformation. Besides the hopping integral $t$ and the nearest-neighbour exchange interaction  $J$, we have taken into account an additional further-neighbour exchange interaction between the  nodal Ising spins $J'$. The ground-state analysis as well as thermodynamic study have been performed for the $F$ and $AF$ further-neighbour interactions. It has been shown that the ground-state phase diagrams strongly depend on the type of further-neighbour interaction. As  expected, the non-zero value of further-neighbour coupling changes the $P$ phase (detected for $J'/J=0$) to spontaneously long-range ordered $F$ or $AF$ phases with respect to the type of the further-neighbour interaction.
The $AF$ further-neighbour interaction  stabilizes the $AF$ phase, while the $F$ one produces two new phases determined by the value of $t/J$. If the hopping term of the mobile electrons is smaller than the nearest-neighbour interaction  ($t/J<1$), the electron subsystem always prefers an electron distribution with   one parallel oriented electron per each decorating site with respect to its neighbouring Ising spin.  In the opposite limit ($t/J>1$), the electron distribution is not fundamentally influenced even though  the localized Ising spins change their orientation to the parallel one. It has been found from our analysis that  these two new phases become dominant in the phase diagram for the $F$ regime with relatively large $J'/J$ with a complete  absence of the $AF$ phase.

Our special interest has been devoted to reentrant phase transitions, which represent a highly debated problem at present.  The  reentrant phase transitions  where observed and investigated in a variety of different physical systems, e.g., binary liquid mixtures~\cite{McEwan,Narayan}, spin glasses~\cite{Binder}, superconductors~\cite{Fisher}, liquid crystals~\cite{Cladis} and   intermetallic compounds~\cite{Kolmakova,Venturini}. It is known that  different intermetallic rare-earth compounds can produce  different types of magnetic reentrant transitions~\cite{Venturini}. For example,  the manganese subsystem of the  RMn$_2$Ge$_2$ (R= Pr, Nd) compounds undergoes at first a transition from the $P$ to an $AF$ state and then to the $F$ one with  decreasing temperature for light rare earths, e.g., Pr or Nd. On the other hand, SmMn$_2$Ge$_2$ compound exhibits a phase transition from the $P$ to the $AF$ state and then from the $AF$ to the $F$ state as the temperature is lowered. 
It is evident that the spectrum of the magnetic reentrant transitions is very rich and theoretical  models for a description of this remarkable phenomenon are therefore highly desirable. 

In our previous work~\cite{cenci1} we have introduced a relative simple model describing the physics of the interacting many-body system composed of the localized Ising spins and mobile electrons. In spite of some simplifications, the model surprisingly described the existence of the reentrant behaviour, separately for the $F$ or $AF$ state. However, the model has not been able to explain an existence of  mixed reentrant phase transitions (from the $AF$ through the $P$ state to the $F$ state or vice versa) as found in many rare-earth compounds. Nevertheless, it was shown in the present paper that a little modification of this model, namely taking into consideration the further-neighbour  interaction between the localized  Ising spins, can also describe the above mentioned  mixed reentrant phase transitions. Indeed, a few novel types of reentrant transitions  have been determined upon the value of the  electron concentration $\rho$. The non-zero value of the further-neighbour interaction can produce  unique  reentrant transitions  with three consecutive critical points, namely $AF-P-F-P$ for $J'/J>0$ and $\rho\approx 2$ or $F-P-AF-P$ for $J'/J<0$ and $\rho\approx 1$, which has been detected as an effect of the additional further-neighbour Ising interaction $J'/J$ and cannot be observed for $J'/J=0$. Due to the annealed nature of the electron distribution, the specific heat presents finite-cusp singularities at the critical temperature, both for integer-valued as well as fractional electron concentrations. Finally,  the obtained result unveiled a competition between the hopping integral, the Ising interaction between nearest-neighbour Ising spin and mobile electrons and the further-neighbour   $AF$  interaction between the localized Ising spins.
 As a result, a rich  variety of temperature dependencies of magnetization  and specific heat have been presented for the coupled spin-electron system under investigation, whereas many of them are quite reminiscent of that observed experimentally in various magnetic systems (e.g. intermetallic compounds~\cite{Kolmakova,Venturini,Ghimire}). In addition,  it has turned out that the critical exponents are from the standard Ising universality class except that for the specific heat, which always shows at a phase transition a finite cusp instead of logarithmic divergence.

\vspace{0.5cm}
This work was supported by the Slovak Research and Development Agency (APVV)
under Grant APVV-0097-12 and ERDF EU Grant under the contract No. ITMS26110230097 and No. ITMS26220120005. Partial financial support from the Brazilian agencies CAPES (Coordena\c{c}\~ao de Aperfei\c{c}oamento de Pessoal de N\'{\i}vel Superior) and CNPq (Conselho Nacional de Desenvolvimento Cient\'{\i}fico e Tecnol\'ogico) is also acknowledged.

%
\begin{landscape}
\noindent\begin{minipage}[h]{1.3\textwidth}
\setcounter{table}{1}
%
\begin{center}
\resizebox{1\textwidth}{!} {
\begin{tabular}{l||l|l|l|l}
phase & $J'$& Eigenvalue (${E})$& Eigenvector & border expression\\
\hline\hline
 $\begin{array}{c}
    0
   \end{array}$       
&
 $\begin{array}{l}
    J'=0\\\rowcolor[gray]{0.8}J'<0\\\rowcolor[gray]{0.8}\\J'>0\\ 
   \end{array}$       
&
$\displaystyle\left.
\begin{array}{l}
\\ \\ \\\\
\end{array}\right\}
\begin{array}{l}
    {E}(\mbox{0})=-J'\\
   \end{array}   $
&
$\begin{array}{l}
|\mbox{0}\rangle=\prod_{k=1}^{Nq/2}|\pm1\rangle_{\sigma_{k1}}\otimes |0,0\rangle_{k}\otimes|\pm1\rangle_{\sigma_{k2}}\\
\rowcolor[gray]{0.8}
|\mbox{0}\rangle=\prod_{k=1}^{Nq/2}|1\rangle_{\sigma_{k1}}\otimes |0,0\rangle_{k}\otimes|-1\rangle_{\sigma_{k2}}\\
\rowcolor[gray]{0.8}\\
|\mbox{0}\rangle=\prod_{k=1}^{Nq/2}|1\rangle_{\sigma_{k1}}\otimes |0,0\rangle_{k}\otimes|1\rangle_{\sigma_{k2}}\\
\end{array}$
&  
$\begin{array}{ll}
\mu<-\lambda & (\mbox{0-I})\\
\rowcolor[gray]{0.8}\mu<-\lambda-2J' & (\mbox{0-I})\\
\rowcolor[gray]{0.8}\mu<-\omega & (\mbox{0-II})\\
\mu<-\lambda & (\mbox{0-I})
\end{array}$\\
\hline
 $\begin{array}{c}
    \mbox{I}
   \end{array}$       
&
 $\begin{array}{l}
    J'=0\\\rowcolor[gray]{0.8}J'<0\\J'>0\\\\\\
   \end{array}$       
&
$\displaystyle\left.
\begin{array}{l}
\\ \\ \\ \\ \\
\end{array}\right\}
\begin{array}{l}
    {E}(\mbox{I})=-\lambda-\mu-J'
   \end{array}   $
&
$\displaystyle\left.
\begin{array}{l}
\\ \\ \\ \\ \\
\end{array}\right\}
\begin{array}{l}
    |\mbox{I}\rangle=\prod_{k=1}^{Nq/2}|1\rangle_{\sigma_{k1}}\otimes \frac{1}{\sqrt{2}}\left(\hat{c}^\dagger_{k1,\uparrow}+\hat{c}^\dagger_{k2,\uparrow}\right)|0,0\rangle_k\otimes|1\rangle_{\sigma_{k2}}
   \end{array}   $
&  
$\begin{array}{ll}
-\lambda<\mu<\lambda-2\omega & (\mbox{0-I-II})\\
\rowcolor[gray]{0.8}-\lambda-2J'<\mu<\lambda-2\omega+2J' & (\mbox{0-I-II})\\
-\lambda<\mu<\lambda-2\omega+2J' & (\mbox{0-I-II$_1$})\\
-\lambda<\mu<-\xi & (\mbox{0-I-II$_2$})\\
-\lambda<\mu<\xi & (\mbox{0-I-II$_3$})
\end{array}$\\
\hline
 $\begin{array}{c}
    \mbox{II}
   \end{array}$       
&
 $\begin{array}{l}
    J'=0\\\rowcolor[gray]{0.8}J'<0\\J'>0\\ 
   \end{array}$       
&
$
\displaystyle\left.
\begin{array}{l}
\\ \\ \\ 
\end{array}\right\}
\begin{array}{l}
      {E}(\mbox{II/II$_1$})=-2\omega-2\mu+|J'|\\
   \end{array}   $
&
$\displaystyle\left.
\begin{array}{l}
\\ \\ \\ 
\end{array}\right\}
\begin{array}{lcl}
    |\mbox{II/II$_1$}\rangle&=&\prod_{k=1}^{Nq/2}|1\rangle_{\sigma_{k1}}\otimes \left[a(\hat{c}^\dagger_{k1,\uparrow}\hat{c}^\dagger_{k2,\downarrow})+b(\hat{c}^\dagger_{k1,\downarrow}\hat{c}^\dagger_{k2,\uparrow})\right.\\
&+&\left.c(\hat{c}^\dagger_{k1,\uparrow}\hat{c}^\dagger_{k1,\downarrow})+ d(\hat{c}^\dagger_{k2,\uparrow}\hat{c}^\dagger_{k2,\downarrow})\right]|0,0\rangle_k\otimes|-1\rangle_{\sigma_{k2}}
   \end{array}   $
&  
$\displaystyle\left.
\begin{array}{l}
\\ \\ \\ 
\end{array}\right\}
\begin{array}{ll}
\lambda-2\omega+2J'<\mu<-\lambda+2\omega-2J' & (\mbox{I-II/II$_1$-III})\\
\rowcolor[gray]{0.8}-\omega<\mu<\omega & (\mbox{0-II-IV})
\end{array}$\\
 $\begin{array}{c}
   \\\\
   \end{array}$       
&
 $\begin{array}{l}
 \\\\\\
   \end{array}$       
&
$\begin{array}{l}
      {E}(\mbox{II$_2$})=-2J-2\mu-J'\\
      {E}(\mbox{II$_3$})=-2t-2\mu-J'\\\\
    
   \end{array}   $
&
$\begin{array}{l}
\\ \\ \\
\end{array}
\begin{array}{lcl}
 |\mbox{II}_2\rangle&=&\prod_{k=1}^{Nq/2}|1\rangle_{\sigma_{k1}}\otimes\left(\hat{c}^\dagger_{k1,\uparrow}\hat{c}^\dagger_{k2,\uparrow}\right)|0,0\rangle_k\otimes|1\rangle_{\sigma_{k2}}\\
|\mbox{II}_3\rangle&=&\prod_{k=1}^{Nq/2}|1\rangle_{\sigma_{k1}}\otimes\frac{1}{2}\left[\hat{c}^\dagger_{k1,\uparrow}\hat{c}^\dagger_{k2,\downarrow}- \hat{c}^\dagger_{k1,\downarrow}\hat{c}^\dagger_{k2,\uparrow}\right.\\
&+&\left.\hat{c}^\dagger_{k1,\uparrow}\hat{c}^\dagger_{k1,\downarrow}+ \hat{c}^\dagger_{k2,\uparrow}\hat{c}^\dagger_{k2,\downarrow}\right]|0,0\rangle_k\otimes|1\rangle_{\sigma_{k2}}  
     \end{array}   $
&  
$\displaystyle
\begin{array}{l}
\\ \\ \\
\end{array}
\begin{array}{ll}
-\xi<\mu<\xi & (\mbox{I-II$_2$-III})\\
\xi<\mu<-\xi & (\mbox{I-II$_3$-III})\\\\
\end{array}$\\
\hline
 $\begin{array}{c}
    \mbox{III}
   \end{array}$       
&
 $\begin{array}{l}
    J'=0\\\rowcolor[gray]{0.8}J'<0\\J'>0\\\\\\
   \end{array}$       
&
$\displaystyle\left.
\begin{array}{l}
\\ \\ \\ \\ \\
\end{array}\right\}
\begin{array}{l}
    {E}(\mbox{III})=-\lambda-3\mu-J'
   \end{array}   $
&
$\displaystyle\left.
\begin{array}{l}
\\ \\ \\ \\ \\
\end{array}\right\}
\begin{array}{lcl}
    |\mbox{III}\rangle&=&\prod_{k=1}^{Nq/2}|1\rangle_{\sigma_{k1}}\otimes\frac{1}{\sqrt{2}}\left[\hat{c}^\dagger_{k1,\uparrow}\hat{c}^\dagger_{k1,\downarrow}\hat{c}^\dagger_{k2,\uparrow} \right.\\
&-&\left.\hat{c}^\dagger_{k1,\uparrow}\hat{c}^\dagger_{k2,\uparrow}\hat{c}^\dagger_{k2,\downarrow}\right]|0,0\rangle_k\otimes|1\rangle_{\sigma_{k2}}
   \end{array}   $
&  
$
\begin{array}{ll}
-\lambda+2\omega<\mu<\lambda & (\mbox{II-III-IV})\\
\rowcolor[gray]{0.8}-\lambda+2\omega-2J'<\mu<\lambda+2J' & (\mbox{II-III-IV})\\
-\lambda+2\omega-2J'<\mu<\lambda & (\mbox{II$_1$-III-IV})\\
\xi<\mu<\lambda & (\mbox{II$_2$-III-IV})\\
-\xi<\mu<\lambda & (\mbox{II$_3$-III-IV})
\end{array}$\\
\hline
 $\begin{array}{c}
    IV
   \end{array}$       
&
 $\begin{array}{l}
    J'=0\\\rowcolor[gray]{0.8}J'<0\\\rowcolor[gray]{0.8}\\J'>0\\ 
   \end{array}$       
&
$\displaystyle\left.
\begin{array}{l}
\\ \\ \\\\
\end{array}\right\}
\begin{array}{l}
    {E}(\mbox{IV})=-4\mu-J'
   \end{array}   $
&
$\begin{array}{l}
|\mbox{IV}\rangle=\prod_{k=1}^{Nq/2}|\pm1\rangle_{\sigma_{k1}}\otimes 
\left[\hat{c}^\dagger_{k1,\uparrow}\hat{c}^\dagger_{k1,\downarrow}\hat{c}^\dagger_{k2,\uparrow}\hat{c}^\dagger_{k2,\downarrow} \right]
|0,0\rangle_{k}\otimes|\pm1\rangle_{\sigma_{k2}}\\
\rowcolor[gray]{0.8}
|\mbox{IV}\rangle=\prod_{k=1}^{Nq/2}|1\rangle_{\sigma_{k1}}\otimes\left[\hat{c}^\dagger_{k1,\uparrow}\hat{c}^\dagger_{k1,\downarrow}\hat{c}^\dagger_{k2,\uparrow}\hat{c}^\dagger_{k2,\downarrow} \right] |0,0\rangle_{k}\otimes|-1\rangle_{\sigma_{k2}}\\
\rowcolor[gray]{0.8}\\
|\mbox{IV}\rangle=\prod_{k=1}^{Nq/2}|1\rangle_{\sigma_{k1}}\otimes\left[\hat{c}^\dagger_{k1,\uparrow}\hat{c}^\dagger_{k1,\downarrow}\hat{c}^\dagger_{k2,\uparrow}\hat{c}^\dagger_{k2,\downarrow} \right] |0,0\rangle_{k}\otimes|1\rangle_{\sigma_{k2}}\\
\end{array}$
&  
$\begin{array}{ll}
\mu>\lambda & (\mbox{III-IV})\\
\rowcolor[gray]{0.8}\mu>\lambda+2J' & (\mbox{III-IV})\\
\rowcolor[gray]{0.8}\mu>\omega & (\mbox{II-IV})\\
\mu>\lambda & (\mbox{III-IV})
\end{array}$\\
\\
\hline
\label{tab2}
\end{tabular}
}\captionof{table}{The list of eigenvalues, eigenvectors and border expressions for different phases from the ground-state phase diagrams corresponding to the investigated spin-electron model (\ref{eq1}). Constants $a$, $b$, $c$ and $d$ used in the notation of eigenvectors II and II$_1$ have the following form: $a=\frac{1}{2}\left(\frac{J}{\sqrt{J^2+t^2}}+1\right)$, $b=\frac{1}{2}\left(\frac{J}{\sqrt{J^2+t^2}}-1\right)$ and $c=d=\frac{t}{2\sqrt{J^2+t^2}}$. The notation $\lambda$, $\omega$ and $\xi$ occurring in expressions for eigenvalues and border expressions are equal to  $\lambda=J+t$, $\omega=\sqrt{J^2+t^2}$ and $\xi=J-t$.}
\end{center}
\end{minipage}
\end{landscape}
\end{document}